\documentclass[letter,12pt]{article}
\pdfoutput=1
\usepackage{amsfonts}
\usepackage{color}
\usepackage[top=0.8in, bottom=0.6in, left=1.2in, right=1.2in]{geometry}
\usepackage{bbm}
\usepackage{stmaryrd}
\usepackage{graphicx}
\usepackage{float}
\graphicspath{{./figure_tex/}}
\usepackage{authblk}
\usepackage{appendix}

\newcommand{\beq}{\begin{equation}}
\newcommand{\eeq}{\end{equation}}
\newcommand{\beqr}{\begin{eqnarray}}
\newcommand{\eeqr}{\end{eqnarray}}
\newcommand{\beqrn}{\begin{eqnarray*}}
\newcommand{\eeqrn}{\end{eqnarray*}}
\newcommand{\beqn}{\begin{equation*}}
\newcommand{\eeqn}{\end{equation*}}
\newcommand{\bei}{\begin{itemize}}
\newcommand{\beii}{\begin{itemize} \item}
\newcommand{\eei}{\end{itemize}}
\newcommand{\ben}{\begin{enumerate}}
\newcommand{\een}{\end{enumerate}}
\newcommand{\bes}{\begin{small}}
\newcommand{\ees}{\end{small}}
\newcommand{\bec}{\begin{center}}
\newcommand{\eec}{\end{center}}

\newcommand{\om}{\omega}

\newcommand{\eps}{\epsilon}

\newcommand{\IF}{IF }

\usepackage{amsmath, amsthm, amssymb}
\usepackage{graphics}
\usepackage{graphicx}
\usepackage{epstopdf}

\usepackage{rotating}
\usepackage{fullpage}
\usepackage{natbib}  
 \usepackage{color}

\newcommand{\EV}{\mathbf{E}} 
\newcommand{\bfy}{\mathbf{y}} 
\newcommand{\bfI}{\mathbf{I}} 
\newcommand{\bfK}{\mathbf{K}} 
\newcommand{\bfJ}{\mathbf{J}} 
\newcommand{\bfC}{\mathbf{C}}
\newcommand{\bfA}{\mathbf{A}}
\newcommand{\bfW}{\mathbf{W}}
\newcommand{\bfX}{\mathbf{X}}
\newcommand{\bfY}{\mathbf{Y}}
\newcommand{\bfH}{\mathbf{H}}
\newcommand{\bfL}{\mathbf{L}}
\newcommand{\bfD}{\mathbf{D}}
\newcommand{\bfM}{\mathbf{M}}
\newcommand{\bfU}{\mathbf{U}}
\newcommand{\bfu}{\mathbf{u}}
\newcommand{\bfone}{\mathbf{1}}

\newcommand{\bfrho}{{\boldsymbol\rho}}
\newcommand{\bfTheta}{{\boldsymbol\Theta}}
\newcommand{\rhoout}{\rho^{\mathrm{output}}}

\newcommand{\yt}{\tilde{y}}
\newcommand{\At}{\tilde{A}}
\newcommand{\Ft}{\tilde{F}}

\newcommand{\St}{\tilde{S}}
\newcommand{\Ct}{\tilde{C}}

\newcommand{\bfKt}{\mathbf{\tilde{K}}} 
\newcommand{\bfCt}{\mathbf{\tilde{C}}}
\newcommand{\bfAt}{\mathbf{\tilde{A}}}
\newcommand{\bfJt}{\mathbf{\tilde{J}}}
\newcommand{\bfFt}{{\mathbf {\tilde{F}}}}

\newcommand{\cov}{\mathrm{cov}}
\newcommand{\var}{\mathrm{var}}

\newcommand{\EVs}[1]{\EV\left[#1\right]} 
\newcommand{\EVE}[1]{\EV_e\hspace{-0.04in}\left[#1\right]} 

\newcommand{\dout}{d^{\mathrm{out}}}
\newcommand{\din}{d^{\mathrm{in}}}
\newcommand{\qdiv}{q_{\mathrm{div}}}
\newcommand{\qch}{q_{\mathrm{ch}}}
\newcommand{\qcon}{q_{\mathrm{con}}}
\newcommand{\rhoavg}{\rho^{\mathrm{avg}}}
\newcommand{\rhoin}{\rho^{\mathrm{input}}}

\newcommand{\Qdiv}{Q_{\mathrm{div}}}
\newcommand{\Qch}{Q_{\mathrm{ch}}}
\newcommand{\Qcon}{Q_{\mathrm{con}}}

\newcommand{\bfQdiv}{\mathbf{Q}_{\mathrm{div}}}
\newcommand{\bfQch}{\mathbf{Q}_{\mathrm{ch}}}
\newcommand{\bfQcon}{\mathbf{Q}_{\mathrm{con}}}

\newcommand{\ER}{Erd\"{o}s-R\'{e}nyi}

\newtheorem{theorem}{Theorem}[section]
\newtheorem{lemma}[theorem]{Lemma}
\newtheorem{proposition}[theorem]{Proposition}

\begin{document}
\title{Motif Statistics and Spike Correlations in Neuronal Networks}
\date{\today}
\author[1]{Yu Hu}
\author[2]{James Trousdale}
\author[2,3]{Kre\v{s}imir Josi\'{c}}
\author[1,4]{Eric Shea-Brown}
\affil[1]{Department of Applied Mathematics, University of Washington, Seattle, WA 98195}
\affil[2]{Department of Mathematics, University of Houston, Houston, TX 77204-5001}
\affil[3]{Department of Biology and Biochemistry, University of Houston, Houston, TX 77204-5001}
\affil[4]{Program in Neurobiology and Behavior, University of Washington, Seattle, WA 98195}

\maketitle

\begin{abstract}
Motifs are patterns of subgraphs of complex networks. We studied the impact of such patterns of connectivity on the level of correlated, or synchronized, spiking activity among pairs of cells in a recurrent network model of integrate and fire neurons.  For a range of network architectures, we find that the pairwise correlation coefficients, averaged across the network, can be closely approximated using only three statistics of network connectivity.  These are the overall network connection probability and the frequencies of two second-order motifs:  diverging motifs, in which one cell provides input to two others, and chain motifs, in which two cells are connected via a third intermediary cell.  Specifically, the prevalence of diverging and chain motifs tends to increase correlation. Our method is based on linear response theory, which enables us to express spiking statistics using linear algebra, and a resumming technique, which extrapolates from second order motifs to predict the overall effect of coupling on network correlation.  Our motif-based results seek to isolate the effect of network architecture perturbatively from a known network state. 
\end{abstract}

\tableofcontents

\section{Introduction}

Neural networks are highly interconnected:  a typical neuron in mammalian cortex receives on the order of a thousand inputs.  The resulting collective spiking activity is characterized by {\it correlated} firing of different cells. Such correlations in spiking activity are the focus of a great deal of theoretical and experimental work.  This interest arises because correlations can strongly impact the neural coding of information, by introducing redundancy among different (noisy) neurons, allowing noise cancellation effects, or even serving as additional ``channels" of information \citep{Zoh:94,Sin+95,Shadlen:1998,PanzeriSTR99,abbott:99,sompolinsky01,Pan+01,Sch+03,Lat+05,JosicSDR08,Beck:2011}.  Correlations can also help gate the transfer of signals from one brain area to another~\citep{Salinas:2000,Fries:2005,Bruno:2011}, and determine the statistical vocabulary of a network in terms of the likelihood that it will produce a particular spike patterns in its repertoire~\citep{Sch:2006,Shlens:2006}.  

Thus, the possible effects of correlations on the encoding and transmission of signals are diverse.  Making the situation richer still, correlations can depend on characteristics of the ``input" signals (or {\it stimuli}) themselves.  The result is an intriguing situation in which there is no simple set of rules that determines the role of correlations -- beneficial, detrimental, or neutral -- in neural computation.  Moreover, despite major progress, the
pattern and strength of correlations in neuronal networks~\emph{in vivo} remain under debate~\citep{Ecker:2010,Cohen:2011}.
This demands tools that will let us predict and understand correlations and their impact in specific types of neural circuits.

In this paper, we take one step toward this goal:  We determine how simple statistics of network  connectivity contribute to the average level of correlations in a neural population.  While other factors such as patterns of correlations in upstream neural populations and the dynamical properties of single cells contribute in important ways, we seek to isolate the role of connection statistics in the most direct way possible.  We return to the question of how our results on connectivity might be combined with these other factors at several places in the text, and in the Discussion.

We define the statistics of network connectivity via {\it motifs}, or subgraphs that are the building blocks of complex networks.  While there are many ways to characterize connectivity, we choose the motif-based approach for two main reasons.  First, we wish to follow earlier work in theoretical neuroscience in which the frequency of small motifs is used to define low-order statistics of network adjacency matrices~\citep{Zhao:2011}.  Second, recent experiments have characterized the frequencies with which different motifs occur in biological neural networks, and found intriguing deviations from what we would expect in the case of unstructured, random connectivity~\citep{Song:2005, Haeusler:2009}.  

\begin{figure}[H] 
	\centering     
	\includegraphics[width=1.5in]{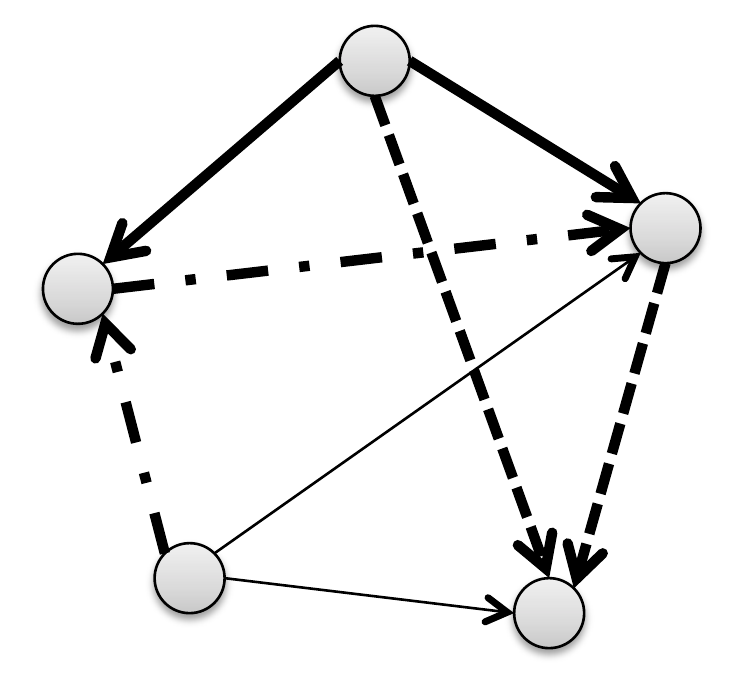} 
	\includegraphics[width=4in]{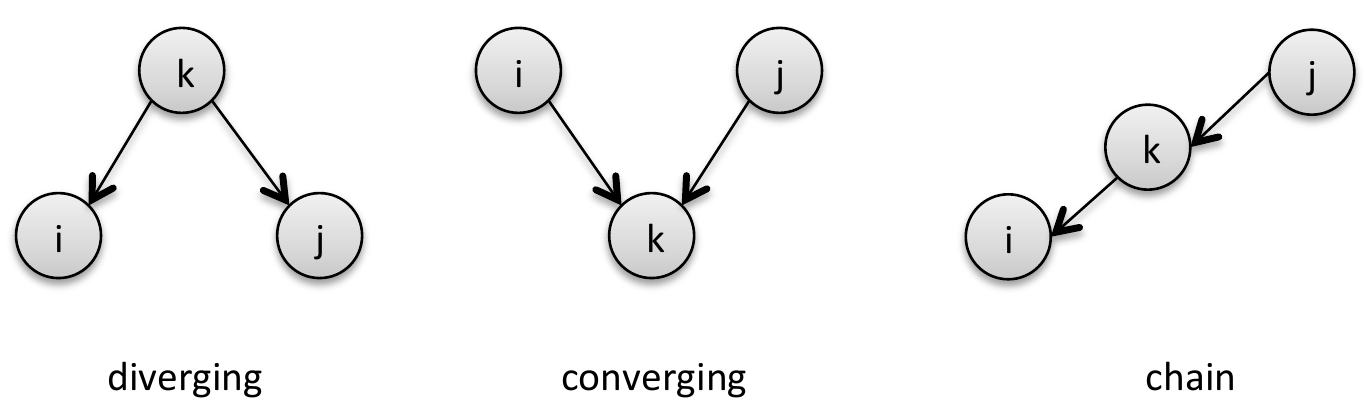}  
		\caption{(Left) Counting motifs in a network. Example of a diverging motif in the network is shown in bold solid line. Similarly a converging and a chain motif are shown in dashed line and dash-dotted lines, respectively. In total, the network has 8 connections, 6 diverging, 7 converging and 5 chain motifs. (Right) The different types of second order motifs.}
		\label{motif2}
\end{figure}

Fig.~\ref{motif2} depicts the motifs we will consider: a single cell projecting to two downstream cells (diverging motif), a single cell receiving inputs from two upstream cells (converging motif), and a series of two connections in and out of a central cell (chain motif).  To assess how prevalent these motifs are in a given network, we count their occurrences, and compare the observed motif counts with those expected in a random graph~\citep{Song:2005}. This is a {\it regular} network which has the same total number of connections, and in which each cell has the same, evenly divided number of incoming and outgoing connections -- i.e, the same {\it in and out degree}. (Importantly, the (relative) motif counts for such regular graphs agree with those in the classical model of random graph, the \ER{} model in  the limit of large network size; thus, when we refer to the prevalence of network motifs, this means in comparison to either a regular or \ER{} graph.) 

\begin{figure}[H]
	\centering     
	\includegraphics[width=3in]{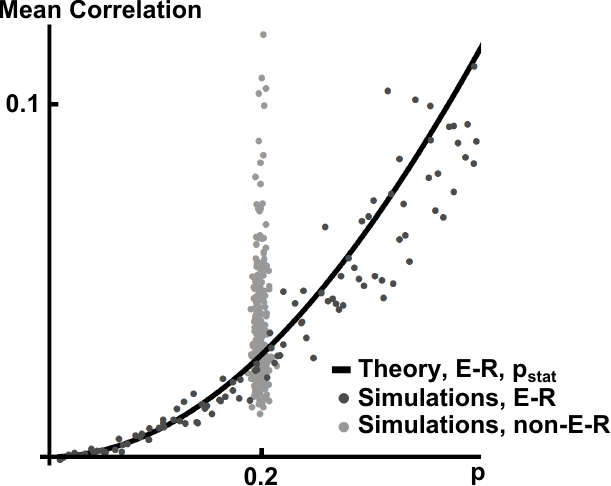}
	\caption{Impact of changing motif frequencies on mean correlation in non-\ER{} networks (gray dots) compared with the effect of changing connection probability in \ER{} networks (black dots). Mean correlation coefficients (averaged over all cell pairs in the network) are plotted against connection probability $p_{\mathrm{stat}}$ in the \ER{} model. Keeping $p_{\mathrm{stat}}=0.2$, and varying second order motif frequencies strongly affects average correlations. The curve presents the theoretical predictions from Eq.~\eqref{E:twopop_nl_resumming_ER}.  For network parameters see Fig.~\ref{first figure}.}  
		\label{figure0}
\end{figure}

Fig.~\ref{figure0} illustrates the importance of network motifs in determining the average correlation across the network.  Here, we simulate 265 different networks of excitatory and inhibitory, exponential integrate and fire cells~\citep{Day+01}. Importantly, we bias each neuron so that it fires with the same rate, regardless of its connectivity. We explain in more detail below that this is important to isolate the effects of network connectivity alone.
The black dots show the average correlation for \ER{} networks that have different connection probabilities $p$.   As expected, correlations increase with connection probability.  Next, the grey dots show the average correlation in networks that all have the same connection probability ($p=0.2$), but with a different prevalence of motifs compared to the corresponding \ER{} model.  Interestingly, the range of correlation values obtained at this fixed connection probability $p$ is as large as that obtained in the \ER{} networks over a range of $p$ values that reaches to roughly twice the connectivity.  Thus, motifs -- over and above net connectivity -- play a strong role in determining the strength of correlations across a network.

\begin{figure}[H] 
	\centering     
	\includegraphics[width=6in]{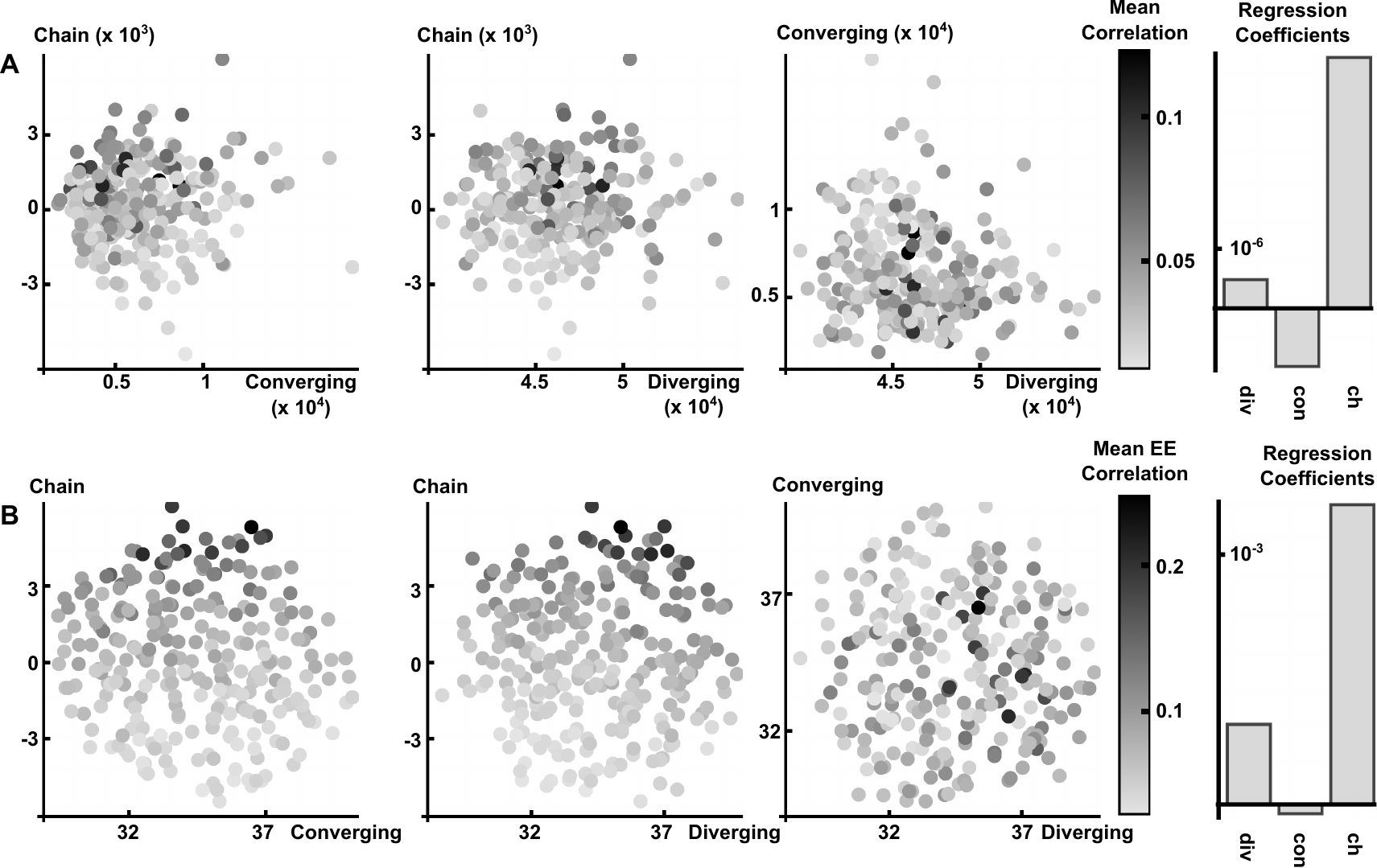}  
		\caption{Mean correlation as a function of motif frequency in networks of excitatory and inhibitory neurons. {\bf A} Correlation coefficient averaged over \emph{all} cell pairs as a function of total motif counts.  {\bf B} Correlation coefficients averaged over pairs of two excitatory cells, as a function of weighted sums of subgroup motif counts (see Fig.~\ref{f:motif_ei}  and Eq.~\eqref{E:twopop_resummed_exp}). The linear combination coefficients are determined by the resumming theory (see Sec.~\ref{EI_resum}). The network consists of 51 excitatory neurons and 49 inhibitory neurons with excitatory and inhibitory coupling strengths set at 22.82 mV$\cdot$ms or -22.82 mV$\cdot$ms. All neurons have the same uncoupled cellular dynamic parameters (see Eq.~\eqref{IF}): $\tau_{i}=20, v_{th}=20, v_{r}=-60, \tau_{ref}=2, E_{L,i}=-60, E_{i}=8, \sigma_i^2=12, v_T=-53,\Delta_T=1.4,\tau_{S,i}=5, \tau_{D,i}=10$. Spike count correlation coefficients were calculated using a 500ms window size. These graphs are re-sampled from 4096 degree distribution generated graphs using Latin hypercube sampling based on parameter space defined by axis in panel {\bf{B}} over a reliably sampled region.  In panel {\bf{A}}, some of the motifs are given a ``positive" count -- i.e., if they involve two excitatory or two inhibitory connections in a chain -- and others are given a negative count.  For example a converging motif with one excitatory presynaptic cell and one inhibitory presynaptic cell will contribute negatively.}
	\label{first figure}
\end{figure}

But which motifs contribute?  And are the second-order motifs of Fig.~\ref{motif2} sufficient, or must higher-order motifs (involving four or more cells) be included to understand average correlations?  Fig.~\ref{first figure} suggests part of the answer. Here we used the same 265 network samples represented by the gray dots in Fig.~\ref{figure0}. Connection probability was kept at $p \approx 0.2$, but frequencies of second order motifs differed.  First, panel {\bf{A}} of Fig.~\ref{first figure} shows the dependence of  average network correlations on the total motif counts in the network.  Disappointingly, no clear relationship is evident.

A more careful approach is presented in panel {\bf{B}} of Fig.~\ref{first figure}. First,  we plot only the mean correlation among cell pairs of a given type -- here, excitatory cell pairs.  Second, we do not lump together all motifs shown in Fig.~\ref{motif2}.  Instead, we calculate weighted motif counts that are classified according to the types of constituent neurons. The three panels now exhibit a much clearer trend:  Correlation levels vary systematically with weighted motif counts for the diverging and chain motifs, while there is no clear dependence on the converging motif.  This improved linear fit is quantified by the coefficient of determination $R^2$ (as usual, the squared correlation coefficient), for the mean correlation among excitatory cells regressed against the motif counts.

Our goal in the balance of this paper is to explain two aspects of the relationship suggested in panel {\bf{B}} of Fig.~\ref{first figure}.  First, we show how to derive analytically a relationship between motif counts and network-wide correlation that successfully identifies these trends.  We previously built on the work of~\cite{Lindner:2005} to derive an explicit expression for the pairwise correlation 
between cells in a network~\citep{Trousdale:2012}; close analogs of this expression have been derived via related approaches for Hawkes processes~\citep{Hawkes:1971-2,Pernice:2011,Pernice:2012}. This expression shows how patterns of network connectivity, together with 
the dynamical properties of single cells, shape correlations.   Here we extend our previous work to show how the same expression can reveal
the impact of motifs on correlations.

Secondly, we use this analytical approach to obtain a result that is not readily apparent from Fig.~\ref{first figure}.  
We show that -- for a range of different network models (see Sec.~\ref{s:graph} ) -- the average network correlation can be closely approximated using the connection probability and frequencies of motifs that involve {\it only two} connections between cells (See Fig.~\ref{motif2}).  Specifically, we show that the prevalence of diverging and chain motifs tends to increase average correlation (in excitatory only networks), while converging motifs have no effect (in either single population or multi-population networks) . 

\par
The paper is structured as follows.
We explain in Sec.~\ref{S:linresponse} our  setup and introduce the linear response theory which is the basis of our analysis. Sec.~\ref{s:graph} defines the three types of motifs and their frequencies, and discusses the classes of networks to which we test and apply our theory. For simplicity, we first demonstrate our analysis in a case of a population of a single population of excitatory cells in Sec.~\ref{S:single_pop_theory}, and then generalized to interacting populations of excitatory and inhibitory cells  in Sec.~\ref{S:two_pop_theory}. We conclude by discussing a crucial assumption of our network model (Sec.~\ref{heter_theory}), and compare our theory with simulations of integrate and fire neuron networks (see Sec.~\ref{s:sim_theory}).  Biological interpretations, limitations, and extensions are covered in our Discussion, and an appendix and table of notation follow.

\section{Neuron models, cross-spectra, and measures of collective network activity} \label{S:linresponse}

In this section we describe a model spiking network composed of integrate and fire (IF) neurons.   We introduce the measures that we will use to quantify the 
level of correlations between cell pairs in the network.  The analytical 
approximations of these correlations given in Eq.~\eqref{E:lindner} will be the basis for the subsequent analysis. 
While we develop the theory for networks of \IF neurons, the main
ideas are applicable to many other models (e.g. Hawkes model).  We return to this point in the Discussion.

\subsection{Networks of integrate-and-fire neurons}

In a network of {\it integrate-and-fire} (IF) units, the dynamics of each cell are described by a single membrane voltage variable $v_i$, which satisfies
\begin{equation} \label{IF}
\tau_i\dot{v_i} = -(v_i-E_{L,i}) + \psi(v_i) + E_i + \sqrt{\sigma_i^2\tau_i}\xi_i(t) + f_i(t) \;.
\end{equation}
Here, $E_{L,i}$ is the leak reversal potential, $\psi(v_i)$ is a spike generating current, and
$E_i$ is the mean external synaptic input from sources not modeled.   In numerical simulations we use the exponential integrate-and-fire (EIF) model~\citep{FourcaudTrocme:2003}, so that  
$\psi(v) \equiv \Delta_T \exp[(v-v_T)/\Delta_T]$.  Each cell has independent fluctuations due to internal noise and external inputs (e.g., from a surrounding network that is not explicitly modeled). We describe such effects by additive terms, $\sqrt{\sigma_i^2\tau_i}\xi_i(t)$, which are Gaussian white noise processes, ~\citep{White:2000,Renart:2004}. Synaptic input to cell $i$ from other cells in the network is denoted by $f_i(t)$ (see below).

When the membrane potential reaches a threshold, $v_{th}$, an action potential, or {spike}, is generated; the membrane potential is then reset to a lower voltage
$v_r$ and held there for an absolute refractory period $\tau_r$.  We denote the time at which the $j^{th}$ neuron fires its $k^{th}$ spike as $t_{j,k}$; taken together, these define this neuron's {\it spike train} 
$
y_j(t) = \sum_k \delta(t-t_{j,k}) \;.
$
Importantly, synaptic interactions among neurons are initiated at these spike times, so that the total synaptic input to the $i^{th}$ cell is
\begin{equation*} 
f_i(t) = \sum_j ( \bfJ_{ij} * y_j)(t),\qquad\text{where}\qquad \bfJ_{ij}(t) = \begin{cases} 
\bfW_{ij} \left(\frac{t-\tau_{D,j}}{\tau_{S,j}^2}\right)\exp\left[-\frac{t-\tau_{D,j}}{\tau_{S,j}}\right]&\qquad t \geq \tau_{D,j} \\ 
0 & \qquad t < \tau_{D,j}
\end{cases}.
\end{equation*}
In the absence of 
a synaptic connection from cell $j$ to cell $i$, we set $\bfW_{ij} =0$. Hence,  
the $N \times N$ matrix of synaptic weights, $\bfW$, defines a directed, weighted network. 

In simulations we used the parameters  given in  the caption of Fig.~\ref{first figure}, unless stated otherwise.

\subsection{Measure of network correlation}\label{cc_def}

Dependencies between the responses of cells in the network may be quantified using the spike train auto- and cross--correlation functions~\citep{Gabbiani:2010}. For a pair
of spike trains $y_i(t),y_j(t)$, the cross-covariance function is given by
$$
\bfC_{ij}(\tau) = \EVs{(y_i(t)-r_i) (y_j(t + \tau)-r_j)},
$$
where $r_i$ and $r_j$ are the firing rates of the two cells. Here and throughout the paper we assume that the spike trains form a (multivariate) stationary process.  The auto-correlation function is the cross-correlation of the output of the cell and itself, and $\bfC(t)$ is the matrix of cross-correlation functions. The matrix $\bfCt(\omega)$ with entries defined by
$$
\bfCt_{ij}(\omega) = \EVs{\yt_i(\omega)\yt_j^*(\omega)},
$$
is the matrix of cross-spectra (see next section). The cross-spectrum of a pair of cells is equivalent to the Fourier transform of their cross-correlation function.~\citep{Stratonovich:1967}

Denote by $N_{y_i}(t_1,t_2) = \int_{t_1}^{t_2}y_i(s)ds$ the spike count of cell $i$ over a time window $[t_1,t_2]$. The spike count correlation $\bfrho_{ij}(T)$ over windows of length $T$ is defined as
$$
\bfrho_{ij}(T) = \frac{\cov\left(N_{y_i}(t,t+T),N_{y_j}(t,t+T)\right)}{\sqrt{\var\left(N_{y_i}(t,t+T)\right) \var\left(N_{y_j}(t,t+T)\right)}} \; .
$$
We will make use of the total correlation coefficient $\bfrho_{ij}(\infty) = \lim_{T\rightarrow\infty} \bfrho_{ij}(T)$ which captures dependencies between the processes $y_i,y_j$ over arbitrarily long timescales, but may also describe well the nature of dependencies over reasonably short timescales~\citep{Bair:2001,delaRocha:2007,SheaBrown:2008,Rosenbaum:2011}.
The spike count covariance is related directly to the cross-correlation function by~\citep{Tetzlaff:2008}
$$
\cov\left(N_{y_i}(t,t+T),N_{y_j}(t,t+T)\right) = \int_{-T}^T \bfC_{ij}(\tau)(T - |\tau|)d\tau.
$$
Thus, total correlation may be defined alternatively in terms of the integrated cross-correlations (equivalently, the cross-spectra evaluated at $\omega = 0$):
\beq \label{rho_def}
\bfrho_{ij}(\infty) = \frac{\bfCt_{ij}(0)}{\sqrt{\bfCt_{ii}(0)\bfCt_{jj}(0)}}.
\eeq

Throughout this paper, we will use $\bfrho_{ij}(\infty)$ as the measure of correlation and use the above equation to calculate it from the cross-spectra matrix~\citep{Pernice:2011}. However, our analysis can be similarly applied to study $\bfCt_{ij}(\om),$ and hence the entire correlation function in time~\citep{Trousdale:2012}.

\subsection{Linear response approximation of cell response covariance}
\label{s:linear_response_1}

Linear response theory~\citep{Gabbiani:2010,Risken:1996} can be used 
to approximate the response of single cells, and the joint response of cells in a network~\citep{Lindner:2001,Brunel:2001,Lindner:2005,Trousdale:2012}.  Consider an \IF neuron obeying Eq.~\eqref{IF}, but with the mean of the inputs $f(t)$ absorbed into the constant $E$.  We denote the remaining, \emph{zero-mean} input by $\epsilon X(t)$, so that
\begin{equation}\label{E:v_theory}
\tau \dot{v} = -(v-E_L)+ \psi(v) + E + \sqrt{\sigma^2\tau} \xi(t) + \epsilon X(t) \;.
\end{equation}
For fixed input fluctuations $\epsilon X(t)$, the output spike train will be different for each realization of the noise $\xi(t)$, and each initial condition $v(0)$. The 
time-dependent firing rate is obtained by averaging the resulting spike train over noise realizations and a stationary distribution of initial conditions.  For all values of $\epsilon$, this stationary distribution is taken to be the one obtained when $\eps=0$.  We denote the resulting averaged firing rate as
$r(t) = \langle y(t) \rangle$.  Linear response theory approximates this firing rate as
\begin{equation*}
r(t) = r_0 + (A*\epsilon X)(t),
\end{equation*}
where $r_0$ is the firing rate in the absence of input ($\eps=0$), and the linear response kernel, $A(t),$ characterizes  the  response to first order in $\epsilon$.  This  approximation is remarkably accurate over a wide range of parameters; for example, see~\citep{Ostojic:2009,Richardson:2009}.

Next, we turn to the problem of approximating the output of a cell on a single trial, rather than the average across trials.  We denote the Fourier transform of a function $f$ by $\tilde{f} = \mathcal{F}(f)$. However, for spike trains we adopt a convention that $\tilde{y}_i(\om)$ is the Fourier transform of the mean subtracted spike train $y_i(t)-r_i$.  Following~\citep{Lindner:2001,Lindner:2005,Trousdale:2012} we approximate the spiking output of a cell self-consistently by
\begin{equation} \label{E:ansatz}
\yt_i(\omega) \approx \yt_i^0(\omega) + \At_i(\omega) \left(\sum_j \bfJt_{ij}(\omega) \yt_j(\omega) \right),
\end{equation}
where $\yt_i^0(\omega)$ is a realization of the output of cell $i$ in the absence of input. Defining the 
interaction matrix $\bfKt$ with entries $\bfK_{ij}(t) \equiv (A_i*\bfJ_{ij}) (t)$, 
we can use Eq.~\eqref{E:ansatz} to solve for the vector of Fourier transformed spike train approximations
\beq
\label{e.linsol}
\yt(\omega) = (\bfI - \bfKt(\omega))^{-1}\yt^0(\omega),
\eeq
and  matrix of cross-spectra
\begin{equation} \label{E:lindner}
\begin{split}
\bfCt (\omega) \approx \bfCt^\infty(\omega) & = (\bfI - \bfKt(\omega))^{-1} \langle \yt^0(\omega) \yt^{0*}(\omega) \rangle (\bfI - \bfKt^*(\omega))^{-1} \\
&= (\bfI- \bfAt \bfW \bfFt)^{-1}\bfCt^0(\bfI- \bfFt^* \bfW^T \bfAt^* )^{-1} \; .
\end{split}
\end{equation}
Here $\bfAt, \bfCt^0, \bfFt$ are diagonal matrices: $\bfAt_{ii}(\omega) = \At_i(\omega)$ is the linear response
of cell $i$,  $\bfCt^0_{ii}(\omega) = \Ct^0_i(\omega) = \langle \yt_i^0(\omega) \yt_i^{0*}(\omega)\rangle$ is its ``unperturbed" (i.e., without coupling) power spectrum, and $\bfFt_{ii}(\omega) = \Ft_i(\omega)$ is the Fourier transform of the synaptic coupling
kernel from cell $i$. As noted later, our results and analysis will hold at all frequencies and thus can be used to study correlations at all timescales. The weighted connectivity matrix $\bfW$, defines the structure of the network.

To simplify the exposition, we initially assume certain symmetries in the network. For instance, we consider {\it homogeneous networks} in which cells have identical (unperturbed) power spectra, linear response functions, and synaptic kernels.  In this case the diagonal matrices in Eq.~\eqref{E:lindner} act like scalars.  We slightly abuse notation in this case,  and replace 
$\At_{i} (\om)$ by  $\At(\omega)$, and $\Ct^0_{i}(\omega)$ by $\Ct^0(\omega)$.   This  allows
us to disentangle the effects of network structure from the effects of neuronal responses on network activity.    The resulting cross-spectrum matrix (evaluated at $\omega=0$) is
\begin{equation}\label{cov_hom}
\bfCt^\infty = \Ct^0(\bfI-\At \bfW)^{-1}(\bfI-\At\bfW^{T})^{-1}
\end{equation}
(note that $\Ft_i(0)=1$ by definition).  We use this simpler expression in what follows, and return to heterogeneous networks in Sec~\ref{heter_theory}. 
Since we consider only total correlation, we omit the dependence of the spike count correlation coefficient on window size $T$. Additionally, all spectral quantities are evaluated at $\omega = 0$, so we also suppress the dependence on $\omega$. Finally, we define average network correlation by
\begin{equation}
\label{E:cc_norm}
\rhoavg=\frac{1}{N(N-1)}\sum_{i \neq j}^N \bfrho_{ij} \; \;.
\end{equation}

In subsequent sections, we will examine the  average covariance across the network
\beqn
\langle \bfCt^\infty \rangle=\frac{1}{N^2}\sum_{i j}^N \bfCt_{ij}^\infty \; \;.
\eeqn
This average cannot be directly related to that in Eq.~\eqref{E:cc_norm}, where individual summands
are normalized, and diagonal terms are excluded.  The motif-based theory we develop predicts
 $\langle \bfCt^\infty \rangle$, and gives no information about the specific entries $\bfCt^\infty_{ij}$.  However, $\rhoavg$ can be determined approximately from $\langle \bfCt^\infty \rangle$ alone.  We describe these approximations in Appendix~\ref{s:rho_approx}.

\subsection{Applicability of linear response theory}\label{stability}
\label{s: stability}

Our methods depend on two, related sets of conditions  for their validity.  First, we take  our cells to be driven by a white noise background. This background linearizes the response of the cells to sufficiently weak perturbations, improving the accuracy of the approximation \eqref{E:ansatz}; its presence is our first condition.

Second, turning to network effects, we assume that the spectral radius $\Psi(\bfKt) < 1$, which gives non-singularity of the approximating processes in Eq.~\eqref{E:ansatz} and allows us to make the series expansion we describe in Sec~\ref{s:E_trunc}. In practice,
we have found that the linear approximation to correlations will cease to provide an accurate approximation before this occurs, likely owing in part to
a failure of the perturbative approximation. Furthermore, the approximation seems to be most accurate at weak interaction strengths as characterized by a small radius of the bulk spectrum of $\bfKt$.

For \ER{} networks, \cite{Rajan:2006} derived an asymptotic (large $N$) characterization of the spectral radius of the synaptic weight matrix. 
In particular, for an \ER{} network consisting of only excitatory cells with synaptic weight $w$, there will be a single eigenvalue at $pNw$ with the remaining eigenvalues
distributed uniformly in a circle around the origin of radius $w\sqrt{p(1-p)N}$. In networks with both excitatory and inhibitory populations, there is a single outlier at 
$pN_Ew_E+pN_Iw_I$ and all other eigenvalues will be distributed (non-uniformly) within the circle of radius $\sqrt{p(1-p)(N_E w_E^2+N_I w_I^2)}$. We will use these
expressions for the bulk spectrum to quantify the strength of interactions given by the asymptotic spectral radius of $\bfKt = \At\bfW$ (referred to as the \ER{} spectral radius).

We note that we used \IF simulations to directly confirm the accuracy of the linear response approximation for excitatory-inhibitory networks in Fig.~\ref{sim_nrs}. For a complete discussion of the
performance of the linear response theory, see~\citep{Trousdale:2012}.

\section{Graphical structure of neuronal networks}\label{sec_graph}
\label{s:graph}
Our main goal is to determine how the small-scale statistical structure of directed networks influences the collective dynamics they produce -- namely, the strength of spike correlations in networks of model neurons.
We will quantify network structure using the probability of
connections between pairs and among triplets of cells, organized into {\it network motifs}~\citep{Song:2005,Zhao:2011}.

\subsection{Network motifs and their frequencies}

A motif is a subgraph composed of a small number of cells. We classify motifs according to the number of edges they contain.  We begin by considering directed networks composed of identical cells.  First order motifs contain one connection and hence come in only one type --- two cells with a one-way connection. Second order motifs contain two connections, and therefore involve at most three interacting cells.  These motifs come in three types: diverging, converging and chain  motifs (See Fig.~\ref{motif2})~\citep{Song:2005,Zhao:2011}.  (Note that in our definition, a cell can appear twice in the triplet of cells that define a second order motif.  For example, the chain motif in Fig.~\ref{motif2}
is equivalent to a bidirectionally coupled pair of cells when $i = j$.)

We will consider mainly the impact of second order motifs, over and above first order effects.  The three motifs shown in Fig.~\ref{motif2} arise naturally in our analysis of correlated spiking activity.  In particular, we will show that the {\it frequency} at which each motif occurs in the network can accurately predict levels of correlation across the network.

We next introduce notation that will allow us to make these ideas precise.  Let  $\bfW^0$ be the adjacency matrix, so that  $\bfW^0_{i,j} = 1$ indicates the presence of a directed connection from cell $j$ to cell $i$, and $\bfW^0_{i,j} = 0$ indicates its absence.
To quantify the frequency of a motif in a given graph, we first count the total number of times  the motif occurs, and divide by the total number of possible occurrences in a graph of that size.  For first order motifs this definition gives the  \emph{empirical} connection probability,
\beqn
p= \left( \sum_{i,j} \bfW^0_{i,j} \right)/N^2.
\eeqn

The preponderance of second order motifs is measured in two stages.  First, we similarly normalize the motif count.  Second, we subtract the value expected in a reference graph.  The resulting expressions are:
\beqr \label{motif_def}
\qdiv &=& \sum_{i,j,k}(\bfW^0_{i,k} \bfW^{0}_{j,k})/N^3-p^2 = \left( \sum_{i,j}(\bfW^0  \bfW^{0T})_{i,j}\right)/N^3-p^2,   \label{q1}  \\
\qcon&=& \left( \sum_{i,j}(\bfW^{0T} \bfW^{0})_{i,j}\right)/N^3-p^2,     \label{q2}   \\
\qch&=& \left( \sum_{i,j}(\bfW^0 \bfW^{0})_{i,j}\right)/N^3-p^2,  \label{q3}
\eeqr
where $\bfW^{0T}$ denotes the transpose of $\bfW^{0}$.  Consider the expression defining  $\qdiv$:  the sum in the first equality simply counts the total number of connections from one cell ($k$) to two others ($i$ and $j$), and divides by the total number of possible connections of this type ($N^3$).  This can be written as matrix multiplication followed by a sum over all entries $i,j$, as shown.  In each case we subtract the value $p^2$, which corresponds to the frequency of the motif in a regular graph, as well as the asymptotic frequency in an \ER{} graph as the number of cells,
$N,$ diverges to infinity.  Indeed, for \ER{} graphs any edge is present with chance $p$, and any second-order motif requires the presence of two edges.  Thus, $\qdiv$ corresponds to the propensity for a network to display diverging motifs, over and above expectations from an \ER{} or regular network.  The other measures in Eqns.~(\ref{q1}-\ref{q3}) have similar interpretations.

The quantities in Eqns.~(\ref{q1}-\ref{q3})  can also be interpreted as empirical measures of covariance \citep{Roxin:2011,Zhao:2011}. If we denote by $\EVE{\cdot}$ the empirical average over all entries in a given network, then we may write
$$
\qdiv =  \EVE{\bfW^0_{i,k} \bfW^{0}_{j,k}}- \EVE{\bfW^0_{i,k}}\EVE{\bfW^{0}_{j,k}}.
$$
Equality is attainable for $\qdiv$ and $\qcon$ (but not simultaneously).   We also note that the quantities defined in Eqns.~(\ref{q1}-\ref{q3}) are not independent (but do have three degrees of freedom). If we first sum over indices $i,j$ in Eq.~\eqref{motif_def}, we can rewrite the expression in terms of in and out degrees,  $\{\din_{k}\}, \{\dout_{k}\}$. For example,
\beq \label{div_degree}
\qdiv=\sum_k (\dout_k)^2/N^3-p^2 =\var(\dout)/N^2
\eeq
is the scaled sample variance of the out degree across the network. Similarly,
\beq \label{con_ch_degree}
\qcon=\var(\din)/N^2, \quad \qch =\cov(\dout, \din)/N^2,
\eeq
where $\cov(\cdot,\cdot)$ denotes sample covariance \citep{Zhao:2011}. Therefore,
\beq \label{e:q_dependence}
\qdiv \ge 0, \quad \qcon\ge 0, \quad |\qch | \leq \sqrt{\qdiv \qcon} \; .
\eeq
We further show in Appendix \ref{s:q_range} that
\beq \label{q_range}
| \qdiv |,| \qcon |, | \qch | \leq p(1-p).
\eeq
Eqns.~(\ref{e:q_dependence}-\ref{q_range}) identify the attainable ranges of motif frequencies.  In generating our networks, we compare the extent of motif frequencies that we produce using a particular scheme against this maximum possible range (see also Sec.~\ref{gen_netw}).

\begin{figure}[H]
	\centering
	\includegraphics[width=1.45in]{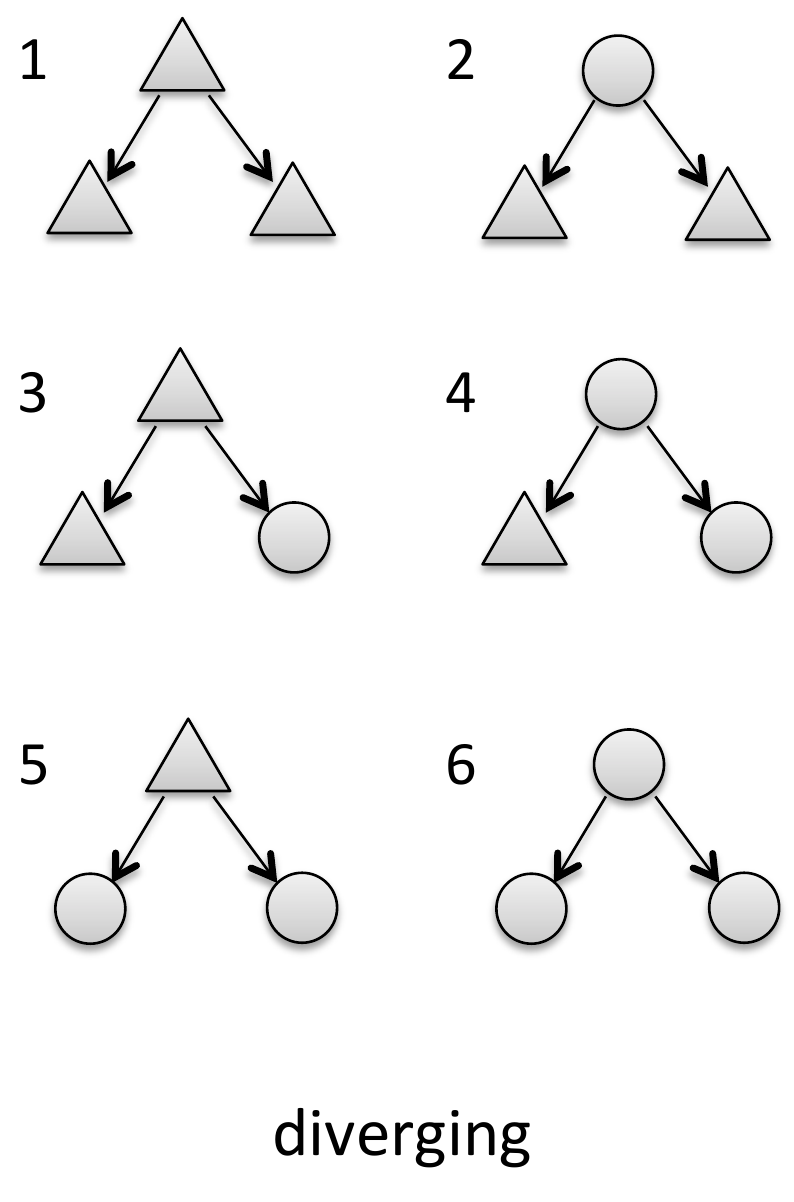}
	\includegraphics[width=1.5in]{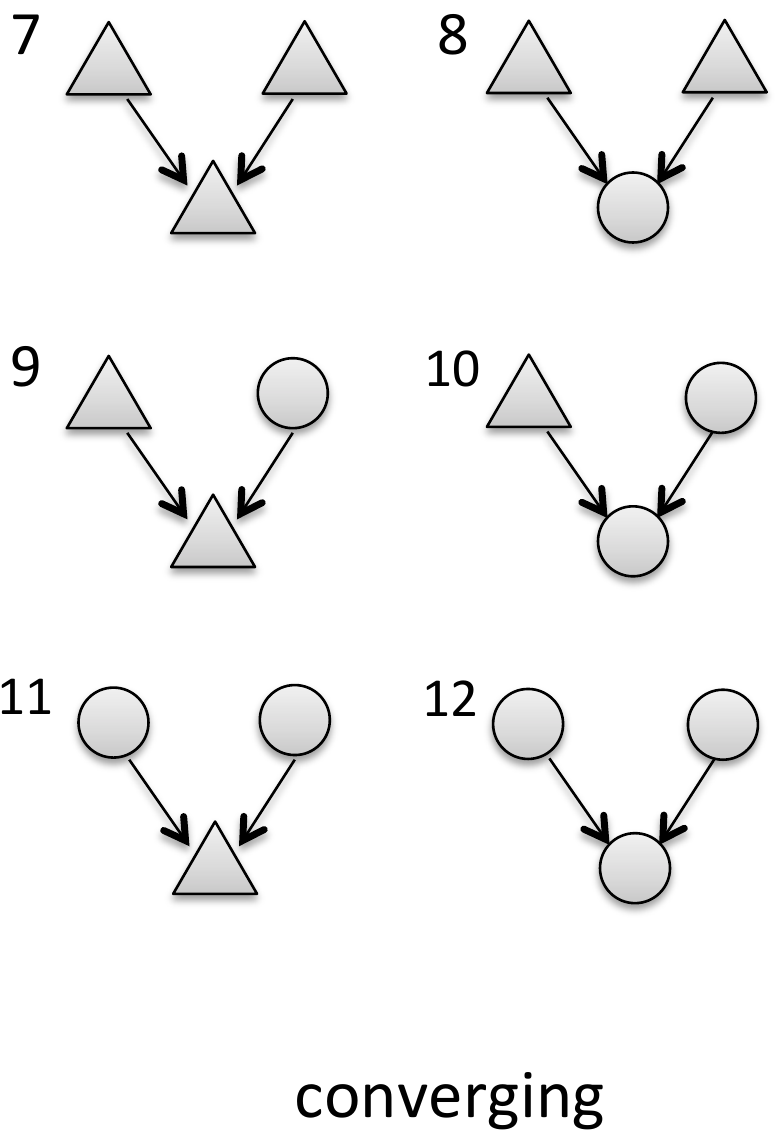}
	\includegraphics[width=1.5in]{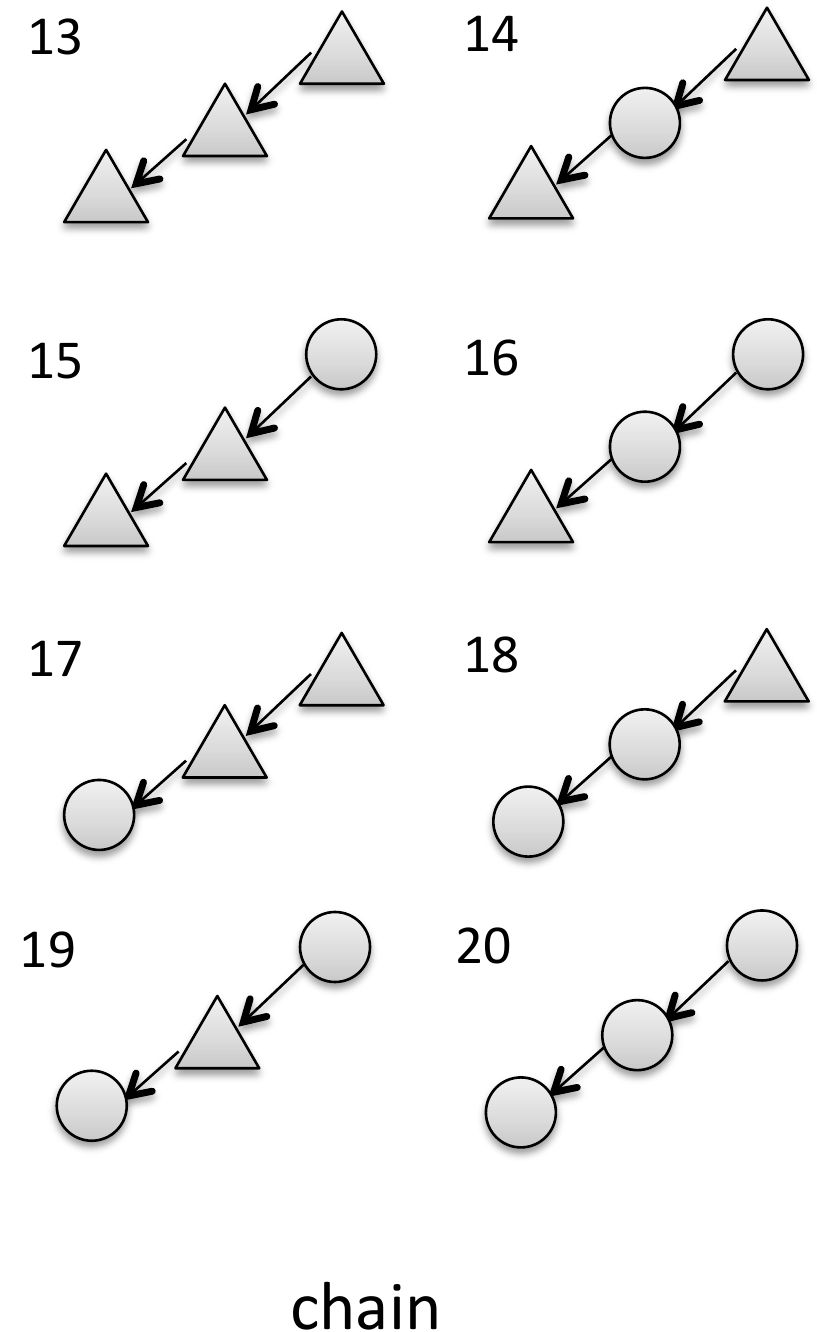}
	\caption{ Second order motifs in populations of excitatory and inhibitory cells: There are 20 subtypes of the 3 main motif types. Triangles represent excitatory neurons and circles for inhibitory neurons.
}
\label{f:motif_ei}
\end{figure}

In networks composed of excitatory and inhibitory cells, we can
represent interactions between cells using a signed
connectivity matrix. Edges emanating from inhibitory neurons are represented by negative entries and those from excitatory neurons by positive entries.
In this case, motifs are further subdivided  according to their constituent cells.  For instance, there are 6 distinct diverging  motifs, since if we list all $2^3$ group types of 3 cells, we see that the motif $E\leftarrow I \rightarrow I$ is the same as $I\leftarrow I \rightarrow E$, and $E\leftarrow E \rightarrow I$  is the same as $I\leftarrow E \rightarrow E$. Similarly, there are 6 distinct converging motifs, and 8 distinct chain motifs, for a total of 20 distinct second order motifs  (See Fig.~\ref{f:motif_ei}).

This will clearly lead to some cumbersome notation!  Therefore, while populations of interacting E and I cells are our ultimate goal, we first describe our ideas in a population of cells of a single type.

\subsection{Generating graphs with given motif frequency}\label{gen_netw}

To numerically examine the impact of motif frequency on dynamics, we need to
generate graphs that are equal in connection probability, but differ in the preponderance of second order motifs. The empirical connection probability in network samples will have small fluctuations around the statistical (i.e. expected) value we fixed. We use two ways of generating such graphs. The first is the degree distribution method \citep{Chung:2002} (related to configuration model \citep{Roxin:2011, Newman:2003,Newman:2001}). Here, following \citep{Zhao:2011} we use a two-sided power law, with various rising and decreasing exponents, peak locations and truncation limits, as expected in- and out-degree distributions. The other is the \emph{second order network} (SONET) method (for details see~\citep{Zhao:2011}). Network samples generated using both methods cover the range of motif frequencies observed experimentally in cortical circuits~\citep{Song:2005,Zhao:2011}. Naturally, this experimentally observed range is smaller  than the full extent of possibly attainable frequencies (see Eqns.~(\ref{e:q_dependence}-\ref{q_range}); however, the SONET method covers this full range as well. Details are given in Appendix~\ref{s:graph_gen}).

We use both methods to generate network samples in the excitatory only case and found similar results; here, we only show data generated using the SONET method as it covers a larger range of motif frequencies. In excitatory-inhibitory networks, we use the degree distribution method.

We emphasize that our approaches below do not \emph{a priori} specify any particular way of generating network samples.  However, their accuracy will depend on how this is done.  In particular, while we find that our approach is quite accurate for the family of networks that we described above, they can break down in networks that have significant additional structure, a point to which we will return.

\section{Impact of second order motifs in networks of excitatory cells}\label{S:single_pop_theory}

As we have shown in Section~\ref{S:linresponse}, linear response theory can be used to approximate cross-correlations between cells in a neuronal network.  We next explore how the key expression, given in Eq.~\eqref{cov_hom}, can be applied to relate the frequency of second order motifs to the average correlation across pairs of cells. For simplicity, we first consider networks consisting only of a single class of excitatory cells.  The results extend naturally to the case of two interacting populations of excitatory and inhibitory cells, as we show in Sec.~\ref{S:two_pop_theory}.

\subsection{Results: linear dependence between mean correlation and motif frequencies}

\begin{figure}[H]
\centering
\includegraphics[width=6in]{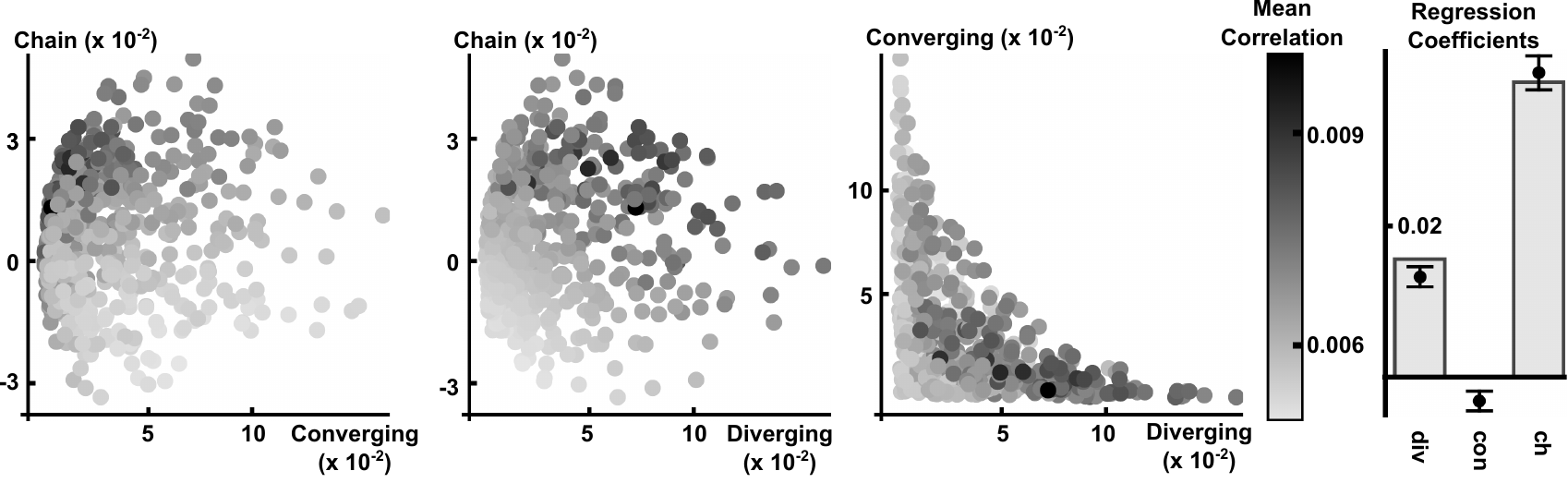}
\caption{ (Left) The relationship between second order motif frequencies and average correlation in purely excitatory networks. Each dot corresponds to a network sample and its shading represents the corresponding average correlation coefficient computed using Eq.~\eqref{E:lindner}. Each axis represents one of the three second order motif types as defined in  Eq.~(\ref{q1} --\ref{q2}). 
The effective coupling strength, characterized by the \ER{} spectral radius, was 0.2 for all networks (See Sec.~\ref{s: stability}). (Right) Bars show the linear regression coefficients calculated from the resumming theory (See Eq.~\eqref{E:single_pop_expansion}) and numerically (points) between motif frequencies and average network correlations. The error bars around each point denote  95\% confidence intervals for the regression coefficients. The coefficient of determination $R^2$  is 0.80. The data was obtained from 512 network samples generated using 
the SONET algorithm (See Sec.~\ref{gen_netw}). }
\label{F:E_bar}
\end{figure}

Fig.~\ref{F:E_bar} illustrates the relationship between second order motif frequency and average correlation in networks of excitatory cells.  In all examples in this section, we use networks of $N=100$ cells, with parameters as given in the caption of Fig.~\ref{first figure}, and with graph structures generated by the SONET algorithm (See Sec.~\ref{gen_netw}). Correlation coefficients between cells are computed using the linear response approximation given in Eq.~\eqref{E:lindner}.  We find that average correlations depend strongly on the frequency of diverging and especially chain motifs, but only weakly on the frequency of converging motifs. To quantify the linear fit, we use the coefficient of determination between the linear fit and the result of Eq.~\eqref{E:lindner}, obtaining $R^2=0.80$.  This high value suggests that the second order motifs are highly predictive of network correlation, in the simplest possible (linear) way (as explained in Appendix~\ref{s:adjust_q}, this prediction can be further improved when we compensate for the fluctuations in empirical connection probabilities due to the finite size of network samples).  In the balance of this section, we explain why this is the case,
derive via a resumming theory a nonlinear predictor
of mean correlation in terms of motif frequencies, and extract from
this an explicit linear relationship between the
probability of observing second order motifs and the mean correlation
in the network.

\subsection{First theory:  second order truncation for network correlation}\label{s:E_trunc}

If  the spectral radius $\Psi(\At \bfW) < 1$,  then the matrix inverses in Eq.~\eqref{cov_hom} can be expanded in a power series in $\At \bfW$ as 
\begin{equation}\label{E:power_series_exp}
\frac{\bfCt^\infty}{\Ct^0}= \sum_{i,j = 0}^\infty \At^{i+j}\bfW^i (\bfW^{T})^j
\end{equation}
\citep{Horn:1990}.
Terms in this expansion correspond naturally to paths through the graph defining the network~\citep{Trousdale:2012,Pernice:2011,Rangan:2009-1,Rangan:2009-2}. For example, the terms $(\At^2\bfW^2)_{ij}=\At^2 \sum_{k} \bfW_{ik}\bfW_{kj}$ give the contributions of length two chains between pairs of cells $i,j$. Entries in the matrices $\At^3\bfW^2\bfW^T$, $\At^3\bfW(\bfW^{T})^2$ give the contributions of diverging motifs of third order consisting of a projection from a single cell to a cell pair.  In this case, one branch of the motif is of length two, while the other is a direct connection.

Let $\bfone_{N_1N_2}$ denote the $N_1 \times N_2$ matrix of ones, and define the $N-$vector $\bfL = \frac{1}{N} \bfone_{N,1}$.  We define the orthogonal projection matrices $\bfH,\bfTheta$ which will play a crucial role in the following analysis:
\begin{equation}\label{E:h_theta_def}
\bfH = N\bfL\bfL^T, \quad \bfTheta = \bfI - \bfH.
\end{equation}

Note that if $\bfX$ is an $N\times N$ matrix, then $\bfL^T\bfX\bfL =: \langle \bfX\rangle$ is the empirical average of all entries in $\bfX$. We first observe that the empirical network connection probability  can be
obtained from the adjacency matrix, $\bfW^0$, as
$$
p  = \bfL^T \bfW^0 \bfL.
$$
We can also express second order motif frequencies in terms of  intra-network averages. For instance,
\begin{equation}\label{E:qdiv_average}
\begin{split}
\qdiv &= \frac{1}{N} \bfL^T \bfW^0\bfW^{0T}\bfL - p^2\\
&=  \frac{1}{N}\bfL^T \bfW^0\left(\bfH+\bfTheta\right)\bfW^{0T}\bfL - p^2\\
&=  \left(\bfL^T \bfW^0\bfL\right)\left( \bfL^T\bfW^{0T}\bfL \right)+  \frac{1}{N}\bfL^T \bfW^0\bfTheta\bfW^{0T}\bfL - p^2\\
&= \frac{1}{N}\bfL^T \bfW^0\bfTheta\bfW^{0T}\bfL.
\end{split}
\end{equation}
Similarly, $\qcon,\qch$ may be expressed as
\begin{equation}\label{E:qcon_average}
\begin{split}
\qcon &=  \frac{1}{N}\bfL^T\bfW^{0T}\bfW^0\bfL - p^2 \\
&= \frac{1}{N}\bfL^T \bfW^{0T} \bfTheta \bfW^0 \bfL,
\end{split}
\end{equation}
and
\begin{equation}\label{E:qch_average}
\begin{split}
\qch &= \frac{1}{N}\bfL^T\bfW^0\bfW^0\bfL - p^2\\
&= \frac{1}{N}\bfL^T\bfW^{0T}\bfW^{0T}\bfL - p^2\\
&=  \frac{1}{N}\bfL^T \bfW^0 \bfTheta \bfW^0 \bfL.
\end{split}
\end{equation}

To relate second-order motif frequencies to mean correlations between pairs of cells, we can truncate Eq.~\eqref{E:power_series_exp} at second order in $(\At\bfW)$, giving
\begin{equation}\label{E:power_series_exp_sot}
\frac{\bfCt^\infty}{\Ct^0} \approx I + \At w \bfW^0 + \At w \bfW^{0T}  + \left(\At w\right)^2 \bfW^0\bfW^{0T}+ \left(\At w\right)^2 \left(\bfW^0\right)^2+ \left(\At w\right)^2 \left(\bfW^{0T}\right)^2.  
\end{equation}
To obtain the empirical average of pairwise covariances in the network, $\langle \bfCt^\infty \rangle$, we multiply both sides of Eq.~\eqref{E:power_series_exp_sot} on the left and right by $\bfL^T$ and $\bfL,$ respectively.  Making use of Eqns.~(\ref{E:qdiv_average}-\ref{E:qch_average}), we obtain
\begin{equation}\label{E:mean_cov_sot}
\frac{\langle\bfCt^\infty\rangle}{\Ct^0}\approx \frac{1}{N} + 2\At w p + 3N\left(\At w\right)^2 p^2  + N\left(\At w\right)^2 \qdiv + 2N\left(\At w\right)^2 \qch. 
\end{equation}

\begin{figure}[H] 
\centering
\includegraphics[width=6in]{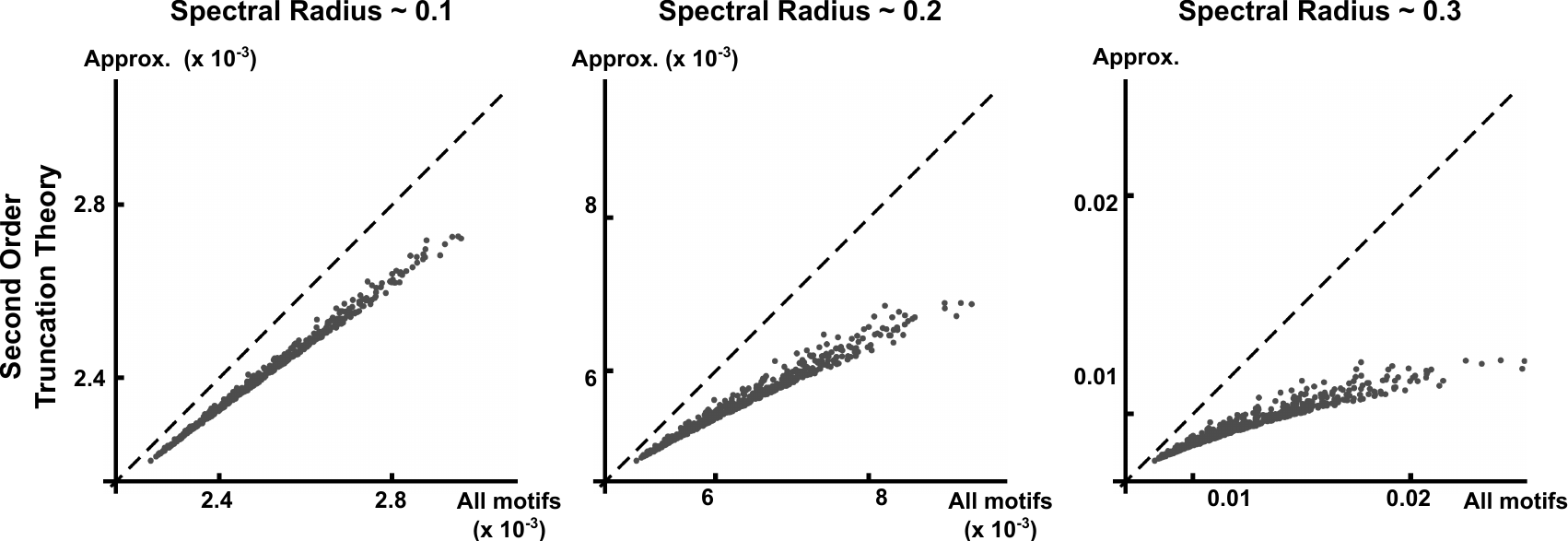}
\caption{Scatter plot comparing the prediction of average network correlation obtained using Eq.~\eqref{cov_hom} (horizontal axes) to the second order truncation in Eq.~\eqref{E:mean_cov_sot} (vertical axes). The diagonal line $y=x$ is plotted for reference. Each panel corresponds to a different coupling strength in the same set of 512 adjacency matrices. The effective coupling strength is characterized by the \ER{} spectral radius shown at the top of each panel (See Sec.~\ref{s: stability}).}
\label{F:E_truck}
\end{figure}

How well does this second-order truncation predict levels of correlation across different neural networks?  Fig.~\ref{F:E_truck} shows that the truncation correctly captures general trends in levels of correlation from network to network, but makes a substantial systematic error.  Here, we plot correlations predicted with the truncated Eq.~\eqref{E:mean_cov_sot} as an approximation of the full expression (i.e., to all orders) for average correlations given by Eq.~\eqref{E:lindner}. Indeed, the truncated expression gives consistent predictions only at very small coupling strengths. We conclude that the terms which were discarded (all terms of order  three and higher  in $\At \bfW$) can have an appreciable impact on  average network correlation, and will next develop methods to capture this impact. 

\subsection{Improved theory:  resumming to approximate higher-order contributions to network correlation} \label{S:single_pop_rs}

A much better approximation of the average network correlation can be obtained by considering the impact of second order motifs on higher
order terms in the expansion given by Eq.~\eqref{E:power_series_exp}.
Note that in an \ER{} network,  every motif of order $m$ occurs with probability $p^m$ (with the exception
of motifs which involve the same connection multiple times), so that on average $\qdiv=\qcon=\qch=0$.
For non--\ER{} networks, the expected values of $\qdiv, \qcon$ and $\qch$ are typically not zero.  As we  show next, the introduction of additional second order structure in the network also affects the 
frequency of motifs of higher order.

Consider the average covariance as determined from the full order linear response approximation of Eq.~\eqref{E:power_series_exp}:
\begin{equation}\label{E:average_expansion}
\frac{\langle \bfCt^\infty \rangle}{\Ct^0} =  \sum_{i,j=0}^\infty \left(\At w\right)^{i+j} \bfL^T \left(\bfW^0\right)^i \left(\bfW^{0T}\right)^j \bfL.
\end{equation}

We will first express every term in the sum given by Eq.~\eqref{E:average_expansion} approximately in terms of first and second order motif frequencies ($p, \qdiv, \qcon$ and $\qch$).
We  illustrate this approximation in two examples, before proceeding to the general calculation.

Consider the term $\bfL^T \left(\bfW^0\right)^3 \bfL$ corresponding to the average number of length three chains connecting a pair of cells.  Using $\bfI=\bfH+\bfTheta$,
we can proceed as in the computation leading to Eq.~\eqref{E:qdiv_average}, 
\beqr \label{E:split1}
\nonumber\bfL^T \left(\bfW^0\right)^3 \bfL &= &\bfL^T \bfW^0 (\bfH+\bfTheta) \bfW^0   (\bfH+\bfTheta) \bfW^0 \bfL\\
&=&\bfL^T \bfW^0 \bfH \bfW^0 \bfH \bfW^0 \bfL+\bfL^T \bfW^0 \bfTheta \bfW^0 \bfH \bfW^0 \bfL\\
&&+
\nonumber\bfL^T \bfW^0 \bfH \bfW^0 \bfTheta \bfW^0 \bfL+\bfL^T \bfW^0 \bfTheta \bfW^0 \bfTheta \bfW^0 \bfL.
\eeqr
We next replace $\bfH$ by $N\bfL\bfL^T$ in the last expression to obtain
\beqr \label{3-chain}
\nonumber\bfL^T \left(\bfW^0\right)^3 \bfL &=&N^2(\bfL^T \bfW^0\bfL)( \bfL^T\bfW^0 \bfL) (\bfL^T \bfW^0 \bfL)+N(\bfL^T \bfW^0 \bfTheta \bfW^0 \bfL) (\bfL^T \bfW^0 \bfL)\\
&&+
N(\bfL^T \bfW^0 \bfL )(\bfL^T \bfW^0 \bfTheta \bfW^0 \bfL)+\bfL^T \bfW^0 \bfTheta \bfW^0 \bfTheta \bfW^0 \bfL.
\eeqr
Here, the first three terms are composed of factors that correspond to the connection probability ($\bfL^T \bfW^0 \bfL$) and second order chain motif frequency ($\bfL^T \bfW^0 \bfTheta \bfW^0 \bfL$). These  terms provide an estimate of the frequency of a length three chain in a graph, \emph{in terms of the frequency of smaller motifs that form the chain}. The last term, $\bfL^T \bfW^0 \bfTheta \bfW^0 \bfTheta \bfW^0 \bfL$, gives the frequency of occurrence of the length three chain in addition to that obtained by chance from second order motifs. Such higher order terms will be gathered separately and denoted by  \emph{h.o.t.} in the approximation.  We therefore obtain, 
\begin{equation} \label{E:lin_sum}
\bfL^T \left(\bfW^0\right)^3 \bfL = N^2(p^3 + 2p\qch) + \mathrm{h.o.t.}
\end{equation}

\begin{figure}[H]
\centering
\includegraphics[width=4in]{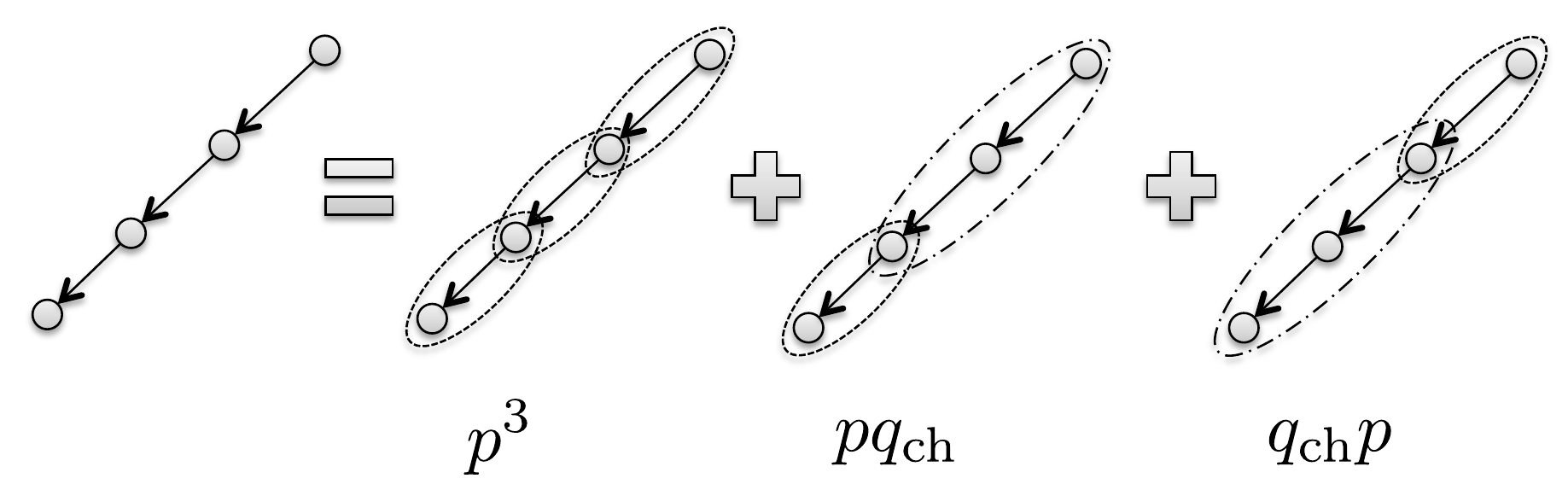}
\caption{Estimating the number of occurrences of a third order chain motif.   The first 3 terms in Eq.~\eqref{3-chain}  correspond to different ways that a chain of length three can  be composed from smaller motifs. As explained in the text, each of the three terms represents the estimated probability that  smaller motifs occur in one of the arrangements on the right.  
\label{F:eg_1}
}
\end{figure}

The exact form of this expression can be understood by referring to Fig.~\ref{F:eg_1}:
The leading $N^2$ denotes the number of possible length three chains between a pair of cells (such a chain can pass through $N^2$ different intermediate pairs of cells) and  $p^3$ represents the probability that one of these length three chains is ``present" in an \ER{} graph. Recall that $\qch$ measures the probability that a length two chain is present, above that expected in an \ER{} network.  Therefore, $p\qch$ represents a second order estimate of the probability above (\ER) chance  that the first two connections in the chain $(\qch)$ and the last $(p)$ are present simultaneously. The prefactor 2 appears because this may also occur if the first and final two connections are present simultaneously.

\begin{figure}[H]
\centering
\includegraphics[width=4in]{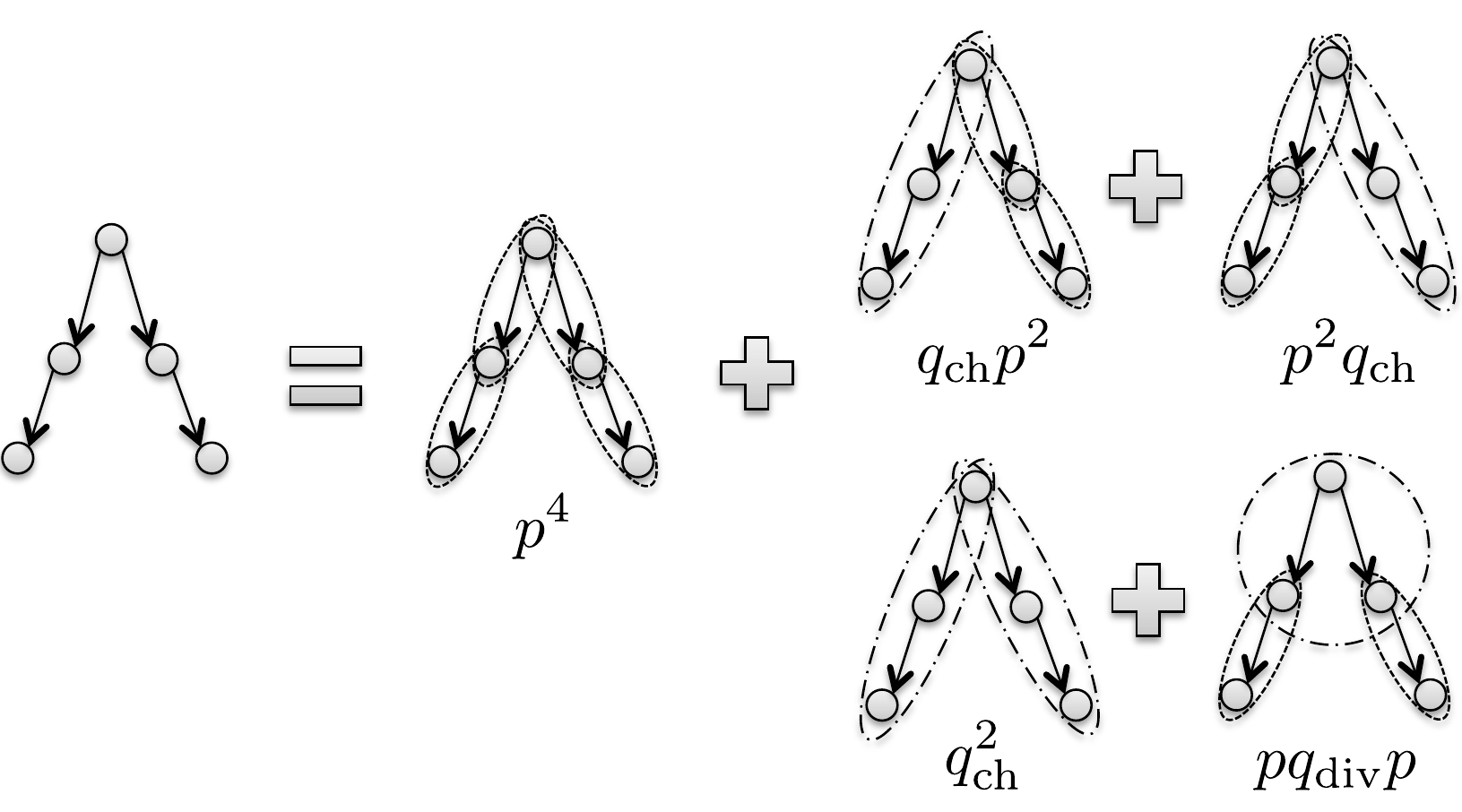}
\caption{Estimating the number of occurrence of a fourth order motif.  Eq.~\eqref{E:lin_sum2} can be understood by decomposing this motif into the constituent first and second order motifs. }
\label{eg_2}
\end{figure}

As a second example, consider the term $\bfL^T \left(\bfW^0\right)^2 \left(\bfW^{0T}\right)^2 \bfL$, corresponding to indirect diverging motifs with two connections on each branch. A computation similar to that used to obtain Eq.~\eqref{E:lin_sum} now gives
\begin{equation} \label{E:lin_sum2}
\bfL^T \left(\bfW^0\right)^2 \left(\bfW^{0T}\right)^2 \bfL = N^3\left[p^4 + p^2\qdiv + 2p^2\qch + \qch^2\right] + \mathrm{h.o.t.}
\end{equation}
The four terms in this sum can be understood with the help of Fig.~\ref{eg_2}:
$p^4$ represents the  chance of observing the motif in an \ER{} network. The product $p^2\qdiv$ is the estimated probability above (\ER) chance that the motif is formed by two connections emanating from the source cell, present simultaneously and independently with two connections emanating from the tips of the branches. The term $p^2\qch$ gives the estimated probability, above (\ER{}) chance, that one branch is present, simultaneously and independently, from two connections which form the other branch (the prefactor 2 concerns the probability of this occurring in each of the two branches).  The last term, $\qch^2$, gives the estimated probability, above the \ER{} chance level, that two length two chains simultaneously emanate from the root cell.

In Eq.~\eqref{E:lin_sum2}, \emph{h.o.t.} denotes the three distinct terms 
\beqrn
&&\bfL^T \bfW^0\bfTheta \bfW^0  \bfTheta  \bfW^{0T} \bfTheta  \bfW^{0T} \bfL,~N(\bfL^T \bfW^0\bfL )(\bfL^T  \bfW^0 \bfTheta  \bfW^{0T} \bfTheta  \bfW^{0T} \bfL)
,\\
&&N(\bfL^T \bfW^0 \bfTheta \bfW^0 \bfTheta  \bfW^{0T} \bfL)( \bfL^T  \bfW^{0T} \bfL).
\eeqrn
These terms contain two or more occurrences of $\bfTheta$ in one factor, and hence correspond to motifs of higher than second order. In general, a factor which contains $m$ occurrences of $\bfTheta$ will depend on the frequency of a motif of order $m+1$, beyond that which is imposed by the frequency of motifs of order $m$.

The idea behind these two examples extends to all terms in the series in Eq.~\eqref{E:average_expansion}, assuming absolute convergence.
Each term in the resulting series contains a factor of the form
  $\bfL^T \left(\bfW^0\right)^i\left(\bfW^{0T}\right)^j \bfL$ corresponding to a 
motif of order $i+j$.  This motif corresponds to two chains of length $i$ and $j$, respectively, emanating from the same root cell.   To understand the impact of
second order motifs we need to decompose this motif as illustrated 
in Figs.~\ref{F:eg_1} and~\ref{eg_2}.  While this is a challenging combinatorial 
problem, we show that the answer can be obtained by rearranging the terms in Eq.~\eqref{E:average_expansion}.

Each  factor of the form   $\bfL^T \left(\bfW^0\right)^i\left(\bfW^{0T}\right)^j \bfL$ in Eq.~\eqref{E:average_expansion} can be split  by inserting $\bfI=\bfH+\bfTheta$ between each occurrence of $\bfW^0$ or $\bfW^{0T}$, as in 
 Eq.~\eqref{E:split1}.  The resulting expression can be used to identify the impact of motifs of
 order $k$ on terms in the expansion of order $i+j \geq k$.   The following Proposition,
 proved in Appendix~\ref{s:prop_proof}, formalizes these ideas.
  
\begin{proposition}
\label{P:E_full}
Let $\bfH$ be the rank-1 orthogonal projection matrix generated by the unit $N$-vector $\bfu$,  $\bfH=\bfu \bfu^T$, $\bfTheta=\bfI-\bfH$. For any $N\times N$ matrix $\bfK$, let
$$
\bfK_n = \left(\bfK\bfTheta\right)^{n-1}\bfK = \underbrace{\bfK \bfTheta \bfK \cdots \bfTheta \bfK}_{n \mathrm{\ factors\ of\ }  \bfK}.
$$
If the spectral radii $\Psi(\bfK) < 1$ and $\Psi( \bfK \bfTheta)<1$, then
\begin{equation} \label{nrs_eq}
\begin{split}
 \bfu^T(\bfI-\bfK)^{-1}&(\bfI-\bfK^T)^{-1} \bfu\\
 &= \left(1 - \sum_{n=1}^\infty  \bfu^T \bfK_n \bfu \right)^{-1}\left(1 +  \sum_{n,m=1}^\infty \bfu^T \bfK_n \bfTheta \bfK_m^T \bfu \right)
\left(1 - \sum_{m=1}^\infty  \bfu^T \bfK_m^{T} \bfu \right)^{-1} .
\end{split}
\end{equation}
\end{proposition}

We will use Prop.~\ref{P:E_full} to derive a  relation between second order motif strengths and mean covariances. Assuming that $\Psi(\At w \bfW^0),~\Psi(\At w\bfW^0\bfTheta)<1$, and setting $\bfu=\sqrt{N} \bfL$ and $\bfK=\At w \bfW^0$, applying Prop.~\ref{P:E_full} to Eq.~\eqref{cov_hom} gives
\begin{equation}\label{E:resumming_motifs}
\begin{split}
 \frac{\langle\bfCt^\infty \rangle}{\Ct^0}
&=\frac{1}{N} \left(1 - \sum_{n=1}^\infty (N\At w )^n \bfL^T \bfW^0_n \bfL \right)^{-1}\left(1 +  \sum_{n,m=1}^\infty  (N\At w )^{n+m} \bfL^T \bfW^0_{n,m}  \bfL \right)\\
&\qquad\qquad\cdot\left(1 - \sum_{m=1}^\infty  (N\At w )^m\bfL^T \bfW^{0T}_m \bfL \right)^{-1}, 
\end{split}
\end{equation}
where 
\beqrn
&&\bfW_n^0 = \frac{1}{N^{n-1}}\underbrace{\bfW^0 \bfTheta \bfW^0 \cdots \bfTheta \bfW^0}_{n \mathrm{\ factors\ of\ } \bfW^0}, \\
&&
\bfW_{n,m}^0 = \frac{1}{N^{n+m-1}}\underbrace{\bfW^0 \bfTheta \bfW^0 \cdots \bfTheta \bfW^0}_{n \mathrm{\ factors\ of\ }   \bfW^0} \bfTheta \underbrace{\bfW^{0T} \bfTheta \bfW^{0T} \cdots \bfTheta \bfW^{0T}}_{m \mathrm{\ factors\ of\ } \bfW^{0T}}.
\eeqrn

Keeping only terms in Eq.~\eqref{E:average_expansion} which can be expressed as polynomials of second order in motif frequency and connection probability is equivalent to keeping only terms involving $\bfW_1^0$ (connection probability), $\bfW_2^0$ (chain motifs) and $\bfW^0_{1,1}$ (diverging motifs) in Eq.~\eqref{E:resumming_motifs}.  This yields an 
expression which involves only first and second order motif frequencies:
\begin{equation}\label{E:singlepop_nl_resumming_so}
  \frac{\langle\bfCt^\infty \rangle}{\Ct^0}
= \frac{1}{N} \frac{1 + \left(N\At w\right)^{2} \qdiv }{\left[1 - \left(N \At w\right) p - \left(N \At w\right)^2 \qch \right]^2}.
\end{equation}

\begin{figure}[H] 
\centering     
\includegraphics[width=6in]{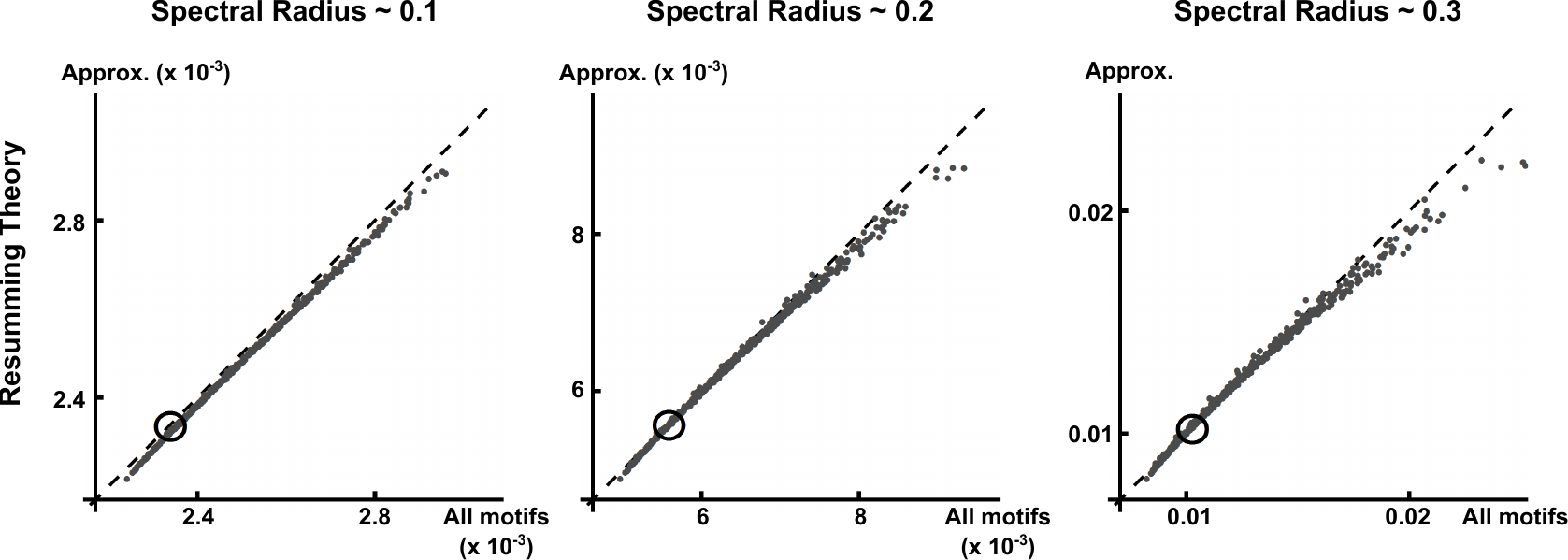}
\caption{A comparison of the  mean correlation obtained using Eq.~\eqref{cov_hom} (horizontal axes) to the resumming approximation in Eq.~\eqref{E:singlepop_nl_resumming_so} (vertical axes). The diagonal line $y=x$ is plotted for reference. Each panel corresponds to a different scaling of coupling strength for the same set of 512 adjacency matrices same as Fig.~\ref{F:E_truck}, which are sampled from SONET (see Sec.~\ref{gen_netw}). The effective coupling strength is characterized by the \ER{} spectral radius, and is recorded at the top of each panel (see Sec.~\ref{s: stability}). Here, the open circle indicates the level of correlation expected from an \ER \; network with the same overall connection probability and strength.}
\label{F:E_nrs}
\end{figure}

Fig.~\ref{F:E_nrs}  shows that the approximation to the covariance given by Eq.~\eqref{E:singlepop_nl_resumming_so} provides a significant improvement over the second order truncation approximation in Eq.~\eqref{E:mean_cov_sot} (See Fig.~\ref{F:E_truck}).   We emphasize that it requires only three scalars that summarize the statistics of the entire connection graph:  the overall connection probability and the propensity of two second-order motifs.  We offer a heuristic explanation for the effectiveness of the resumming theory based on spectral analysis of $\bfW$ in Appendix~\ref{s:spec_intui}. 

If we expand the denominator in Eq.~\eqref{E:singlepop_nl_resumming_so} as a power series in $\left[\left(N\At w\right)p + \left(N\At w\right)^2\qch\right]$, and keep only terms which are linear in $\qdiv,~\qch$, we obtain
\begin{equation}\label{E:single_pop_expansion}
\frac{\langle\bfCt^\infty \rangle}{\Ct^0}
=
  \frac{1}{N(1-N \At w p)^2} + \frac{N(\At w)^2}{(1 - N \At w p)^2} \qdiv
  + \frac{2 N (\At w)^2 }{(1 - N \At w p)^3} \qch+ \mathrm{h.o.t.}
\end{equation}
This linear relation was used to estimate the regression coefficients in Fig.~\ref{F:E_bar} (bar plot).

Finally, we note that similar means may be used to derive versions of Eq.~\eqref{E:singlepop_nl_resumming_so}  that (unlike Eq.~\eqref{E:single_pop_expansion}) retain a nonlinear dependence on motif frequencies, but keep motifs of either higher or lower order.  To approximate the impact of  motifs up to order $r$, we would keep terms with factors $\bfW_n^0, \bfW_{n,m}^0$ where $n,n+m \leq r$ in Eq.~\eqref{E:resumming_motifs}.  For example, if we take $r=1$, we estimate the mean covariance based only on the probability of occurrence of first order motifs -- that is, connection probability. This is equivalent to estimating the covariance in
idealized \ER{} networks where $\qdiv$ and $\qch$, and their higher order analogs are precisely zero.
In Eq.~\eqref{E:resumming_motifs}, we set all terms involving $\bfW$ to zero save $\bfW^0_1$, giving
\beq
\label{e:ER_E}
\frac{\langle\bfCt^\infty \rangle}{\Ct^0}=\frac{1}{N\left(1 - N \At w  p \right)^2}.
\eeq
This predicted mean correlation is indicated by the open dots in Fig.~\ref{F:E_nrs} (which accurately describes \ER{} networks as shown in Fig.~\ref{figure0}). This again demonstrates how motif structures exert a large influence over averaged spike correlations.

\subsection{Correlations in external input}

As is natural for communication among different brain areas or layers~\citep{Sha+98}, or as arises for certain sensory inputs~\citep{Doi+04}, we next consider the case in which the network under study receives an additional noisy input that is {\it correlated} from cell-to-cell; in other words, the cells receive common input from upstream sources~\citep{delaRocha:2007,JosicSDR08,Trong2008}.  We take this input to have total covariance structure ($\omega = 0$) $\bfCt^\eta = \sigma_X^2 \bfI + \sigma_X^2\rhoin(\bfone_{NN}-\bfI)$, so that the variance of such external input (not to be confused with the implicitly modeled external noise $\xi(t)$ term in Eq.~\eqref{E:v_theory}) to each neuron is fixed ($\sigma_X^2$) independent of $\rhoin$ and the correlation coefficient of the inputs to all cell pairs is $\rhoin$~\citep{Lindner:2005,Marinazzo:2007}.  Eq.~\eqref{cov_hom} then has the form (see~\citep{Trousdale:2012} for details, and an improvement applicable when the extra noise source is white):
\begin{equation*}
\begin{split}
\bfCt^\infty &= \left(\bfI - \At \bfW\right)^{-1}\left(\Ct^0\bfI + \At^2\bfCt^\eta\right)\left(\bfI-\At\bfW^T\right)^{-1}\\
&=  \left(\bfI - \At \bfW\right)^{-1}\left[\left(\Ct^0 + \At^2\sigma_X^2\right)\bfI + \left(\At^2\sigma_X^2\rhoin\right)(\bfone_{NN}-\bfI)\right]\left(\bfI-\At\bfW^T\right)^{-1}.
\end{split}
\end{equation*}
In this case, the output variance in the absence of coupling ($\bfW \equiv 0$) predicted by the linear response theory is $(\Ct^0 + \At^2\sigma_X^2)$. Normalizing and multiplying by $\bfL^T,\bfL$ to arrive at an approximation of average correlation, and applying the ideas of Sec.~\ref{S:single_pop_rs}, we find that
$$
\frac{\langle\bfCt^\infty\rangle}{\Ct^0 + \At^2\sigma_X^2} = \frac{B_1[1 + (N\At w)^2\qdiv] + NB_2}{N\left[ 1 - (N\At w)p - (N\At w)^2 \qch\right]^2},
$$
where
$$
B_1 = \frac{\Ct^0 +\At^2\sigma_X^2(1-\rhoin)}{\Ct^0+\At^2\sigma_X^2}, \qquad B_2 = \frac{\At^2\sigma_X^2\rhoin}{\Ct^0 + \At^2\sigma_X^2}
$$
(compare Eq.~\eqref{E:singlepop_nl_resumming_so}).  We might ask how the presence of input correlations affects output correlations by calculating the \emph{change in output correlation} resulting from input correlations of size $\rhoin$:
\begin{equation}
\begin{split}
\Delta \rhoout &= \frac{\langle\bfCt^\infty\rangle}{\Ct^0 + \At^2\sigma_X^2}  - \left. \frac{\langle\bfCt^\infty\rangle}{\Ct^0 + \At^2\sigma_X^2}  \right|_{\rhoin = 0}\\
&= \frac{B_2[N-1 -(N\At w)^2\qdiv]}{N\left[ 1 - (N\At w)p - (N\At w)^2 \qch\right]^2}.
\end{split}
\end{equation}
For simplicity, we can linearize this expression in $\qdiv, \qch$ to examine the interaction between second order motifs and input correlations, giving
\beq
\label{e:rho_out}
\Delta \rhoout = \frac{B_2}{N} \left[ \frac{N-1}{[1 - (N\At w)p]^2} - \frac{(N\At w)^2}{[1 - (N\At w)p]^2} \qdiv + \frac{2(N-1)(N\At w)^2}{[1 - (N\At w)p]^3}\qch\right].
\eeq
If we assume that $w \sim \mathcal{O}\left(\frac{1}{N}\right)$ (so that the \ER{} spectral radius $N\At w \sim \mathcal{O}(1)$, see Sec.~\ref{s: stability}), then, asymptotically, $\qdiv$ only has an order $\mathcal{O}(\frac{1}{N})$ (negative) contribution, while $\qch$ has a order $\mathcal{O}(1)$ contribution to $\Delta \rhoout$ (note that the first term in \eqref{e:rho_out} is also $\mathcal{O}(1)$, which represents the ``base" response to correlated input in a \ER{} network). This implies that in large networks, the chain motif is the most important motif in determining how input correlations will be transferred into network correlations -- which will be ``output" to the next area downstream.

\section{Impact of second order motifs in networks of excitatory and inhibitory neurons}
\label{S:two_pop_theory}

Biological neuronal networks are composed of excitatory and inhibitory neurons (EI networks).  To treat this case, we next show how our theory extends to the case of networks composed of two 
interacting subpopulations.

\subsection{Results: linear dependence between mean correlation and motif frequencies}  

As in the previous section, we start by a numerical exploration of the contribution of motifs to network-averaged correlation in excitatory-inhibitory networks.  We generated 512 networks using the degree distribution method, as described in Sec.~\ref{gen_netw}.  We evaluated network-averaged correlations using the full linear response expression given in Eq.~\eqref{cov_hom}.  We then performed a linear regression analysis of the dependence of network correlations on   motif frequency  in networks of excitatory and inhibitory neurons, for the 20 second order motifs shown in Fig.~\ref{f:motif_ei}.  

The linear regression gives a reasonable fit ($R^2 \ge 0.69$, see caption). Chain and diverging motifs contribute significantly to the network-averaged correlation, with some motifs having a higher effect than others.  Moreover, mean correlations depend only weakly on converging motifs. Note also that these regression coefficients  correspond to how we devise the axes in Fig.~\ref{first figure}{\bf{B}}.  There, we use these regression coefficients (or actually the theory predication for these values based on Eq.~\eqref{E:twopop_resummed_exp}) as weights to linearly combine multiple motif frequencies. Specifically, we take linear combinations of the 6 diverging, 6 converging and 8 chain motif frequencies (see Fig.~\ref{f:motif_ei}) respectively, giving the 3 axes in Fig.~\ref{first figure}{\bf{B}}.

\begin{figure}[H] 
\centering
\includegraphics[width=6in]{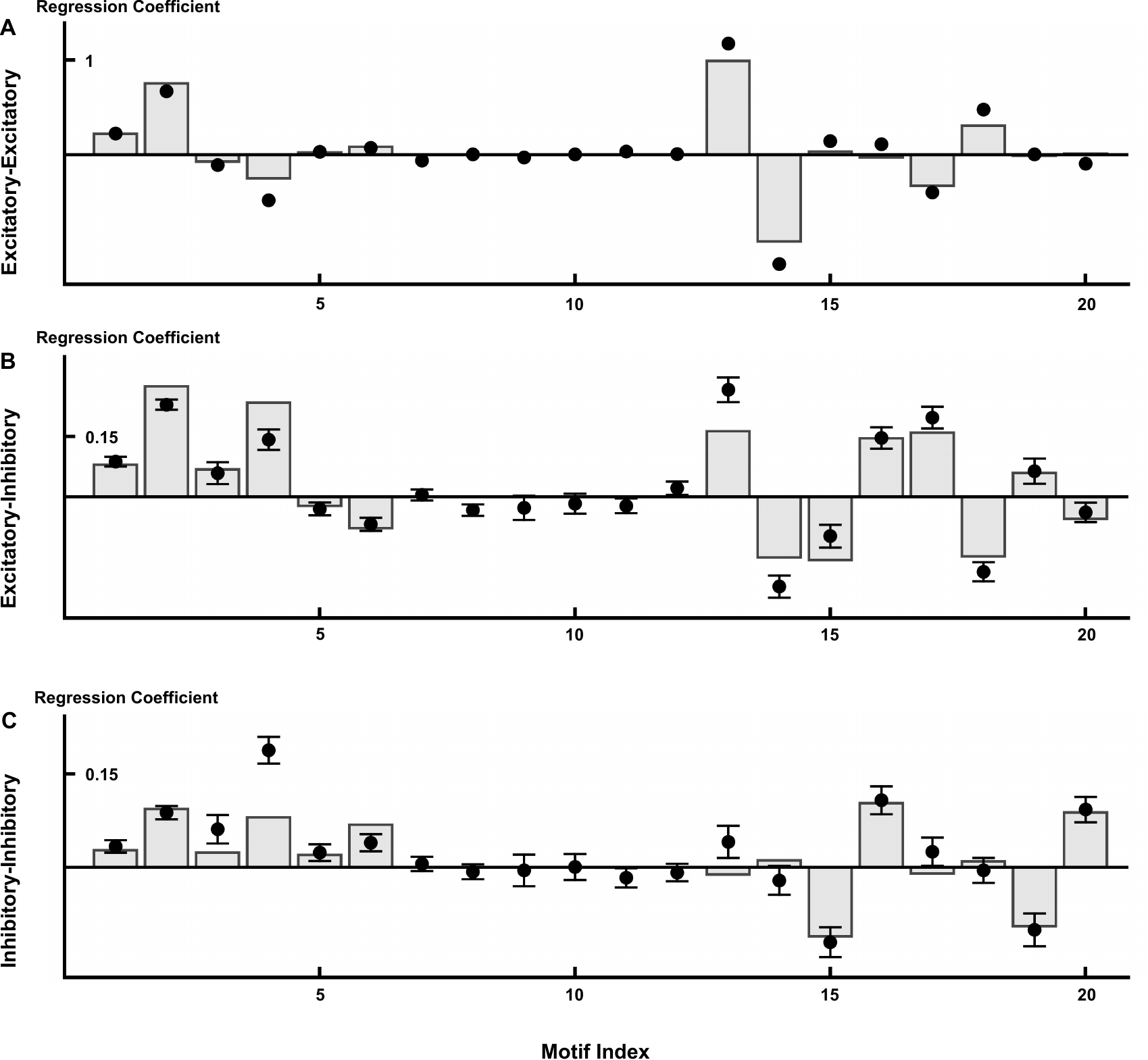}
\caption{Linear regression coefficients between network-averaged correlations within different subpopulations (vertical axes) and motif frequencies (horizontal axes). The three panels correspond to averages across  EE, EI and II cell pairs, from top to bottom. The linear regression coefficient between average correlation and 20 motif frequencies are represented by a dot (See Fig.~\ref{f:motif_ei} for the enumerated list of second order motifs in an EI network). Error bars represent a  95\% confidence intervals. The vertical bars were obtained from the resumming theory described in Sec.~\ref{EI_resum}. 
Here $R^2 =   0.91, 0.86, 0.69$ respectively. The spectral radius of $\At\bfW$ under the \ER{} assumption is 0.25, and $N_E=80$, $N_I=20$, $|w_I|/|w_E|=3.707$ so that $pN_Ew_E+pN_Iw_I\approx 0$ giving approximate balance between average excitatory and inhibitory inputs. Parameters are given in Fig.~\ref{first figure}. } 
\label{F:bars1}
\end{figure}

\subsection{First theory: second order truncation for network correlation}
\label{S:EI_setup}

To extend the theory developed in Sec.~\ref{S:single_pop_theory} to EI networks, we need to take into account distinct second order motifs. For instance, there are eight different types of three-cell chains (See Fig.~\ref{f:motif_ei}). Although this makes the notation more burdensome, the main ideas are the same. Indeed, following a similar approach, the theory can be extended to an arbitrary number of subpopulations. 

Consider a network of size $N=N_E+N_I$, consisting of $N_E$ excitatory and $N_I$ inhibitory neurons. Excitatory (resp. inhibitory) connections have weight $w_E$ (resp. $w_I$), so that $w_E > 0$ and $w_I < 0$. The connection probability from class $X$ to class $Y$ is $p_{YX}$.  Note that the connectivity matrix can now be written as
\begin{equation*}
\ \bfW =  \left(\begin{matrix}\bfW_{EE} & \bfW_{EI} \\ \bfW_{IE} & \bfW_{II} \end{matrix}
\right) = 
\left(\begin{matrix}w_E \bfW^0_{EE} & w_I \bfW^0_{EI} \\ w_E \bfW^0_{IE} & w_I \bfW^0_{II} \end{matrix}
\right),
\end{equation*}
 where,  $\bfW_{XY}$ and $\bfW_{XY}^0$ are respectively, the weighted and unweighted  connection matrices between  cells of class $Y$ to cells of class $X$.
The population sizes, weights, connection probabilities, and firing rates together determine the balance between excitatory and inhibitory inputs that cells receive (see Discussion). Throughout this section we use networks with same parameters and architecture as in Fig. \ref{F:bars1}.

First, define the $N \times 2$ block-averaging matrix $\bfL$ by
\begin{equation*}
\bfL =  \left(\begin{matrix} \bfL_E & 0 \\ 0 & \bfL_I \end{matrix}\right) = \left( \begin{matrix}\bfone_{N_E,1} / N_E & 0 \\ 0 & \bfone_{N_I,1}/N_I  \end{matrix}\right),
\end{equation*}
where $\bfone_{NM}$ is the $N\times M$ matrix of ones. We define the two-population analogs of the orthogonal projection matrices $\bfH$ and $\bfTheta$ by
\begin{equation} \label{E:Hmatrix}
\bfH = \bfL\bfD_2\bfL^T, \qquad \text{and} \qquad \bfTheta = \bfI - \bfH,
\end{equation}
where $\bfD_2$ is the $2\times 2$ matrix
$$
\bfD_2 = \left(\begin{matrix}N_E & 0 \\ 0 & N_I \end{matrix}\right).
$$
Direct matrix multiplication shows that if $\bfX$ is a matrix with block form
$$
\bfX = \left(\begin{matrix}\bfX_{EE} & \bfX_{EI} \\ \bfX_{IE} & \bfX_{II} \end{matrix}\right),
$$
where $\bfX_{YZ}$ is an $N_Y \times N_Z$ matrix, then $\bfL^T \bfX \bfL$ is a $2 \times 2$ matrix of block-wise averages of $\bfX$, that is,
$$
\langle \bfX \rangle_B \stackrel{\mathrm{def}}{=} \bfL^T \bfX \bfL = \left(\begin{matrix} \bfL_E^T \bfX_{EE} \bfL_E & \bfL_E^T \bfX_{EI}\bfL_I \\ \bfL_I^T\bfX_{IE}\bfL_E & \bfL_I^T\bfX_{II}\bfL_I\end{matrix}\right)=
 \left(\begin{matrix} \langle \bfX_{EE} \rangle & \langle\bfX_{EI}\rangle \\ \langle\bfX_{IE} \rangle&\langle \bfX_{II}\rangle \end{matrix}\right).
$$
We will make use of the \emph{empirical average connection strength} matrix $\bfM$ given by
$$
\bfM = \bfL^T \bfW \bfL = \left( \begin{matrix} w_Ep_{EE} & w_I p_{EI} \\ w_E p_{IE} & w_I p_{II} \end{matrix}\right).
$$

To examine the dependence of mean correlations \emph{within a block} on the frequency of second order motifs, we consider the block-wise average of the covariance matrix $\bfCt^\infty$,
$$
\langle \bfCt^{\infty} \rangle_B = \bfL^T \bfCt^{\infty} \bfL = \left(\begin{matrix} \langle \bfCt^{\infty}_{EE} \rangle & \langle\bfCt^{\infty}_{EI}\rangle \\ \langle\bfCt^{\infty}_{IE} \rangle&\langle \bfCt^{\infty}_{II}\rangle \end{matrix}\right).
$$

Excitatory and inhibitory connection weights need not be equal.  It is therefore necessary to consider motif frequencies simultaneously with connection weights. For instance, the contributions of a length two chain passing through an excitatory or inhibitory intermediary cell, such as motifs 13 and 14 in Fig.~\ref{f:motif_ei}, are not necessarily equal and opposite in sign.  Their contributions are dependent on the ratio $w_E/w_I$. To account for this, we define motif {\it strength matrices} $\bfQdiv, \bfQcon,$ and $\bfQch$ as follows.  

The strength of diverging motifs expected in an \ER{} network is given by
\begin{equation*}
\begin{split}
\bfQdiv^{\mathrm{ER}} &= \bfM\bfD_2\bfM^T=(\bfL^T \bfW \bfL)\bfD_2(\bfL^T \bfW^T \bfL)\\
&=\bfL^T \bfW \bfH \bfW^T \bfL\\
&= \left(\begin{matrix} N_E(w_E p_{EE})^2 + N_I(w_Ip_{EI})^2 & N_Ew_E^2 p_{EE}p_{IE} + N_Iw_I^2 p_{EI}p_{II} \\ N_Ew_E^2 p_{EE}p_{IE} + N_Iw_I^2 p_{EI}p_{II} & N_E(w_E p_{IE})^2 + N_I(w_Ip_{II})^2
\end{matrix}\right).
\end{split}
\end{equation*}
Here multiplication by $\bfD_2$ converts average individual motif strength (e.g. $w_E^2p_{EE}p_{IE}$) to average total motif strength. The empirical average of the strength of diverging motifs  is given by \begin{equation*}
\begin{split}
\bfQdiv^{\mathrm{total}} &= \bfL^T \bfW \bfW^T \bfL \\
&= \left( \begin{matrix} \langle \bfW_{EE}\bfW_{EE}^T\rangle + \langle \bfW_{EI}\bfW_{EI}^T \rangle &
 \langle \bfW_{EE}\bfW_{IE}^T\rangle + \langle \bfW_{EI}\bfW_{II}^T \rangle \\
 \langle \bfW_{IE}\bfW_{EE}^T\rangle +  \langle \bfW_{II}\bfW_{EI}^T \rangle &
 \langle \bfW_{IE}\bfW_{IE}^T\rangle +  \langle \bfW_{II}\bfW_{II}^T \rangle \\
 \end{matrix}\right).
\end{split}
\end{equation*}
Hence, we can write the expected total strength of diverging motifs in excess of that expected in an \ER{} network as 
\begin{equation}\label{E:twopop_qdiv_def}
\begin{split}
\bfQdiv &= \left(\begin{matrix} \Qdiv^{EE} & \Qdiv^{EI} \\ \Qdiv^{IE}&\Qdiv^{II}\end{matrix}\right) 
\stackrel{\mathrm{def}}{=} \bfQdiv^{\mathrm{total}} - \bfQdiv^{\mathrm{ER}} \\
&{=} \left( \begin{matrix} N_Ew_E^2\qdiv^{EE,E} + N_Iw_I^2\qdiv^{EE,I} & N_Ew_E^2\qdiv^{EI,E} + N_Iw_I^2\qdiv^{EI,I} \\ N_Ew_E^2\qdiv^{EI,E} + N_Iw_I^2\qdiv^{EI,I} & N_Ew_E^2\qdiv^{II,E} + N_Iw_I^2\qdiv^{II,I}\end{matrix}\right)\\
&= \bfL^T \bfW \bfW^T \bfL - \bfL^T \bfW \bfH \bfW^T \bfL\\
&= \bfL^T \bfW(\bfH+\bfTheta) \bfW^T \bfL - \bfL^T \bfW \bfH \bfW^T \bfL\\
&= \bfL^T\bfW\bfTheta\bfW^T\bfL.
\end{split}
\end{equation}
where
$$
\qdiv^{XY,Z} = \frac{1}{N_Z}\langle \bfW_{XZ}^0\bfW_{YZ}^{0T} \rangle - p_{XZ}p_{YZ}
$$
represents the probability of observing a diverging motif to cells of classes $X,Y$ from a cell of class $Z$ in excess of that expected in an \ER{} network.

It is important to note that the matrix $\bfQdiv$ contains motif \emph{strengths} in excess of what would be expected in an \ER{} network (i.e., number of occurrences, scaled by connection weights and probability of occurrence), while the scalars $\qdiv$ still correspond to probabilities.

The strengths and frequencies of converging motifs can be expressed similarly, giving
\begin{equation}\label{E:twopop_qcon_def}
\begin{split}
\bfQcon  &= \left(\begin{matrix} \Qcon^{EE} & \Qcon^{EI} \\ \Qcon^{IE} & \Qcon^{II}\end{matrix}\right) 
\stackrel{\mathrm{def}}{=} \bfL^T\bfW^T\bfW\bfL - \bfM^T\bfD_2\bfM
= \bfL^T \bfW^T \bfTheta \bfW \bfL \\
&= \left( \begin{matrix} N_Ew_E^2\qcon^{EE,E} + N_Iw_E^2\qcon^{EE,I} & N_Ew_Ew_I\qcon^{EI,E} + N_Iw_Ew_I\qcon^{EI,I} \\ N_Ew_Ew_I\qcon^{EI,E} + N_Iw_Ew_I\qcon^{EI,I} & N_Ew_I^2\qcon^{II,E} + N_Iw_I^2\qcon^{II,I}\end{matrix}\right),\\
\end{split}
\end{equation}
where
$$
\qcon^{XY,Z} = \frac{1}{N_Z}\langle \bfW^{0T}_{XZ}\bfW_{YZ}^0 \rangle - p_{ZX}p_{ZY},
$$
represents the probability of observing a converging motif to cells of class $Z$ from cells of classes $X,Y$ in excess of that expected in an \ER{} network.

Finally, for chain motifs, 
\begin{equation}\label{E:twopop_qch_def}
\begin{split}
\bfQch &= \left(\begin{matrix} \Qch^{EE} & \Qch^{EI} \\ \Qch^{IE} & \Qch^{II}\end{matrix}\right) 
\stackrel{\mathrm{def}}{=} \bfL^T\bfW^2\bfL - \bfM\bfD_2\bfM
 =  \bfL^T \bfW\bfTheta \bfW\bfL\\
&= \left( \begin{matrix} N_Ew_E^2\qch^{EEE} + N_Iw_Ew_I\qch^{EIE} & N_Ew_Ew_I\qch^{IEE} + N_Iw_I^2\qch^{IIE} \\ N_Ew_E^2\qch^{EEI} + N_Iw_Ew_I\qch^{EII} & N_Ew_Ew_I\qch^{IEI} + N_Iw_I^2\qch^{III}\end{matrix}\right),\\
\end{split}
\end{equation}
where
$$
\qch^{ZYX} = \frac{1}{N_Y}\langle \bfW_{ZY}^0\bfW_{YX}^0 \rangle - p_{ZY}p_{YX}
$$
represents the probability of observing a length two chain motif beginning at a cell of type $X$ and terminating at a cell of type $Z$, passing through a cell of type $Y$, in excess of what would be expected in an \ER{} network.

A truncation of Eq.~\eqref{E:power_series_exp} at second order gives an initial approximation of the block-wise average $\langle\bfCt(0)\rangle_B$:
\begin{equation}\label{E:twopop_sot}
\begin{split}
\langle \bfCt^{\infty}\rangle_B/\Ct^0 &= \bfL^T\left[\bfI + \At\left( \bfW +  \bfW^T\right) + \At^2 \left(\bfW^2 + \bfW^{2T} +   \bfW\bfW^T\right)\right]\bfL + \mathrm{h.o.t.}\\
&\approx  \left[\bfL^T\bfL + \At\left(\bfM + \bfM^T\right) + \At^2 \left(  \bfM\bfD_2\bfM+ \bfM^T\bfD_2\bfM^T + \bfM\bfD_2\bfM^T \right) \right.\\ 
& +\At^2 \left.\left(  \bfQch + \bfQch^T +  \bfQdiv  \right) \right].
\end{split}
\end{equation}
Terms not involving a second order motif matrix correspond to the contributions of motifs up to second order in a purely \ER{} network --- for example, $\bfM\bfD_2\bfM$ gives the expected contribution of length two chains in an \ER{} network. 

\begin{figure}[H] 
\centering     
\includegraphics[width=6in]{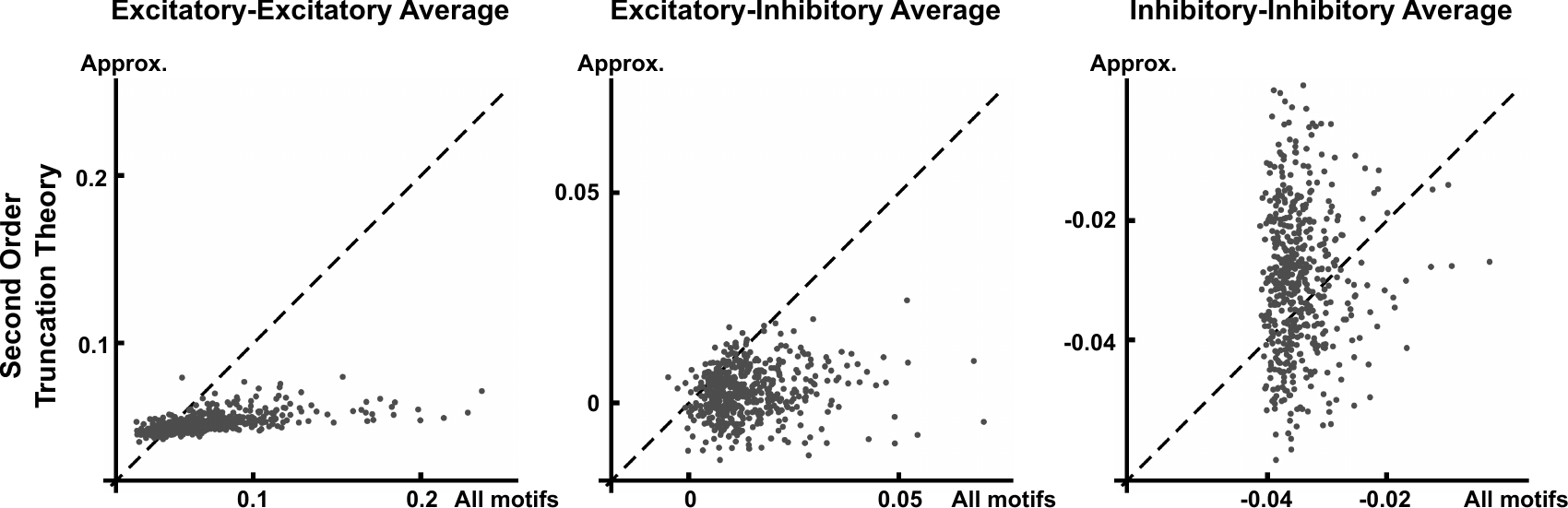}
\caption{Block-wise average correlations obtained from the second order truncation of Eq.~\eqref{E:twopop_sot}. On the horizontal axis are the average correlation between cells from the given classes (calculated from Eq.~\eqref{cov_hom}), and the approximate value obtained from the truncation is given on the vertical axis. The diagonal line $y=x$ corresponds to perfect agreement between the true value and the approximation.  The spectral radius of $\At\bfW$ under the \ER{} assumption is 0.33 (see Sec.~\ref{s: stability}).  Other network parameters are the same as in Fig.~\ref{F:bars1}, \ref{F:EI_nrs} and \ref{hom_heter}. Here $R^2= 0.33, 0.02, 0.0005$  respectively for the three panels, confirming that the truncation approach gives a poor prediction of network correlation.   Here, the open circle indicates the level of correlation expected from an \ER \; network with the same overall connection probabilities and strengths.}
\label{F:truncEI}
\end{figure}

Fig.~\ref{F:truncEI} compares the second order truncation given by Eq.~\eqref{E:twopop_sot} with the mean correlations obtained from the entire series in Eq.~\eqref{E:power_series_exp}.  The correlations in the network of inhibitory and excitatory cells can be appreciable, even when the spectral radius of matrix $\At \bfW$ is much smaller than one. However, the second order truncation of Eq.~\eqref{E:twopop_sot} gives a poor approximation of network-averaged correlations.

\subsection{Improved theory: resumming to approximate higher-order contributions to network correlation}\label{EI_resum}

As in the case of a single population, we can improve our prediction of mean correlations by accounting for the contributions of second order motifs to all orders 
in connection strength. The equivalent of Eq.~\eqref{E:average_expansion} 
has the form
\begin{equation}\label{E:twopop_average_exp}
\frac{\langle \bfCt^\infty \rangle_B}{\Ct^0}= \frac{1}{\Ct^0}\bfL^T\bfCt^{\infty} \bfL =\sum_{i,j=0}^\infty \At^{i+j} \bfL^T \bfW^i (\bfW^{T})^j \bfL,
\end{equation}
where $\bfW$ and $\bfL$ are as defined in the previous section.

 First, we generalize Prop.~\ref{P:E_full} to the case of two populations (for a full proof, see Appendix~\ref{s:prop_proof_2})
\begin{proposition}
\label{EI_full}
Let $\bfH=\bfU\bfU^T$ be a orthogonal projection matrix generated by a $N\times M$ matrix $\bfU$, whose columns are orthonormal vectors. Define $\bfTheta=\bfI-\bfH$. For any $N\times N$ matrix $\bfK$, define $\bfK_n$ as
$$
\bfK_n =\underbrace{\bfK \bfTheta \bfK \cdots \bfTheta \bfK}_{n \mathrm{ \ factors \ of \ } \bfK}.
$$
If spectral radius $\Psi(\bfK),~\Psi( \bfK \bfTheta)<1$, we have
\beqr \label{EI_nrs_eq}
\nonumber &&\bfU^T(\bfI-\bfK)^{-1}(\bfI-\bfK^T)^{-1} \bfU\\
 &&= \left(\bfI - \sum_{n=1}^\infty  \bfU^T \bfK_n \bfU \right)^{-1}\left(\bfI +  \sum_{n,m=1}^\infty \bfU^T \bfK_n \bfTheta \bfK_m^T \bfU \right)
\left(\bfI - \sum_{m=1}^\infty  \bfU^T \bfK_m^{T} \bfU \right)^{-1}. 
\eeqr
\end{proposition}

\begin{figure}[H] 
\centering     
\includegraphics[width=6in]{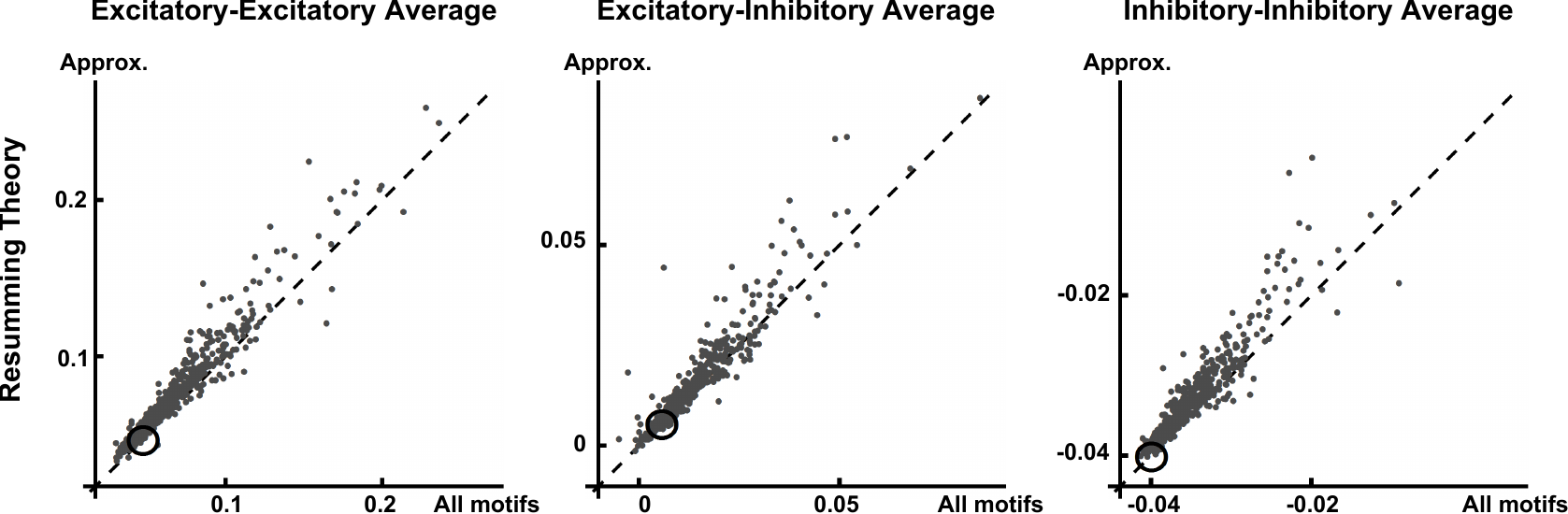}
\caption{Predicting block-wise average correlations from resumming theory. The  horizontal-axis is the average correlation (in a certain block) from original linear response expression of covariance matrix (Eq.~\ref{cov_hom}); the vertical axis the quantity from resumming theory (Eq.~\eqref{E:twopop_nl_resumming_so}). The diagonal line $y=x$ is plotted for reference. The spectral radius of $\At\bfW$ under \ER{} assumption is 0.33 (see Sec.~\ref{s: stability}).  Other network parameters are the same as in Fig.~\ref{F:bars1}. Here $R^2 = 0.93, 0.88, 0.8$7 respectively for the three panels.The open circle indicates the level of correlation expected from an \ER \; network with the same overall connection probability and strength.}
\label{F:EI_nrs}
\end{figure}

We now apply this Proposition to the expression in Eq.~\eqref{E:twopop_average_exp}. Let $\bfU=\bfL \bfD_2^{1/2}$ and $\bfU^T \bfX \bfU=\bfD_2^{1/2} \bfL ^T\bfX \bfL\bfD_2^{1/2}$ for any matrix $\bfX,$  so that $\bfH$ has the form given in Eq.~\eqref{E:Hmatrix}.  In addition,  let $\bfK=\At \bfW$, and assume that $ \Psi(\At\bfW),~\Psi(\At\bfTheta\bfW)<1$.
Then Prop.~\ref{EI_full} gives
\begin{equation}\label{E:twopop_nl_resumming_all}
\begin{split}
\langle \bfCt^{\infty} \rangle_B/ \Ct^0 &= \bfL^T(\bfI-\At \bfW)^{-1}(\bfI-\At\bfW^T)^{-1}\bfL\\
&= \left(\bfI - \sum_{n=1}^\infty \At^n  \bfL^T\bfW_n\bfL\bfD_2 \right)^{-1}\left(\bfD_2^{-1}+ \sum_{n,m=1}^\infty \At^{n+m} \bfL^T \bfW_n  \bfTheta \bfW^{T}_m \bfL \right)\\
&\qquad\cdot\left(\bfI - \sum_{m=1}^\infty \At^m \bfD_2\bfL^T\bfW_m^{T}\bfL \right)^{-1},
\end{split}
\end{equation}
where
$$
\bfW_n = \underbrace{\bfW \bfTheta \bfW \cdots \bfTheta\bfW}_{n \mathrm{ \ factors \ of \ } \bfW}.
$$
As in the single population case,  we can discard all terms in Eq.~\eqref{E:twopop_nl_resumming_all} which do not correspond to  second order motif frequencies. This means that in the first and third set of brackets in Eq.~\eqref{E:twopop_nl_resumming_all}, we discard any terms containing $\bfW_n$ with $n\ge 3$, and for the middle set of brackets, we discard terms containing $\bfW_n \bfTheta \bfW^T_{m}$ for $n+m\ge 3$. This gives
\begin{equation}\label{E:twopop_nl_resumming_so}
\langle \bfCt^{\infty}\rangle_B/ \Ct^0 \approx  \left(\bfI - \At \bfM\bfD_2 - \At^2  \bfQch \bfD_2 \right)^{-1}\left(\bfD_2^{-1} + \At^2 \bfQdiv \right)\left(\bfI - \At \bfD_2\bfM^T - \At^2 \bfD_2 \bfQch^T\right)^{-1}.
\end{equation}
Figure~\ref{F:EI_nrs} illustrates that this  approximation is a great improvement over that
given by truncating the expansion
at second order (compare with Fig.~\ref{F:truncEI}).  Again, we note that our approximation requires only knowledge of the overall connection probabilities among excitatory and inhibitory cells, and the frequency of second-order motifs.

Expanding the inverses in Eq.~\eqref{E:twopop_nl_resumming_so} in a power series, we can again obtain an approximation to block average correlations to linear
order in $\bfQch$, and $\bfQdiv$,
\begin{equation}\label{E:twopop_resummed_exp}
\begin{split}
\langle\bfCt^\infty\rangle_B /\Ct^0
&\approx
\left(\bfI - \At\bfM\bfD_2\right)^{-1}\left[ \bfD_2^{-1} + \At^2\bfQch \left(\bfI - \At \bfD_2\bfM\right)^{-1}+ \left(\bfI-\At\bfM^T\bfD_2\right)^{-1}\At^2\bfQch^T\right.\\
&\left.\vphantom{ \left(\bfI - \At \bfD_2\bfM\right)^{-1}}     \qquad\qquad+ \At^2\bfQdiv  \right]\left(\bfI - \At\bfD_2\bfM^T\right)^{-1}.   
\end{split}
\end{equation}
As in  the single population case, each entry of the $2 \times 2$ matrix on the right hand side of Eq.~\eqref{E:twopop_resummed_exp} gives an approximation
to the block-averaged correlations expressed in terms of the scalars $\qch^{ZYX},\qdiv^{XY,Z}$, providing an analytical estimate to the regression coefficients plotted
in Fig.~\ref{F:bars1}.  An example with uniform connection probability is given in Appendix \ref{s:C_EE}. We note that in general all sub-types of diverging and converging motifs affect each block-wise average correlation.  This is not predicted by the second order truncation in Eq.~\eqref{E:twopop_sot}. Somewhat counterintuitively, $\qdiv^{EI,E}$ and $\qdiv^{EI,I}$ (index 3 and 4 in Fig.~\ref{f:motif_ei}) can contribute negatively to $\langle \bfCt_{EE} \rangle$ as shown in Fig.~\ref{F:bars1}.

As in the single population case, we can also retain contributions to mean correlation which can be expressed as (nonlinear) functions of first order motifs (connection probabilities) only. This follows from setting all $\mathbf {Q}$ terms in Eq.~\eqref{E:twopop_nl_resumming_so} to zero, yielding the following approximation for mean correlation in the two-population analog of \ER{} networks:
\begin{equation}\label{E:twopop_nl_resumming_ER}
\langle \bfCt^{\infty}\rangle_B/ \Ct^0 =  \left(\bfI - \At \bfM\bfD_2 \right)^{-1}\bfD_2^{-1}\left(\bfI - \At \bfD_2\bfM^T\right)^{-1}.
\end{equation}
This predicted mean correlation for an \ER{} network is shown by an open dot in Fig.~\ref{F:EI_nrs}; the deviations from this value illustrate that motif structures can both significantly increase and decarese average network-correlations (see also Fig.~\ref{figure0} and \ref{F:E_nrs}).

\section{Heterogeneous networks} \label{heter_theory}

For simplicity  in the previous sections we assumed a homogeneous network of neurons composed of   cells with identical firing rates, power spectra and response properties.    As a result, the diagonal matrices $\bfCt^0$ and $\bfAt$ in our key expression for correlations, Eq.~\eqref{E:lindner}, were  scalar matrices,  $\tilde C^0 I$ and $\tilde A I$, and could be  factored  -- leaving the connectivity structure of the network to determine the correlation value.  In particular, this structure impacts correlations via the matrix products of adjacency matrices in  Eq.~\eqref{E:power_series_exp}.

Biological neural networks are heterogeneous.  Even if the neurons are  identical before they are coupled, any heterogeneities in the coupling structure will lead to different neurons firing with different rates and hence power spectra (cf.~\cite{Zhao:2011}, which we return to in the Discussion).  Moreover, as a consequence, they will also have different levels of responsivity.  In sum, the matrices $\bfCt^0$ and $\bfAt$ will not have identical entries on the diagonal.

Consequently, the equivalent of the expansion Eq.~\eqref{E:power_series_exp} takes the form 
\begin{equation}\label{E:power_series_exp2}
\bfCt^\infty= \sum_{i,j = 0}^\infty (\bfAt \bfW)^i \bfCt^0 (\bfW^T \bfAt^*)^j.
\end{equation}
As a result, a purely graph theoretic interpretation of the terms in the expansion -- that is, one based on the connectivity {\it alone} -- is no longer possible. Recall that as in Eqns.~(\ref{q1}-\ref{q3}), motif frequencies are associated with terms like $\bfW^i(\bfW^T)^{j}$.  Here the powers $(\bfAt \bfW)^i$ are of  a``weighted" connection matrix (with weights corresponding to the responsivity of different cells). Moreover, $\bfCt^0$ is no longer a scalar matrix and in general does not commute with $\bfW$, introducing additional complications that we address below.

In this section we discuss how to 
extend our results to the case of  heterogeneous neural populations.

\subsection{Performance of the homogeneous approximation}

A first attempt at coping with heterogeneity is to hope that it is unimportant, and to apply the homogeneous theory of Eq.~\eqref{E:power_series_exp} naively.  To do this, we need to choose approximate (scalar) values for the power spectrum $\Ct^0$ and responsivity $\At$ that we will apply to all of the cells, by plugging into the homogeneous network formula given by Eq.~\eqref{E:power_series_exp}.  We choose the $\Ct^0$ as the unadjusted power spectrum (due to the normalization \eqref{rho_def}, the actual value of $\Ct^0$ is not important), rather than using one homogeneous value for all cells. For $\At$, we use the geometric mean of the $\At_i$, which are found from self consistent equations similar to Eq.~\eqref{e.linsol} (see \citep{Trousdale:2012} for details). The results of this naive application of the homogenous theory are shown in panel {\bf A} of Fig.~\ref{hom_heter}. Although general trends are  captured, this approach does not give an accurate approximation.

We note that here coupling is relatively strong (\ER{} formula for spectral radius giving 0.4, see Sec.~\ref{s: stability}). With weaker coupling (spectral radius=0.3), and hence less network-driven heterogeneity, the homogeneity assumption provides improved approximations in the heterogeneous case ($R^2$ measure 0.78, 0.67, 0.67 for EE, EI and II average correlation respectively, data not shown). To quantify the heterogeneity of cellular dynamics across the networks, we compute the coefficient of variation $CV$ of $\At_i$ and $\Ct^0_i$ averaged over network samples ($CV=\langle \mbox{standard dev.} / \mbox{mean} \rangle$). For spectral radius=0.4, the CV is 0.35 and 0.38 for $\At_i$ and $\Ct^0_i$, respectively, while  at spectral radius=0.3, the CV is 0.23 and 0.26 for $\At_i$ and $\Ct^0_i$.  At such levels of heterogeneity, we clearly need a more systematic approach, and we develop this next.

\subsection{Heterogeneous theory}

We are prevented from applying Proposition~\ref{P:E_full} to  Eq.~\eqref{E:power_series_exp2} in the heterogeneous case,
due to the presence of the factor $\bfCt^0$ in the middle of the terms on the right
hand side.  To deal with this difficulty we use the substitution $\bfCt^0 = (\bfCt^0)^{1/2}(\bfCt^0)^{1/2},$ which is possible because the power spectrum is non-nonnegative.  We can  then rewrite Eq.~\eqref{E:lindner} as
\beqr
\nonumber \bfCt (0) &\approx  &   (\bfCt^0)^{1/2}(\bfCt^0)^{-1/2}(\bfI-\bfAt \bfW )^{-1}(\bfCt^0)^{1/2}(\bfCt^0)^{1/2}(\bfI-\bfW^T\bfAt)^{-1}(\bfCt^0)^{-1/2}(\bfCt^0)^{1/2} \\
& =&
 (\bfCt^0)^{1/2}\left(\bfI-(\bfCt^0)^{-1/2}\bfAt \bfW (\bfCt^0)^{1/2} \right)^{-1}  \left(\bfI-(\bfCt^0)^{1/2}\bfW^T\bfAt (\bfCt^0)^{-1/2} \right)^{-1}(\bfCt^0)^{1/2}, \nonumber \\ \label{e.het}
\eeqr
where we are again evaluating all quantities at $\om=0$, so that $\bfFt=\bfI$.

Let $\At_0$ be the geometric mean of $\At_i$, which we choose to normalize responsivity so that weighted quantities will have the same units (see below). We can then define an effective or ``functional" connection matrix, which has the same units as $\bfW:$
\begin{equation}\label{}
   \hat{\bfW}=(\bfCt^0)^{-1/2}\bfAt \bfW  (\bfCt^0)^{1/2}/ \At_0 \;,
\end{equation}
so that Eq.~\eqref{e.het} becomes  
\begin{equation}\label{s pull out}
  (\bfCt^0)^{1/2}(\bfI-\At_0\hat{\bfW})^{-1}(\bfI-\At_0\hat{\bfW}^T)^{-1}(\bfCt^0)^{1/2} \;.
\end{equation}
This expression will be much easier to study, as the diagonal matrix $\bfCt^0$ no longer appears in the middle.  We can thus expand the two terms, $(\bfI-\At_0\hat{\bfW})^{-1}$ and $(\bfI-\At_0\hat{\bfW}^T)^{-1}$
 to obtain an expression that has a form analogous to that of Eq.~\eqref{E:power_series_exp}. 

The only difference in the present case of a heterogeneous network is that the definition of motif frequency and connection probability will involve weighted averages. For example, the entries of $\hat{\bfW}$ are scaled version of entries of the connection matrix $\bfW$:  
$$\hat{W}_{ij} =  \frac{\At_i}{\At_0}\frac{\sqrt{\St_j}}{\sqrt{\St_i}} W_{ij} \; .$$  
As an example, a diverging motif $i \leftarrow k \rightarrow j$ has a weighted contribution of the form
\begin{equation}\label{E:motif_weighted}
\frac{\At_i \At_j}{\At_0^2}\frac{\Ct^0_k}{\sqrt{\Ct^0_i \Ct^0_j}}  W_{ik}W_{jk}.
\end{equation}
The ratio $(\At_i\At_j)/\At_0^2$ in Eq.~\eqref{E:motif_weighted} quantifies the relative responsivity of the recipient cells.  Hence, a particular diverging motif will be weighted more strongly (and provide a greater contribution to average correlation) if the recipient cells are more responsive to inputs. Similarly, $\Ct_i^0,\Ct_j^0,\Ct_k^0$ corresponds to the variance of spike counts in long time windows for the uncoupled cells: the weight is determined by the variance of the projecting cell divided by the geometric mean of the variance of the recipient cells. These observations agree well with intuition:  more responsive cells will be more strongly correlated by a common input, and ``source cells" with larger total variance ($\Ct^0_k$) will lead to a diverging motif with larger impact. 

The  motif frequencies $\qdiv$ and $\qch$, are defined by Eq.~\eqref{E:qdiv_average} and \eqref{E:qch_average} upon substituting $\bfW^0$ with $\hat{\bfW^0}=\hat{\bfW}/w$  in the single population case. In the case of two populations, the matrices $\bfQdiv$ and $\bfQch$  are defined by Eq.~\eqref{E:twopop_qdiv_def} and \eqref{E:twopop_qch_def}, with $\bfW$ replaced by $\hat{\bfW}$. 
These weighted motif frequencies could be estimated experimentally in an active network, via recordings from neurons known to participate in three cell motifs. The advantage of our resumming theory remains that everything needed is based only on the statistics of a relatively small number of cell motifs, rather than higher order information about the connectivity graph.

Due to the normalization in Eq.~\eqref{rho_def},  the two outside diagonal matrix factors in Eq. \eqref{s pull out} cancel in the definition of $\bfrho$, and we can use the following matrix to calculate correlation coefficients (noting that we continue to approximate $\bfCt_{ii}$ by $\Ct_i^0$ when performing the normalization in the denominator of Eq.~\eqref{rho_def}):
\beq\label{E:heter_renorm}
(\bfI-\At_0\hat{\bfW})^{-1}(\bfI-\At_0\hat{\bfW}^T)^{-1} \; .
\eeq

Eq.~\eqref{E:heter_renorm} is the heterogeneous analog of the expression for the correlation $\bfCt^\infty / \Ct^0$ studied in Sections~\ref{S:single_pop_theory} and \ref{S:two_pop_theory} (see Eq.~\eqref{cov_hom}), including diagonal contributions. Applying the motif resumming theory exactly as in these sections  yields corresponding approximations of mean correlation for these systems.  In particular, for the two population case, the mean correlation $ \rhoavg_{EE},\rhoavg_{EI},\rhoavg_{II}$ will be estimated  based on Eq.~\eqref{E:twopop_nl_resumming_so} with the substitution of weighted motif counts $\bfQdiv$ and $\bfQch$ (note that one must still adjust the approximation to account for diagonal terms --- see Appendix~\ref{s:rho_approx}).  In panel {\bf B} of Fig.~\ref{hom_heter}, we plot this prediction of $ \rhoavg_{EE},\rhoavg_{EI},\rhoavg_{II}$ compared with the full expression for correlation in the heterogeneous case (via Eqns.~\eqref{rho_def} and~\eqref{E:lindner}).  Accounting for dynamical heterogeneity across the network by defining {\it weighted} second order motifs provides a reasonably accurate prediction of average correlation.

\begin{figure}[H]
  \center
      \includegraphics[width=6in]{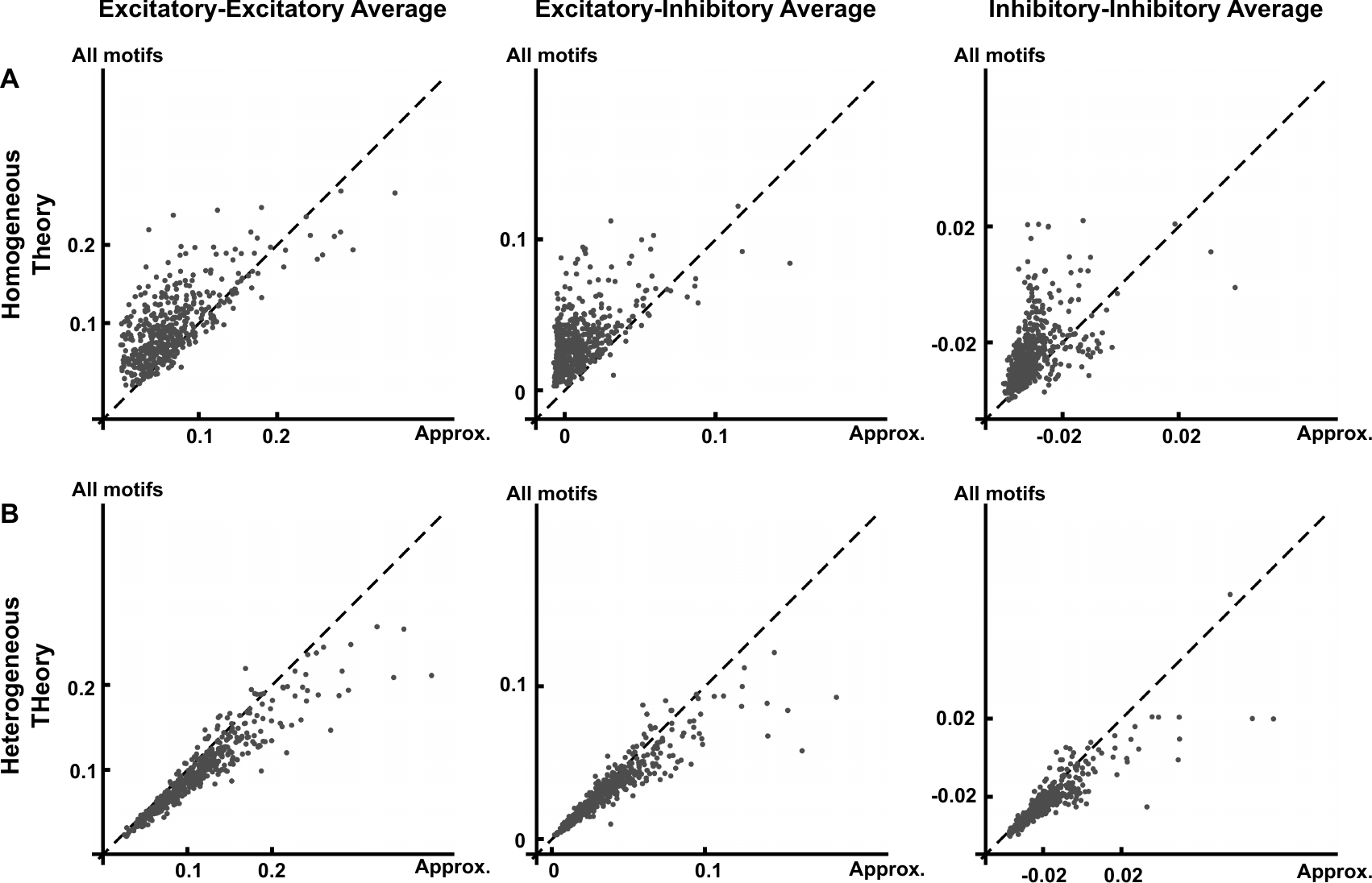}  
      \caption{ {\bf A} Mean correlation from homogeneous resumming theory (horizontal axis) comparing with that from full linear response theory (Eq.~\eqref{cov_hom}) (vertical axis). {\bf B} Mean correlation from heterogeneous resumming theory (horizontal-axis) comparing with that from full linear response theory (vertical-axis). The diagonal line $y=x$ is plotted for reference. The spectral radius of $\At\bfW$ under \ER{} assumption is 0.40 (see Sec.~\ref{s: stability}). Other network parameters are given in the caption of Fig.~\ref{F:bars1}. The coefficients of determination, $R^2,$ are 0.56, 0.44, 0.35 in panel {\bf A} and 0.88, 0.81, 0.73 for panel {\bf B}. }
\label{hom_heter}
\end{figure}

\section{Comparisons with IF simulations}\label{s:sim_theory}

In the preceding sections, we have shown how network-wide correlation can be approximated based on coupling probability together with the frequency of second-order motifs.  In assessing the accuracy of this motif-based theory, we have compared the predictions of the theory with the value of correlation given Eq.~\eqref{E:lindner}, which is exact in the sense that it precisely includes the contribution of motifs at all orders and therefore gives an exact description of linearly interacting point process models \citep{Pernice:2011,Hawkes:1971-1}.

\begin{figure}[H]
\center
\includegraphics[width=6in]{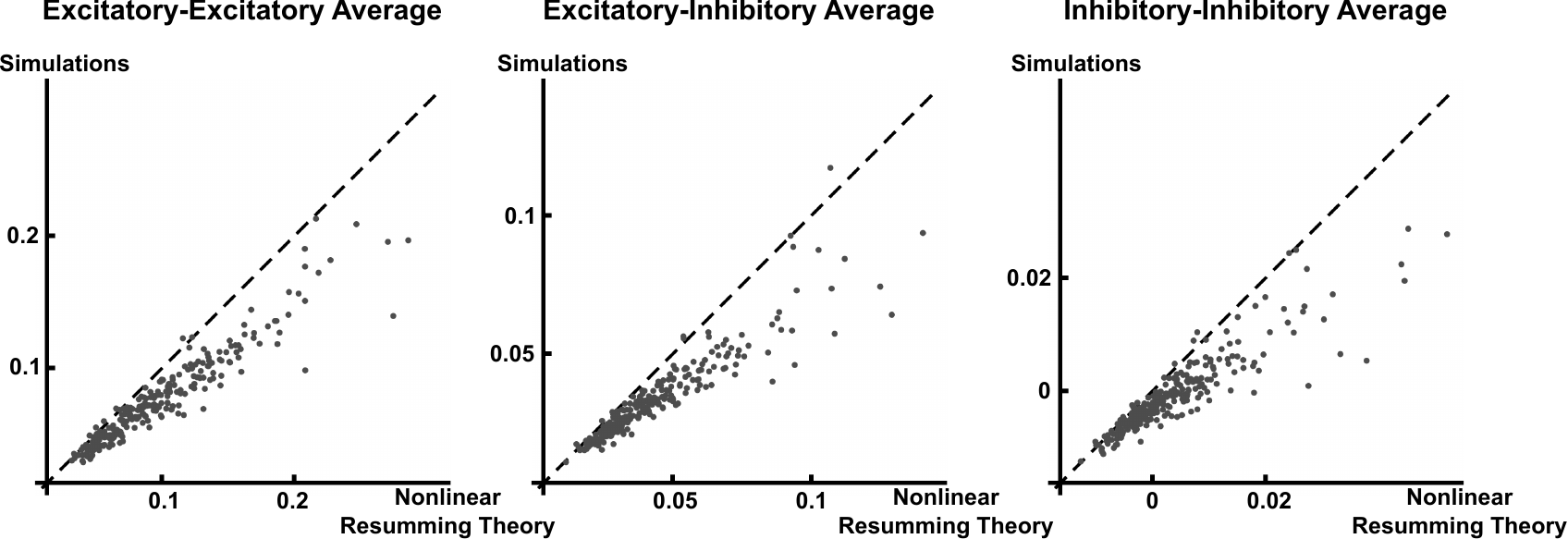}  
\caption{Average correlation from integrate and fire neuron simulations compared with the predictions
of Eq.~\eqref{E:twopop_nl_resumming_so}.  The vertical axis is the mean correlation coefficient calculated from simulations of the same 265 excitatory-inhibitory networks studied in Fig.~\ref{first figure}.  The horizontal axis is the prediction for mean correlation from the resumming theory based on empirical connection probability and second order motif frequencies.  The diagonal line $y=x$ is plotted for reference.   The spectral radius of $\At\bfW$ under the \ER{} assumption is 0.33 (see Sec.~\ref{s: stability}). Coefficients of determination $R^2$ are 0.91, 0.87, 0.79 respectively for the three panels.}
\label{sim_nrs}
\end{figure}

When using our theory to predict the impact of motifs on mean correlation for networks of \IF neurons, we are making an additional approximation in describing integrate-and-fire dynamics with linear response theory~\citep{Trousdale:2012,Ostojic:2009,Lindner:2005}.  We now directly test the performance of our motif-based predictions in \IF networks -- thus probing how errors at each of the two layers of approximation combine.  Specifically, in Fig.~\ref{sim_nrs} we compare the block-wise mean correlations from IF simulation and the predications obtained using Eq.~\eqref{E:twopop_nl_resumming_so}. These simulations are for the same networks used in Fig.~\ref{first figure}. We find that our theory gives a good quantitative match with LIF simulations, despite the multiple approximations involved.  Thus, our theory predicts trends in the impact of motif frequencies on network-wide correlation.

\section{Discussion}

\subsection*{Summary:  Predicting network-wide correlation from three cell motifs}

We studied the impact of the graphical structure of neural networks -- characterized by connectivity motifs -- on their dynamical structure, characterized by correlated activity between cells. As shown in Fig.~\ref{figure0}, varying the frequency of such motifs can strongly impact correlation, over and above the overall level of connectivity in a network.  Following \cite{Zhao:2011}, we focus on the three types of motifs that involve two connections each: the diverging, converging and chain motifs.

We chose a standard spiking neuron model, the integrate-and-fire (IF) neuron, in constructing our recurrent networks (Sec.~\ref{S:linresponse}). For IF neurons~\citep{Lindner:2005,Trousdale:2012}, LNP neurons~\citep{Beck:2011}, and other neuron models such as linearly interacting point process model~\citep{Hawkes:1971-1,Hawkes:1971-2}, one can apply linear response approximations (or, in the case of Hawkes models, solve exact equations -- see ``Neuron models" below) to get an explicit expression for the pairwise correlations depending explicitly on the connectivity matrix (see Eq.~\eqref{E:lindner}).

We expand this expression in a series, where each term has clear correspondence to certain graphical structures~\cite{Pernice:2011,Rangan:2009-1}. In particular, second order terms correspond to second order motifs. Importantly, we show that contributions of higher order terms can be estimated using the frequencies of second order motifs along with the connection probability.

For systems with correlations well-approximated by Eq.~\eqref{E:lindner} and assuming homogeneous cellular dynamics, we find that the frequency of converging motifs, over and above that expected in \ER{} networks, will have no effect on mean correlation in systems (if the connection probability and other motif frequencies are fixed).  Meanwhile diverging and chain motifs will contribute positively. In networks of excitatory and inhibitory neurons,  the three types of motifs are subdivided according to the  type of the constituent neurons.  However, 
average correlations between cells of a given type are still given in terms of
diverging and chain motifs (see Fig.~\ref{first figure}).

We first analyzed networks in which we made a strong homogeneity assumption on the dynamical properties of the uncoupled neurons. The resumming theory we develop approximates the contributions of higher order motifs in terms of the frequency of second order motifs. In Sec.~\ref{heter_theory}, we extended our theory to heterogeneous networks. In such cases, the contribution of one instance of a certain motif to the total motif frequency will be additionally weighted by the relative responsiveness of the neurons composing the motif which \emph{receive} input, as well as the baseline variance of the cells. Overall, our results can be regarded as a general estimator of mean correlation given motif statistics up to second order.

To test our theory numerically, we generated random networks of IF neurons with fixed statistical (i.e., expected) connection probability, but different second order connectivity statistics (see Sec.~\ref{sec_graph}). Simulations  show that the theory provides an accurate description of network correlations (see Fig.~\ref{sim_nrs}), despite the additional error introduced by the linear response approximation of activity (Eq.~\eqref{E:ansatz}). We also compared the resumming theory to the direct evaluation of the linear response theory (Eq.~\eqref{E:lindner}) which takes account of the full graphical structure. The close match between the two (Fig.~\ref{F:E_nrs}, Fig.~\ref{F:EI_nrs}) shows that second order motifs capture much of the dependence of mean correlation on network connectivity.  Moreover, Figs.~\ref{figure0}, \ref{F:E_nrs}, \ref{F:EI_nrs} demonstrate that variations in the frequency of second order motifs produce changes in network correlation that dominate those expected from the small variation in empirical (i.e., realized) connection probability from one network to the next.

Beyond the resumming theory, we also considered two other ways of simplifying the expansion of Eq.~\eqref{E:lindner}. The first was a truncation in connection strength at second-order (in powers of $\At$ and $\bfW$; Secs.~\ref{s:E_trunc}, \ref{S:EI_setup}).  This eliminates all contributions due to motifs of length two or less.  Except in very weakly connected networks,
this is a poor approximation:  Although the contributions of higher order motifs decay exponentially in interaction strength $\At w$, their number also grows exponentially with $Np$.  Thus,  motifs of all orders could, in principle, contribute substantially to average correlations.  Our second truncation was the \ER{} approximation of mean correlation.  This yielded  predictions for mean correlation that included contributions from motifs of all orders.  These predictions therefore depend only on connection probability (See Eqns.~(\ref{e:ER_E}, \ref{E:twopop_nl_resumming_ER})), and 
should be valid when there is very little structure in the connectivity graph compared to an idealized \ER{} network.

\subsection*{Neuron models}

Our motif-based theory can be applied to a variety of different neuron models.
The starting point of our theory, the expression for pairwise cross-spectra given by Eq. \eqref{E:lindner}, arises in a number of settings.

The main tool in our approach is linear response theory.  This connects our methods to \IF models (see Sec.~\ref{s:linear_response_1}).  Importantly, Eq. \eqref{E:lindner} also arises as an {\it exact} expression for linearly interacting point process, or {\it Hawkes} models~\citep{Hawkes:1971-2, Pernice:2011,Pernice:2012}.  In this case,
$$\label{Hawkes_cov}
\bfCt (\omega) = (\bfI-\At(\omega) \bfW)^{-1}\bfY(\bfI-\At(\omega) \bfW^{T})^{-1} \; ,
$$ where $\At(\omega)$ is the Fourier transform of the interaction filters and $\bfY$ is a diagonal matrix of firing rates. Therefore our analysis can be directly and exactly applied to this setting.  We note that the  Hawkes process is a linear-Poisson (LP) model.  Such models are commonly applied in theoretical neuroscience~\citep{Day+01}.

\subsection*{Relationship to other studies on motifs and networks correlation}

Our methodology is very similar to the previous study of~\cite{Zhao:2011}. They considered the impact of diverging, converging and chain motifs (along with the reciprocal connection motif) on the ability of an excitatory recurrent network to stay synchronized. They found that prevalence of the converging motif decreases synchrony, prevalence of the chain motif increases synchrony, and that the diverging motif has no significant effect on synchrony. The difference between their results and ours can be understand from the different dynamical regimes considered. ~\cite{Zhao:2011} considered perturbations from two extreme cases: perfect synchrony, and evenly distributed asynchronous oscillators.  In contrast, we studied the asynchronous regime but allowed the activity of the cells to be correlated.  Hence, our methods also differ:  we use a linear response approach valid for strong internal noise, and weak to intermediate coupling, while~\cite{Zhao:2011} performed a linear stability analysis of coupled oscillator equations.

Many other studies have also examined the relationship between graphical features and spike correlations in networks. \cite{Roxin:2011} studied how spike-time correlation changes when one interpolates the degree distribution of a network from the binomial distribution (corresponding to \ER{} networks) to a truncated power law distribution. He found that increasing the variance of the out-degree distribution will increase the cross-correlation, which is consistent with our results for the diverging motif (see Eq.~\eqref{div_degree}).

\cite{Pernice:2011} and \cite{Trousdale:2012} studied the influence of connection structures on pairwise correlation for a number of network classes. \cite{Pernice:2011}  used the Hawkes neuron model, which as noted above, leads to very similar expressions for correlation as those studied here and by~\cite{Trousdale:2012}. Both approaches relate pairwise correlations  to certain graphical structures of increasing order (i.e., motif size).  In particular, \cite{Pernice:2011}  obtained an expression for average network correlation in terms of the mean input and common input (Eq. (22) in \citep{Pernice:2011}) for regular networks with uniform connection probability, while~\cite{Trousdale:2012} (Eq.~(25) in \cite{Trousdale:2012}) considered correlations in networks where only in-degree was fixed. Both are special cases of our resumming theory  (Note that according to Eq.~\eqref{div_degree} and \eqref{con_ch_degree}, a fixed in (resp. out) degree is equivalent to $\qcon=0$ (resp. $\qdiv=0$) and $\qch=0$). 

A major contribution of the present study is to show that effects of ``higher order" graphical interactions (i.e., motifs including more than three cells) can be approximated in terms of the frequency of second order motifs and the overall connection probability.  This allows a systematic treatment of network-averaged correlation for a broader range of network connectivities.

\subsection*{Limitations}
When applied to integrate and fire neuron networks, our our analysis relies on the validity of the linear response approximation.  In Section~\ref{s:sim_theory}, we demonstrated this validity for a particular firing regime, for the class of random networks studied here.  For more on this issue, see~~\citep{Trousdale:2012} and~\citep{Ostojic:2011}.  We note that one avoids this issue entirely when considering Hawkes processes (see above).

We also assumed that the spectral radius of the total coupling matrix is less than one, in order to expand Eq.~\eqref{E:lindner} into a series, for which each term can be attributed to a different motif.  From this point, our methods rely on our ability to predict the impact of higher order motifs on network correlation based on the frequency of second order motifs.  We demonstrated that our resumming theory can successfully make this prediction for classes of networks generated in two ways:  via two-sided power law distributions, and via the SONET method (see Sec.~\ref{gen_netw}).  However, for certain connectivity matrices, our resumming theory can produce  large errors. An example pointed out to us by Chris Hoffman [personal communication] is $\bfW^0$ containing a $\sqrt{p}N\times \sqrt{p}N$ block of 1 entries and but taking value 0 everywhere else; here, $p$ is the overall connection probability. Note that this matrix corresponds to a graph with one fully connected group and another fully isolated group.  Such disconnected architecture and strong inhomogeneity may be features that produce large spectral radii $ \Psi( \bfW \bfTheta)$, and hence large errors when applying our theory (see also Appendix~\ref{s:spec_intui}; and for a study on the dynamics of inhomogeneous clustered networks, \citep{Watts:1998}).

Three other factors limit the generality of our results.  First, in order to separate the effect of network structure from that of cellular dynamics we initially assumed homogeneous firing rates and identical neurons.  For many real neural networks, this may be a poor approximation.  Thus, in our analysis of heterogeneous networks, we calculate weighted motif strengths, {\it assuming} ``{\it a priori}" knowledge of the (heterogeneous) cellular firing rates and responsivities.  A full theory would rather begin by {\it predicting} this heterogeneity based on network properties.  Intriguingly, \citep{Zhao:2011} show that certain motif frequencies can be a powerful source of such heterogeneity, a potentially important fact that we neglect.  

Second, for simplicity we only considered long-time window spike count correlation. However, our analysis can be easily generalized to study correlation at any time scale, and for any $\omega$ in Eq.~\eqref{E:lindner}. A third, and final limitation of our analysis is that we only studied  pairwise correlation coefficients. This may not adequately describe the network dynamics or reveal the full importance of certain motifs.

\subsection*{Extensions and connections with neurobiology and computation}
Intriguingly, experiments have shown that motif statistics in actual neural networks deviate from the \ER{} network model, opening a possible role of multicellular motifs.  For example,~\cite{Song:2005} found such deviations in connectivity between excitatory cells in visual cortex. \cite{Sporns:2004} also suggest that increased frequencies of certain motifs may be a common feature of neural circuits. Moreover, \citep{Haeusler:2009} studied the different motif statistics from two experimental data of connectivity structures of laminae in cortex, showing deviations from expectations under the \ER{} model without the laminar structure (implying as a possible origin of non-trivial motif frequencies).   Our study can be applied to suggest a possible consequence of these experimental findings for levels of spike-time correlation (in the latter case generalizing it to apply to multi-group networks corresponding to the different laminae).

\citep{Renart:2010} studied pairwise correlations in balanced excitatory-inhibitory \ER{} networks, and found that balanced neuronal networks will have average correlations that decrease rapidly to zero with network size . Our results suggest that, with increased propensity of certain second order motifs, EI networks can potentially produce significantly larger correlations value (see Fig.~\ref{figure0}). However, to apply our findings to the case considered by \citep{Renart:2010}, we would need to make several extensions, including allowing full heterogeneity arising from recurrent connections and checking that linear response theory works in their strongly coupled, balanced dynamical states. In our setting, ``balanced" can be defined as that the total mean input received for each cell is 0, exactly or on average across realizations of the network \citep{Rajan:2006}. This needs to be compared with the dynamic balance concept in \citep{Renart:2010}.

For future studies, we also hope to develop a theory that predicts not just the mean correlation strength across a network, but its variance across cell pairs.  This variance has been studie using theoretical and experimental approaches~\citep{Renart:2010, Ecker:2011}, and it would be interesting to describe how it depends on connection motifs. We will also try to predict the heterogeneity in cellular dynamics caused by motif frequencies, as referred to above -- incorporating, for example, the result emphasized in \citep{Zhao:2011} that variability in input to different neurons depends on converging motifs (Eq.~\eqref{con_ch_degree}).  Finally, we note the important connections between correlations and coding efficiency \citep{Zoh:94,abbott:99,Salinas:2000,JosicSDR08,Sch:2006} (see also Introduction). The recent work of~\citep{Beck:2011} set up a direct connection between the graphical structure of networks and their coding efficiency, using a similar linear response calculation of covariance matrix for linear nonlinear Poisson (LNP) neurons.  The present results cold be used with the approach of~\cite{Beck:2011} to further link the statistics of motifs to properties of signal transmission in neural networks.

\appendix
\section{Approximating $\rhoavg$ from $\langle \bfCt^\infty \rangle$}
\label{s:rho_approx}
 The  average covariance across the network,
$\langle \bfCt^\infty \rangle$ can be used to approximate $\rhoavg$.  Here we describe two such approximations.
 First, when the uncoupled neurons have equal firing rates, and the perturbation from recurrent coupling is weak, the diagonal terms $\bfCt_{ii}^\infty$ will be close to the unperturbed values, $\Ct^0$. In this case, subtracting diagonal terms from the average, we have that
\beq\label{rho_1}
\rhoavg \approx \left( \frac{\langle \bfCt^\infty  \rangle}{\Ct^0}-\frac{1}{N} \right)\cdot \frac{N}{N-1}.
\eeq

In a second, more accurate approximation, we assume permutation symmetry between neurons within one population.
Also, in our networks, self-connections are allowed and occur with the same probability as other connections.
These will lead to identical (marginal) distributions for each entry in the covariance matrix $\bfCt^\infty$, excepting the diagonal entries which are shifted by a constant of $\Ct^0$ due to each neuron's own unperturbed variance (this corresponds to the term proportional to $\bfI$ if one were to expand Eq.~\eqref{cov_hom} as a series --- see Eq.~\eqref{E:power_series_exp}).  
This suggests that $\bfCt^\infty$ has the form
\beqn
 \bfCt^\infty \approx \tilde C_0 \bfI+c \bfone_{NN},
\eeqn
where $\bfone_{NN}$ is the $N\times N$ matrix of all ones, and $c$ is a constant.
If this holds, then, using diagonal entries of this matrix as normalization, we obtain 
\beq \label{rho_2}
\rhoavg \approx \frac{\langle \bfCt^\infty \rangle-\Ct_0/N}{\Ct_0+ \langle \bfCt^\infty \rangle-\Ct_0/N}.
\eeq

The two approximations given in Eqns.~(\ref{rho_1},\ref{rho_2}) are approximately equal for small correlations. We will use Eq.~\eqref{rho_1} to exhibit the linear dependence between mean correlation coefficient and motif frequency (such as linear weights in Fig.~\ref{first figure} Part B) and Eq.~\eqref{rho_2} for quantitative predications (all numerical plots).

For networks consisting of an excitatory and inhibitory population, we have analogs of Eqns.~(\ref{rho_1},\ref{rho_2}) for each of the 4 population blocks of the covariance matrix,
\beqr
\nonumber \rhoavg_{XX} &\approx& \left( \frac{\langle \bfCt_{XX}^\infty  \rangle}{ \Ct^0}-\frac{1}{N_X} \right)\cdot \frac{N_X}{N_X-1},\\
\rhoavg_{XY} &\approx& \langle \bfCt_{XY}(0)  \rangle/ \Ct(0)  \label{e:r_ei_1},
\eeqr
and
\beqr
\nonumber \rhoavg_{XX} & \approx& \frac{\langle \bfCt_{XX} \rangle-\Ct_0/N_{X}}{\Ct_0+ \langle \bfCt_{XX} \rangle-\Ct_0/N_X},\\
 \rhoavg_{XY} & \approx& \frac{\langle \bfCt_{XY} \rangle}{\sqrt{(\Ct_0+ \langle \bfCt_{XX} \rangle-\Ct_0/N_X)(\Ct_0+ \langle \bfCt_{YY} \rangle-\Ct_0/N_Y)}},     \label{e:r_ei_2}
\eeqr
where $X\neq Y\in \{E, I\}$.

\section{Proof of bound on $\qdiv,\qcon,\qch$ for one population}
\label{s:q_range}

We here prove the inequalities in Eq.~\eqref{q_range}.
Noting that we may write
$$
\qdiv = \EVE{\bfW^0_{i,k} \bfW^{0}_{j,k}} - \EVE{\bfW^0_{i,k} }\EVE{\bfW^{0}_{j,k}},
$$
(with similar expressions holding for $\qcon,\qch$) it is sufficient to show $\EVE{\bfW^0_{i,k} \bfW^{0}_{j,k}} \leq p$. Note
\beq
\label{e.intermediate}
\bfH \bfW^0 (\bfW^0)^T \bfH=\left(N^2\EVE{\bfW^0_{i,k} \bfW^{0}_{j,k}}\right)\bfH,
\eeq
where $\bfH$ is defined in  Eq.~\eqref{E:h_theta_def}. Let $\lVert \cdot \rVert_{\mathrm{F}}$ be the Frobenius norm, which is sub-multiplicative. We then have
\beq \label{F_norm}
\lVert\bfH \bfW^0 (\bfW^0)^T \bfH \rVert_{\mathrm{F}}\leq
\lVert\bfH \bfW^0\rVert_{\mathrm{F}} \lVert(\bfW^0)^T \bfH \rVert_{\mathrm{F}} \leq
\lVert\bfH\rVert_{\mathrm{F}} \lVert \bfW^0\rVert_{\mathrm{F}} \lVert(\bfW^0)^T\rVert_{\mathrm{F}}  \lVert\bfH \rVert_{\mathrm{F}}.
\eeq
Since $\lVert \bfH \rVert_{\mathrm{F}}=1$, $\lVert \bfW^0 \rVert_{\mathrm{F}}=\sqrt{\sum_{i,j}(W^0_{i,j})^2}=N \sqrt{p}$, the above inequality, together with \eqref{e.intermediate}, gives $\EVE{\bfW^0_{i,k} \bfW^{0}_{j,k}} \leq p$.  With the fact that $\EVE{\bfW^0_{i,k} }\EVE{\bfW^{0}_{j,k}} = p^2$, we have the bound in Eq.~\eqref{q_range}.

In the second inequality of \eqref{F_norm}, we have used $\lVert\bfH \bfW^0\rVert_{\mathrm{F}}\leq \lVert\bfH\rVert_{\mathrm{F}} \lVert \bfW^0\rVert_{\mathrm{F}}$. The necessary and sufficient condition for achieving equality is the equality condition in the following Cauchy inequalities for all $i,j$
\beqn
\left(\sum_{k}H_{i,k}W^0_{k,j}\right)^2\leq \left(\sum_{k}H_{i,k}^2\right)\left(\sum_{k}(W^0_{kj})^2\right).
\eeqn
Equality holds exactly when $W^0_{k,j}=W^0_{l,j}$ for all $k,l$ -- that is, each column of $\bfW^0$ has same values (either all ones or all zeroes). The equality condition for the first inequality in Eq.~\eqref{F_norm} is identical. To generate an example graph that achieves equality for a specified overall connection $p$, we simply set a fraction $p$ of the columns to be all ones, and set the remaining columns to zero.

Similarly, for converging motifs we can show $\EVE{\bfW^0_{k,i} \bfW^{0}_{k,j}} \leq p$ with the same equality condition as above, and for chain motifs, $\EVE{\bfW^0_{i,k} \bfW^{0}_{k,j}} \leq p$. However, the equalities for $\qdiv,\qcon$ cannot hold simultaneously, and the equality for $\qch$ cannot be achieved.

\section{Graph generation methods}
\label{s:graph_gen}

Here we present more details on how we generated network samples with fixed connection probability,
but different frequencies of second order motifs.  We used two methods.

First, the degree distribution method consists of initially generating a sample of in and out degrees from a truncated power law distribution with density, following \citep{Zhao:2011}, 
\beq
\label{e.marginals}
f(d)=\left\{
\begin{array}{cc}
C_1 d^{\gamma_1}& 0 \leq d \leq L_1 \\
C_2 d^{\gamma_2}& L_1\leq d \leq L_2 \\
0& \text{otherwise}
\end{array}
 \right. ,
\eeq
where $d$ is the in or out degree (see also the configuration model \citep{Roxin:2011,Newman:2003,Newman:2001}).
The two marginal distributions of in and out degree are then coupled using a Gaussian copula with correlation coefficient $\rho$ to generate the in and out degree lists. The parameters $\rho$, $L_1/L_2,L_2$, $\gamma_1>0,\gamma_2<0$ are independently and uniformly sampled for each network, separately for in and out degrees (their ranges are listed in Table~\ref{t:deg_range}). $C_1$ and $C_2$ are chosen so that $f(d)$ is continuous at $L_1$ and the mean of the degree distribution is normalized to $Np$ (the same value for both in and out degrees), where $p$ the fixed connection probability across network samples and $N$ is the network size. 

We then use the degree lists to calculate a probability for each possible connection from cell $j$ to cell $i$, which is proportional to $\din_i \dout_j$. All these $N^2$ probabilities are scaled so that the resulting average for the total number of connections is the same as the quantity $N^2p_{\mathrm{stat}}$. Here $p_{\mathrm{stat}}$ is the target connection probability that we aim to achieve in these samples; we recall that the ``empirical" connection probability achieved in a given graph is notated as $p$. 
\begin{table}[H]
\begin{center}
\begin{tabular}{|c|c|c|c|c|c|}
\hline
 &$\rho$ &$\gamma_1$ & $\gamma_2$ & $L_1/L_2$ & $L_2/N$  \\ \hline
min & -1 & 0.25 & -2.25  & 0 & 0.7  \\ \hline
max & 1 & 2.25 & -0.25  & 1 & 1  \\ \hline
\end{tabular}
\caption{Range of parameters of the truncated power law degree distribution.}
\label{t:deg_range}
\end{center}
\end{table}

For the excitatory-inhibitory case, we generate four degree lists $\din_{E},\dout_{E},\din_{I},\dout_{I}$, again according to the marginal distributions \eqref{e.marginals}.  We therefore need a four dimensional Gaussian copula. Again, parameters for the power law distribution and the correlation coefficient matrix of the copula are randomly chosen. Using these degree list, we can then generate each of the four blocks (defined by cell type) of adjacency matrix in the same way as in the single population case. This method allows us to sample from the whole extent of motif parameters (3 in single population case, 20 in two population case).

For the single population networks studied in Sec.~\ref{S:single_pop_theory}, we generated additional network samples via the SONET method~\citep{Zhao:2011}. The idea of the algorithm is similar to maximum entropy models. Given only the connection probability and second order motif frequency, we try to generate the most ``random'' network satisfying these moment constraints. However, instead of using the Gibbs distribution for the connections as in maximum entropy method, we use a dichotomized ($N^2$ dimensional) Gaussian distribution. We then generate samples of excitatory only networks from the complete range of possible motif frequencies, as described in Eqns.~\eqref{e:q_dependence}-\eqref{q_range}, using the SONET algorithm.

Network samples generated using both methods cover the range of motif frequency observed experimentally in cortical circuits~\citep{Song:2005,Zhao:2011} as shown by Table~\ref{t:motif_range}.   Here, we list motif frequency values as $(\qdiv,\qcon,\qch)/(p(1-p))$ for excitatory only networks, and $(\qdiv^{EE,E},\qcon^{EE,E},\qch^{EEE})/(p_{EE}(1-p_{EE}))$ in excitatory-inhibitory networks (since \citep{Song:2005} only recorded from excitatory neurons, see definition at Eq.~\eqref{E:twopop_qdiv_def}) .

\begin{table}[H]
\begin{center}
\begin{tabular}{|c||c||c|c||c|c||c|c||}
\hline
& &\multicolumn{2}{|c||}{SONET: E only}& \multicolumn{2}{c||}{degree: E only } & \multicolumn{2}{c||}{degree: EI} \\ \hline
 & experiment  & min & max & min & max & min &max \\ \hline
diverging    & 0.033 & 0.001 & 0.913 & 0.015 & 0.181 & 0.018 & 0.295 \\ \hline
converging & 0.044 & 0.001 & 0.838 & 0.016 & 0.189 & 0.022 & 0.279 \\ \hline
chain           & 0.022 &-0.192 & 0.248 &-0.068 & 0.095 &-0.082 & 0.110 \\ \hline
\end{tabular}
\caption{Range of motif frequencies in network samples. ``E only" ( or ``EI") means excitatory only networks (or excitatory-inhibitory networks). ``SONET" and ``degree" refer to the two methods of generating networks.}
\label{t:motif_range}
\end{center}
\end{table}

\section{Compensating for fluctuations in empirical connection probability}
\label{s:adjust_q}
 In order to isolate the impact of higher order motifs, all network samples should have the same empirical connection probability $p$, \emph{i.e.} the same number of connections, as we now explain. The algorithms used to generate networks produced samples with slightly different empirical connection probabilities.  These fluctuations impact the linear relationship between motif frequencies and average correlation, as $p$ factors into the regression coefficients (See Eq.~\eqref{E:single_pop_expansion}). When attempting to determine the regression coefficients from data, fluctuations in $p$ affect the linear trend.

\begin{figure}[H]
\centering
\includegraphics[width=6in]{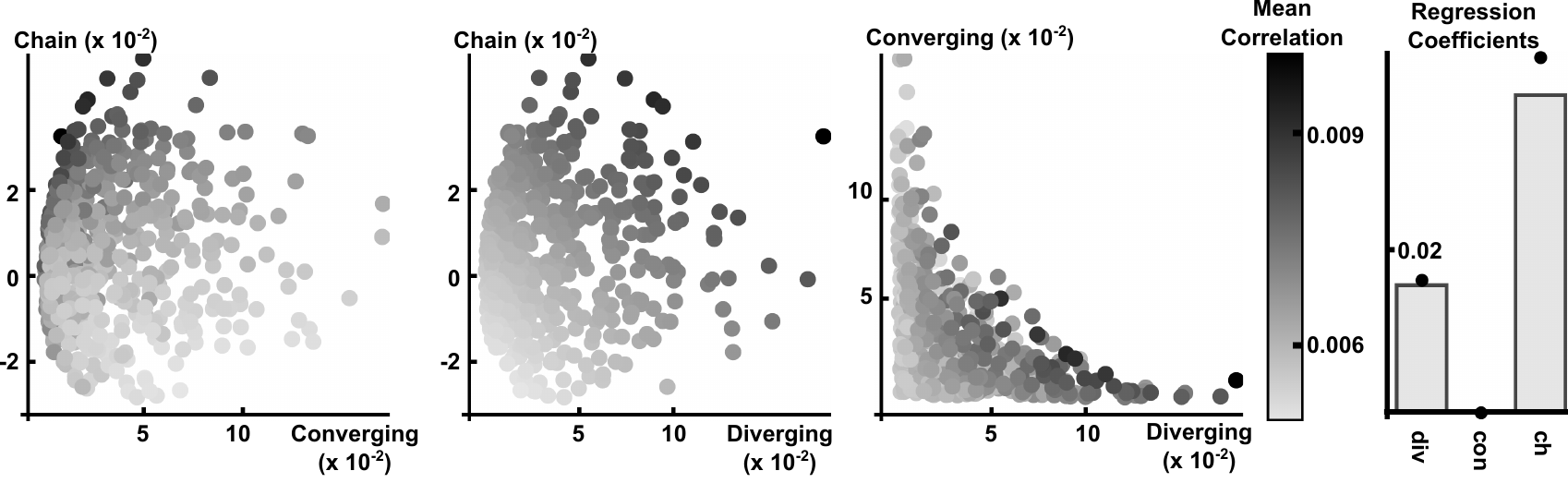}
\caption{Reproduction of Fig.~\ref{F:E_bar} where we have scaled motifs to account for fluctuations in $p$, as described in the text.}
\label{F:E_bar_2}
\end{figure}

To address this issue, we can scale motif frequencies by weighting them so that the contribution of a motif to mean correlation does not have a regression coefficient depending on $p$. For example,  suppose that the theory predicts a linear  relation of the form
$$
\frac{\langle \bfCt^\infty \rangle}{\Ct^0} = f_0(p) + f_{ch}(p)\qch + f_{div}(p)\qdiv.
$$
Then we may define auxillary motif frequencies
$$
q_x' = \frac{f_x(p)}{f_x(p_{stat})}q_x,
$$
and also replace $f_0(p)$ with $f_0(p_{stat})$ so that the theoretically-predicted regression relationship becomes
$$
\frac{\langle \bfCt^\infty \rangle}{\Ct^0} = f_0(p_{stat}) + f_{ch}(p_{stat}) \qch' + f_{div}(p_{stat})\qdiv'.
$$
Here, we have adjusted the definition of a motif frequency in order to account for finite-size fluctuations in the empirical connection probability $p$. The linear regression fit to the quantities $\qch',\qdiv'$ is much improved, achieving an $R^2$ measure of 0.99, up from 0.8 in the case where we did not account for such fluctuations. In Fig.~\ref{F:E_bar_2}, we show the scatter plots exploring the relationship between motifs and mean correlation after performing this scaling, and observe the same key trends of mean correlation as in the unadjusted Fig~\ref{F:E_bar}: strong and positive dependence on chain and diverging motifs, and minimal dependence on converging motif. These trends are now presented even more clearly.

\section{Proof of Proposition~\ref{P:E_full}}
\label{s:prop_proof}

We will make use of the following lemma, which may be verified by direct computation.
\begin{lemma} \label{L:lemma}
Let $\{ x_n \}_{n\ge 1},~\{ y_m \}_{m\ge 1},~\{ z_{nm} \}_{n,m \ge 1}$ be sequences which converge absolutely when summed, and also satisfy
$$
\left|\sum_{n=1}^\infty x_n\right| < 1,\quad \left|\sum_{m=1}^\infty y_m\right|<1.
$$
Then,
\beqn
\sum_{i=1}^\infty \sum_{(n_1,\ldots,n_k)\in \{i\}} \left(\prod_{s=1}^k x_{n_s}\right) = \sum_{i=1}^\infty \left(\sum_{n=1}^\infty x_n\right)^i,
\eeqn
\beqrn
&&
\sum_{i,j=1}^\infty \sum_{\substack{(n_1,\ldots,n_{k+1})\in \{i\} \\ (m_1,\ldots,m_{l+1})\in \{j\}}} \left[ \left(\prod_{s=1}^k x_{n_s}\right)  z_{n_{k+1}m_{l+1}}  \left(\prod_{t=1}^l y_{m_t}\right) \right] 
\\
&& = \left[\sum_{i=0}^\infty \left(\sum_{n=1}^\infty x_n\right)^i\right] \left(\sum_{n,m=1}^\infty z_{nm} \right) \left[\sum_{j=0}^\infty \left(\sum_{m=1}^\infty y_m\right)^j\right].
\eeqrn
where the sum over $(n_1,\ldots,n_k)\in \{i\}$ denotes a sum over all ordered partitions of $i$, $(n_1,\ldots,n_k)$, of length $1 \leq k \leq i$,  with each $n_l \geq 1$.
\end{lemma}

A general result is that, for any matrix $\bfA$, there exists a sub-multiplicative matrix norm ($ \lVert \bfX \bfY \rVert \leq  \lVert \bfX \rVert  \lVert \bfY \rVert $) arbitrarily approaching the spectral radius $\Psi(\bfA)$, that is $ \lVert \bfA \rVert<\Psi(\bfA)+\epsilon$~\citep{Horn:1990}. Since $\Psi(\bfK\bfTheta)<1$, we can choose a sub-multiplicative matrix norm $\lVert \cdot \rVert_\lambda$ satisfying $\lVert\bfK\bfTheta\rVert_\lambda < 1$. As all matrix norms are equivalent, there also exists a constant $c$ satisfying $\lVert\cdot\rVert_2 \leq c \lVert\cdot \rVert_\lambda$~\citep{Horn:1990}. For $n \geq 1$, we have
\begin{equation*}
\begin{split}
\left|\bfu^T\bfK_n \bfu\right| &=\left|\bfu^T\left(\bfK\bfTheta\right)^{n-1}  \bfK \bfu\right| \\
&\leq  \lVert\bfu\rVert_2 \cdot \lVert\left(\bfK\bfTheta\right)^{n-1}\rVert_2 \cdot \lVert\bfK\rVert_2\lVert \bfu\rVert_2 \\
&\leq c\lVert\left(\bfK\bfTheta\right)^{n-1}\rVert_\lambda\lVert\bfK\rVert_2\\
&\leq c\lVert\bfK\rVert_2\lVert\bfK\bfTheta\rVert_\lambda^{n-1}.
\end{split}
\end{equation*}

Thus  $\sum_{n=1}^\infty \bfu^T\bfK_n\bfu$ converges absolutely, as it is bounded above in absolute value by a convergent geometric series. Since $\bfu^T \bfK_n \bfu = \bfu^T\bfK_n^T\bfu$, this also implies the convergence of $\sum_{m=1}^\infty \bfu^T\bfK_m^T\bfu$. Absolute convergence of $\sum_{n,m=1}^\infty \bfu^T\bfK_n\bfTheta\bfK_m^T\bfu$ is proved similarly by noting that for $n,m\geq 1$,
\begin{equation*}
\begin{split}
\left|\bfu^T\bfK_n\bfTheta\bfK_m^T\bfu\right| &= \left|\bfu^T \left(\bfK\bfTheta\right)^{n}\bfK^T\left(\bfTheta\bfK^T\right)^{m-1}\bfu\right|\\
&\leq \lVert\bfu\rVert_2\lVert\left(\bfK\bfTheta\right)^n\rVert_2 \lVert\bfK^T\rVert_2 \lVert\left(\bfK\bfTheta\right)^{(m-1)T}\rVert_2\lVert \bfu\rVert_2\\
&\leq c^2\lVert\bfK\rVert_2 \lVert\left(\bfK\bfTheta\right)^n \rVert_\lambda   \rVert\left(\bfK\bfTheta\right)^{m-1}\rVert_\lambda\\
&\leq c^2\lVert\bfK\rVert_2 \lVert\bfK\bfTheta\rVert_\lambda^{n+m-1}.
\end{split}
\end{equation*}

Now, suppose that $\left|\sum_{n=1}^\infty \bfu^T\bfK_n\bfu\right| < 1$. We will remove this assumption shortly.
We can now expand the right hand side of Eq.~\eqref{nrs_eq} as
\beq \label{nrs_expan}
\left[ \sum_{i=0}^{\infty}\left(\sum_{n=1}^{\infty}  \bfu^T \bfK_n \bfu\right)^i\right] \left(1+\sum_{i,j\ge 1}^{\infty} \bfu^T \bfK_i\bfTheta \bfK_j^T \bfu \right) \left[\sum_{j=0}^{\infty}\left(\sum_{m=1}^{\infty}\bfu^T \bfK_m^T \bfu\right)^j\right].
\eeq
Using Lemma~\ref{L:lemma}, we have that
\begin{equation}\label{E:nrs_expan_2}
\begin{split}
\eqref{nrs_expan}&=
\left[\sum_{i=0}^{\infty}\left(\sum_{n=1}^{\infty}  \bfu^T \bfK_n \bfu\right)^i \right] \left[ \sum_{j=0}^{\infty}\left(\sum_{n=1}^{\infty}\bfu^T \bfK_n^T \bfu\right)^j\right]\\
&\quad+\left[ \sum_{i=0}^{\infty}\left(\sum_{n=1}^{\infty}  \bfu^T \bfK_n \bfu\right)^i \right] \left(\sum_{i,j\ge 1}^{\infty} \bfu^T \bfK_i  \bfTheta \bfK_j^T \bfu \right) \left[ \sum_{j=0}^{\infty}\left(\sum_{n=1}^{\infty}\bfu^T \bfK_n^T \bfu\right)^j\right]\\
&=
\left[ 1+\sum_{i=1}^\infty \sum_{(n_1,\ldots,n_k)\in \{i\}} \left(\prod_{s=1}^k \bfu^T \bfK_{n_s} \bfu\right)\right]
\left[ 1+ \sum_{j=1}^\infty \sum_{(n_1,\ldots,n_l)\in \{j\}} \left(\prod_{t=1}^l \bfu^T \bfK_{n_t}^T \bfu\right)\right]\\
&\quad+\sum_{i,j=1}^\infty \sum_{\substack{(n_1,\ldots,n_{k+1})\in \{i\} \\ (m_1,\ldots,m_{l+1})\in \{j\}}}  \left(\prod_{s=1}^k  \bfu^T \bfK_{n_s} \bfu \right)  \bfu^T \bfK_{n_{k+1}}\bfTheta \bfK_{m_{l+1}}^T \bfu \left(\prod_{t=1}^l \bfu^T \bfK_{m_t}^T \bfu\right).
\end{split}
\end{equation}

Next, note that the product term $\prod_{s=1}^k \bfu^T \bfK_{n_s} \bfu$, where $(n_1,\ldots,n_k)\in \{i\}$, is acquired by distributing across sums $\bfH+\bfTheta$ in $\bfu^T\left[\bfK\left(\bfH+\bfTheta\right)\right]^{i-1}\bfK\bfu$ and taking the unique term in this where factors of $\bfH = \bfu \bfu^T$ divide the $i$ factors of $\bfK$ in to $k$ blocks $\bfK_{n_s}$ of size $n_s$ joined by factors of $\bfTheta$. Therefore, summing over all possible ordered partitions, we have
\begin{equation}\label{E:prop_r1}
\bfu^T \bfK^{i} \bfu=\bfu^T(\bfK(\bfH+\bfTheta))^{i-1}\bfK \bfu  = \sum_{(n_1,\ldots,n_k)\in \{i\}} \left(\prod_{s=1}^k \bfu^T \bfK_{n_s} \bfu\right).
\end{equation}
Similarly,
\begin{equation}\label{E:prop_r2}
\begin{split}
\bfu^T\bfK^{i} \bfTheta (\bfK^T)^{j}\bfu  &= \bfu^T(\bfK(\bfH+\bfTheta))^{i-1}\bfK \bfTheta \bfK^T ((\bfH+\bfTheta)\bfK^T)^{j-1}\bfu\\
&=\sum_{\substack{(n_1,\ldots,n_{k+1})\in \{i\} \\ (m_1,\ldots,m_{l+1})\in \{j\}}}  \left(\prod_{s=1}^k  \bfu^T \bfK_{n_s} \bfu \right)  \bfu^T \bfK_{n_{k+1}}\bfTheta \bfK_{m_{l+1}}^T \bfu \left(\prod_{t=1}^l \bfu^T \bfK_{m_t}^T \bfu\right).
\end{split}
\end{equation}
Using Eqns.~(\ref{E:nrs_expan_2}-\ref{E:prop_r2}), we have that
\begin{equation*}
\begin{split}
\eqref{nrs_expan}&=
\left(\sum_{i=0}^{\infty} \bfu^T \bfK^i \bfu \right) \left(\sum_{j=0}^{\infty}\bfu^T (\bfK^T)^j \bfu \right)
+
\sum_{i,j\ge 1}^{\infty} \bfu^T \bfK^i \bfTheta (\bfK^T)^{j} \bfu   \\
&=\sum_{i,j\ge 0}^{\infty} \bfu^T \bfK^i \bfH (\bfK^T)^j \bfu
+
\sum_{i,j\ge 0}^{\infty} \bfu^T \bfK^i \bfTheta (\bfK^T)^{j} \bfu   \\
&=
\sum_{i,j\geq 0}^\infty \bfu^T\bfK^i\left(\bfK^T\right)^j\bfu \qquad \text{using $ \ \bfI = \bfH+\bfTheta$}  \\
&=
\bfu^T(\bfI-\bfK)^{-1}(\bfI-\bfK^T)^{-1} \bfu,		\qquad \text{using $\ \Psi(\bfK) < 1$},
\end{split}
\end{equation*}
where we have used $\bfu^T  \bfTheta (\bfK^T)^{j} \bfu = \bfu^T \bfK^i \bfTheta \bfu = 0$
on the second line of the equation.

Lastly, we will eliminate the assumption $\left|\sum_{n=1}^\infty \bfu^T \bfK_n \bfu \right|<1$, which was used to establish Eq.~\eqref{nrs_eq}. To see that this can be done, let $z$ be a complex number, and replace $\bfK$ by $z\bfK$ in Eq.~\eqref{nrs_eq}, giving
\begin{equation} \label{E:nrs_eq_z}
\begin{split}
 \bfu^T&(\bfI-z\bfK)^{-1}(\bfI-z\bfK^T)^{-1} \bfu\\
 &= \left(1 - \sum_{n=1}^\infty  z^n \bfu^T \bfK_n \bfu \right)^{-1}\left(1 +  \sum_{n,m=1}^\infty z^{n+m} \bfu^T \bfK_n \bfTheta \bfK_m^T \bfu \right)
\left(1 - \sum_{m=1}^\infty z^m \bfu^T \bfK_m^{T} \bfu \right)^{-1}.
\end{split}
\end{equation}
For sufficiently small $0<\delta<1$, $|z|<\delta$ we have that
\begin{equation*}
\left| \sum_{n=1}^\infty z^n  \bfu^T \bfK_n \bfu\right| \leq \delta \sum_{n=1}^\infty \left| \bfu^T \bfK_n \bfu \right| <1.
\end{equation*}
The absolute convergence of the series $\sum_{n=1}^\infty  \bfu^T \bfK_n \bfu$, $\sum_{n=1}^\infty  \bfu^T \bfK_n \bfu$ and $\sum_{n,m=1}^\infty \bfu^T \bfK_n \bfTheta \bfK_m^T \bfu$ then guarantees that Eq.~\eqref{E:nrs_eq_z} holds on $|z|<\delta$, and furthermore, that the right hand side of \eqref{E:nrs_eq_z} is an analytic function of $z$ on $|z|<\delta$.

Finally, note that the left hand side of \eqref{E:nrs_eq_z} is an analytic function of $z$ on $|z|<1/ \Psi(\bfK)$, since the matrix inverses may be expanded as power series on this range. Since the two sides are equal and analytic on $|z|<\delta$, the right hand side must also be defined and analytic on $|z|<1/ \Psi(\bfK)$, and equal the left hand side on this range. Since $1/ \Psi(\bfK)>1$, we may take $z=1$ in Eq.~\eqref{E:nrs_eq_z}, recovering Eq.~\eqref{nrs_eq}.

\section{Proof of Proposition~\ref{EI_full}}
\label{s:prop_proof_2}

The proof of Proposition~\ref{EI_full} is identical to the one dimensional case Prop.~\ref{P:E_full}, except that we need an entrywise argument to eliminate the assumption $\lVert\sum_{n=1}^\infty  \bfU^T \bfK_n \bfU\rVert_2<1$ that enables the expansion of the right hand side of Eq.~\eqref{EI_nrs_eq}. 
Let $z$ be a complex number, and replace $\bfK$ by $z\bfK$ in Eq.~\eqref{EI_nrs_eq}, giving
\begin{equation}\label{EI_nrs_eq_z}
\begin{split}
\bfU^T&(\bfI-z\bfK)^{-1}(\bfI-z\bfK^T)^{-1} \bfU\\
&= \left(\bfI - \sum_{n=1}^\infty z^n  \bfU^T \bfK_n \bfU \right)^{-1}\left(\bfI +  \sum_{n,m=1}^\infty z^{n+m} \bfU^T \bfK_n \bfTheta \bfK_m^T \bfU \right)
\left(\bfI - \sum_{m=1}^\infty z^m \bfU^T \bfK_m^{T} \bfU \right)^{-1}.
\end{split}
\end{equation}
Note that the series $\sum_{n=1}^\infty  \bfU^T \bfK_n \bfU$ absolutely converges in 2-norm. This follows from writing
$$
\lVert \bfU^T \bfK_n \bfU \rVert_2 
\leq \lVert \bfU^T\rVert_2 \left\lVert \left(\bfK\bfTheta\right)^{n-1} \right\rVert_2 \lVert \bfK\rVert_2 \lVert \bfU\rVert_2
\leq c  \lVert \bfU^T\rVert_2  \lVert \bfK\bfTheta\rVert^{n-1}_\lambda \lVert \bfK\rVert_2 \lVert \bfU\rVert_2,
$$
and use the assumption that $\Psi\left(\bfK\bfTheta\right)<1$ and one associated sub-multiplicative norm $\lVert \bfK \bfTheta \rVert_{\lambda}<1$. $c$ is a constant such that $\lVert \cdot \rVert_2  \leq c \lVert \cdot \rVert_\lambda$.
For sufficiently small $0<\delta<1$, $\forall |z|<\delta$, we have that
$$
\lVert \sum_{n=1}^\infty z^n  \bfU^T \bfK_n \bfU\rVert_2
\leq \delta\sum_{n=1}^\infty  \lVert\bfU^T \bfK_n \bfU\rVert_2  <  1.
$$
With this condition, we can establish identity \eqref{EI_nrs_eq_z} in the same way as in the one dimensional case Prop.~\ref{P:E_full} for $|z|<\delta$.
Furthermore, it is similar as above to show all of the matrix series on the right hand side of Eq.~\eqref{EI_nrs_eq_z} are absolute converge in 2-norm and therefore entry-wise absolute converge. Hence, we have that
each entry of the right hand side of Eq.~\eqref{EI_nrs_eq_z} is an analytic function of $z$ on $|z|<\delta$.

On the other hand, the left hand side of Eq.~\eqref{EI_nrs_eq_z} is entry-wise analytic in $z$ (each entry is indeed a rational function of $z$) on $|z|<1/ \Psi(\bfK)$. Since the two sides are equal and analytic on $|z|<\delta$, the right hand side must also be defined and analytic on $|z|<1/ \Psi(\bfK)$. Note that $1/ \Psi(\bfK)>1$, so we may set $z=1$, yielding Eq.~\eqref{EI_nrs_eq}.

\section{Expression of the linear dependence between $\langle \bfCt_{EE} \rangle$ and motifs}
\label{s:C_EE}

For excitatory-inhibitory networks, the (linearized) relationship between motif frequencies and averaged correlations is given by Eq.~\eqref{E:twopop_resummed_exp}.  Here, we evaluate the terms in this expression in a special case, to yield an explicit expression for the linear dependence of correlation on motif frequencies. 
We will express the block-wise average covariances $\langle \bfCt_{EE} \rangle$ in terms of the 20 individual second order motifs. Along with the (approximate) definition of correlation in Eq.~\eqref{rho_1}, this gives the linear weights. We note that this is how we obtain the specific linear weights used in Fig.~\ref{first figure} {\bf{B}}.

\par
For simplicity, assume the special case that all connection probabilities are identical ($p_{XY} =p$ for all $X,Y \in \{E,I\}$). Define the weight of all excitatory (resp. inhibitory) connections into a cell as
$$
\mu_E = pN_Ew_E \quad (\mathrm{resp. } \ \mu_I = pN_Iw_I)
$$
and the new weight of all connections into a cell as
$$
\mu = \mu_E +\mu_I.
$$
In addition, define $\eta$ to be the strength of total common input to a cell pair in an \ER{} network:
$$
\eta = N_Ew_E^2p^2 + N_Iw_I^2p^2.
$$
Noting that we have
$$
\left(\bfM\bfD_2\right)^2 = \mu \left(\bfM\bfD_2\right),
$$
it is simple to show that
$$
\left(\bfI - \At\bfM\bfD_2\right)^{-1} = \bfI + \frac{\At}{1 - \At\mu} \left(\bfM\bfD_2\right).
$$
Similarly
$$
\left(\bfI - \At\bfD_2\bfM\right)^{-1} = \bfI + \frac{\At}{1 - \At\mu} \left(\bfD_2\bfM\right).
$$
Then, from Eq.~\eqref{E:twopop_resummed_exp} we get the linear dependence on motifs (treat $p$ as fixed) as
\begin{equation}\label{E:twopop_resummed_EE}
\begin{split}
\langle \bfCt_{EE} \rangle/ \Ct^0 &\sim
\At^2\left(\frac{1-\At\mu_I}{1-\At\mu}\right)^2\Qdiv^{EE}
+\At^2\frac{\At\mu_I(1-\At\mu_I)}{(1-\At\mu)^2}\Qdiv^{EI}
+\At^2\left(\frac{\At\mu_I}{1-\At\mu}\right)^2\Qdiv^{II}\\
& +2\At^2 \frac{1-\At \mu_I}{1-\At \mu} \frac{(1-\At\mu_I)^2+\At^2\mu_I^2\frac{N_E}{N_I}}{(1-\At\mu)^2}\Qch^{EE}\\
&+2\At^2 \frac{1-\At \mu_I}{1-\At \mu} \frac{(1-\At\mu_E)\At\mu_I+(1-\At\mu_I)\At\mu_E\frac{N_I}{N_E}}{(1-\At\mu)^2}  \Qch^{EI}\\
& +2\At^2 \frac{\At\mu_I}{1-\At \mu} \frac{(1-\At\mu_I)^2+\At^2\mu_I^2\frac{N_E}{N_I}}{(1-\At\mu)^2} \Qch^{IE}\\
&+2\At^2 \frac{\At \mu_I}{1-\At \mu} \frac{(1-\At\mu_E)\At\mu_I+(1-\At\mu_I)\At\mu_E\frac{N_I}{N_E}}{(1-\At\mu)^2} \Qch^{II},
\end{split}
\end{equation}
where $\Qdiv^{XY},\Qch^{XY}$ were defined in Eqns.~(\ref{E:twopop_qdiv_def},\ref{E:twopop_qch_def}) as
$$
\Qdiv^{XY} = N_Ew_E^2 \qdiv^{XY,E} + N_Iw_I^2 \qdiv^{XY,I}, \qquad \Qch^{XY} = N_Ew_Xw_E \qch^{XEY} + N_Iw_Xw_I \qch^{XIY}.
$$
Since each of the quantities $\Qdiv^{XY},\Qch^{XY}$ are clearly linear in second order motif frequencies, Eq.~\eqref{E:twopop_resummed_EE} gives a linear relation between second order motif frequencies and
block-wise averaged correlation in the two population network. Eq.~\eqref{E:twopop_resummed_EE} is the two population analog of Eq.~\eqref{E:single_pop_expansion} in the single population case.

\section{Intuition for why the resumming approach can produce accurate results}
\label{s:spec_intui}
From Eq.~\eqref{E:resumming_motifs} and Eq.~\eqref{E:twopop_nl_resumming_all} , we see that the error of resumming theory is determined by the tail that we dropped in two series.  With respect to the coupling strength order of magnitude ($w$ or $w_E,w_I $), these are geometric series. Therefore the sum of the tail series is controlled by the leading term, that is
\beqn
\At^3 \bfL^T \bfW \bfTheta  \bfW \bfTheta  \bfW \bfL,~~\At^3 \bfL^T \bfW \bfTheta  \bfW^T \bfTheta  \bfW \bfL,~~\At^3 \bfL^T \bfW \bfTheta  \bfW \bfTheta  \bfW^T \bfL,~~
\eeqn
This shows that the resumming theory has third order accuracy in the effective interaction strength $\At w$.  

Another important factor affecting the accuracy of the resumming theory is the spectral radius $ \Psi(\bfTheta \bfW \bfTheta)= \Psi(\bfW \bfTheta)= \Psi(\bfTheta \bfW)$ (equalities follow from writing $\bfW$ under basis of projections $\bfTheta$ and $\bfH$), which is related to how fast the terms in the series converge to 0. There is a simple intuition of  $ \Psi(\bfW \bfTheta)$ for \ER{} networks.  For single population networks, note that (asymptotically, for large $N$) $\bfW^0\bfTheta=\bfW^0-\bfW^0\bfH $ is a matrix with i.i.d. entries of mean 0. According to the Circular Law, all eigenvalues will asymptotically uniformly distributed within a circle about the origin. Comparing with the spectra of $\bfW^0$ in Sec.~\ref{s: stability}, multiplying by $\bfTheta$ effectively removes the single dominant eigenvalue~\citep{Rajan:2006}.

The removal of this dominant eigenvalue will reduce the spectral radius of $\bfW^0\bfTheta$ as compared to $\bfW^0$ by a factor of $\sqrt{N}$ ($w\sqrt{p(1-p)N}$ compared to $pNw$, see Sec.~\ref{s: stability}). Such reductions of $\Psi(\bfW \bfTheta)$ approximately occur in single population networks and at the blocks of $\bfW\bfTheta$ in two population networks, even though those networks are non-\ER{}. This intuition may help understand why resumming theory works much better than truncation theory: the tails of series in $\bfW\bfTheta$ may be much lighter than those in $\bfW$.
\begin{figure}[H]
  \center
      \includegraphics[width=6in]{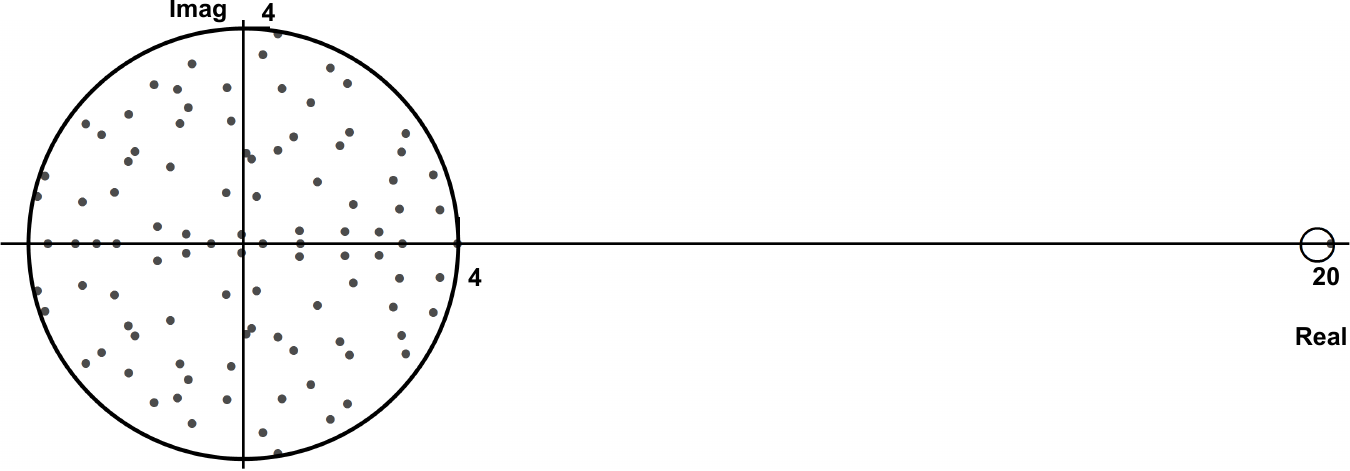}
       \caption{The spectra of single population \ER{} network, the larger circle and small circle on the right are expected locations of the bulk spectra and single eigenvalue calculated from the asymptotic formula in text above and Sec.~\ref{s: stability}. The excitatory only network has 100 neurons, and $p=0.2$, $w=1$.}
   \label{spec_ER}
\end{figure}

\begin{table}[H]
\centering
\begin{tabular}[h]{|l|p{5in}|}
\hline
{\bf Symbol} & {\bf Description}\\
\hline
$v_i, \tau_i, E_{L,i}, \sigma_i$ & Membrane potential, membrane time constant, leak reversal potential, and noise intensity of cell $i$.\smallskip\\

$E_i, \sigma_i$ & Mean and standard deviation of the background noise  for cell $i$. \smallskip\\

$v_{th},v_{r},\tau_{ref}$ & Membrane potential threshold, reset, and absolute refractory period for cells.\smallskip\\

$\psi(v),v_T,\Delta_T$ & Spike generating current, soft threshold and spike shape parameters for the IF model~\cite{FourcaudTrocme:2003}. \smallskip\\

$f_i(t), \eta_i(t)$ & Synaptic input from other cells in the network, and external input to cell $i$. \smallskip\\

$\tau_{S,i}, \tau_{D,i}$ & Synaptic time constant and delay for outputs of cell $i$. \smallskip\\

$y_i(t)$ & Spike train of cell $i$. \smallskip\\

$\bfW_{ij}$ & The $j \rightarrow i$ synaptic weight, proportional to the area under a single post-synaptic current for current-based synapses.\smallskip\\

$\bfJ_{ij}(t)$ & The $j\rightarrow i$ synaptic kernel - equals the product of the synaptic weight $\bfW_{ij}$ and the synaptic filter for outputs of cell $j$.\smallskip\\

$\bfC_{ij}(\tau)$ & The cross-correlation function between cells $i,j$ defined by $C_{ij}(\tau) = \cov(y_i(t+\tau),y_j(t))$.\smallskip\\

$\bfCt=\bfCt(0)$ & The cross-spectrum matrix evaluated at 0 frequency. Unless noted otherwise, all spectral quantities are evaluated at 0 frequency. 
\smallskip\\

$N_{y_i}(t,t+\tau),\rho_{ij}(\tau)$ & Spike count for cell $i$, and spike count correlation coefficient for cells $i,j$ over windows of length $\tau$.\smallskip\\

$r_i, A_i(t), \bfC^0_{ii}$ & Stationary rate, linear response kernel and uncoupled auto-correlation function for cell $i$j.\smallskip\\

$\bfK_{ij}(t)$ & The $j\rightarrow i$ interaction kernel - describes how the firing activity of cell $i$ is perturbed by an input spike from cell $j$. It is defined by $\bfK_{ij}(t) = (A_i * \bfJ_{ij})(t)$.\smallskip\\

$\bfy^n_i(t), \bfC^n_{ij}(t)$ & The $n^{th}$ order approximation of the activity of cell $i$ in a network which accounts for directed paths through the network graph up to length $n$ ending at cell $i$, and the cross-correlation between the $n^{th}$ order approximations of the activity of cells $i,j$.\smallskip\\

$g(t), \tilde{g}(\omega)$ & $\tilde{g}(\omega)$ is the Fourier transform of $g(t)$ with the convention
$$\tilde{g}(\omega) = \mathcal{F}[g](\omega) \equiv \int_{-\infty}^\infty e^{-2 \pi i \omega t}g(t)dt$$ \smallskip\\

$\EVE{\cdot} $& Empirical average, $\frac{1}{N^2}\sum_{i,j}\cdot \text{ or } \frac{1}{N^3}\sum_{i,j,k}\cdot$ depending on context \smallskip\\
$R^2$& coefficient of determination, i.e. the square of correlation coefficient $\rho^2$\smallskip\\
$\bfW^0$ & Adjacency matrix \smallskip\\
$\bfH$ & $\bfH = \frac{1}{N}\bfone_{NN}$ \smallskip\\
$\bfTheta$ & $\bfTheta = \bfI - \bfH$ \smallskip\\

\hline
\end{tabular}
\caption{{\bf Notation used in the text.}}
\label{T:notation}
\end{table}

\section*{Acknowledgements}
We thank Chris Hoffman and Brent Doiron for their helpful insights.  This work was supported by NSF grants DMS-0817649, DMS-1122094, and a Texas ARP/ATP award to KJ, and by a Career Award at the Scientific Interface from the Burroughs Wellcome Fund and NSF Grants DMS-1056125 and DMS-0818153 to ESB.

\bibliographystyle{plainnat}


\end{document}